\documentclass[10pt]{article}  
\usepackage{graphicx}   
\usepackage[a4paper,left=3.0cm,right=3.0cm, bottom=3.0cm, top=3cm]{geometry}
\usepackage{apacite}
\usepackage{ragged2e}
\usepackage{amsmath}
\usepackage{gensymb}

\let\citeA\shortciteA
\let\cite\shortcite
\def\notation{\bgroup
\section*{Notation}
\description
\def\notation##1{\item[\boldmath ##1]}}
\def\endnotation{\enddescription\vskip12pt\egroup}

\usepackage{xcolor}
\newcommand{\MPa}{~\text{MPa}}

\newcommand{\persec}{~\text{s}^{-1}}
\newcommand{\ie}{i.e.}
\newcommand{\eg}{e.g.}
\usepackage{xurl}

\begin{document}

\begin{center}
\LARGE {\bf A brittle constitutive law for long-term tectonic modeling based on sub-critical crack growth\\[12pt]}
\large
Léo Petit$^{1}$, Jean-Arthur Olive$^{1,*}$, Alexandre Schubnel$^{1}$, Laetitia Le Pourhiet$^{2}$, Harsha S. Bhat$^{1}$\\[12pt]

\begin{enumerate}
\small
\setlength\itemsep{0.01em}
\item Laboratoire de G\'eologie, CNRS - \'Ecole normale supérieure - PSL University, Paris, France\\
\item Sorbonne Universit\'e, ISTEP, Paris, France
\item[\large *] Corresponding Author: Jean-Arthur Olive~~olive@geologie.ens.fr
\end{enumerate}
\end{center}

{\textit{\color{gray}{\small Manuscript to appear in Geochemistry, Geophysics, Geosystems}}}

\section*{Key Points:}
\begin{itemize}
\setlength\itemsep{1em}
\item New brittle constitutive law describes the onset of faulting in tectonic simulations.
\item Model is based on sub-critical growth and interaction of micro-cracks.
\item Laboratory-derived model parameters can be used to model crustal-scale faulting.
\end{itemize}

\section*{Abstract}
Adequate representations of brittle deformation (fracturing and faulting) are essential ingredients of long-term tectonic simulations. Such models commonly rely on Mohr-Coulomb plasticity coupled with prescribed softening of cohesion and/or friction with accumulated plastic strain. This approach captures fundamental properties of brittle failure, but is overly sensitive to empirical softening parameters that cannot be determined experimentally.
Here we design a brittle constitutive law that captures key processes of brittle deformation, and can be straightforwardly implemented in standard geodynamic models. In our Sub-Critically-Altered Maxwell (SCAM) flow law, brittle failure begins with the accumulation of distributed brittle damage, which represents the sub-critical lengthening of tensile micro-cracks prompted by slip on pre-existing shear defects. Damage progressively and permanently weakens the rock’s elastic moduli, until cracks catastrophically interact and coalesce up to macroscopic failure. The model’s micromechanical parameters can be fully calibrated against rock deformation experiments, alleviating the need for ad-hoc softening parameters.
Upon implementing the SCAM flow law in 2-D plane strain simulations of rock deformation experiments, we find that it can produce Coulomb-oriented shear bands which originate as damage bands. SCAM models can also be used to extrapolate rock strength from laboratory to tectonic strain rates, and nuance the use of Byerlee's law as an upper bound on lithosphere stresses. We further show that SCAM models can be upscaled to simulate tectonic deformation of a 10-km thick brittle plate over millions of years. These features make the SCAM rheology a promising tool to further investigate the complexity of brittle behavior across scales.
\clearpage

\section{Introduction}
Tectonic plates tend to be almost rigid and primarily deform within narrow boundary zones. In the upper crust (above $\sim 15$ km depth), deformation occurs in the brittle regime through nucleation and growth of fractures and faults, which profoundly affect the shape of geological structures and planetary topography. Accurate descriptions of brittle deformation processes are therefore key to answer fundamental questions such as: How and when does a new fault break? How long can it stay active and under what conditions can tectonic stresses reactivate previously active faults? Which mechanisms promote brittle strain localization and modulate off-fault deformation?

\begin{figure}[t!]
\noindent\includegraphics[width=\textwidth]{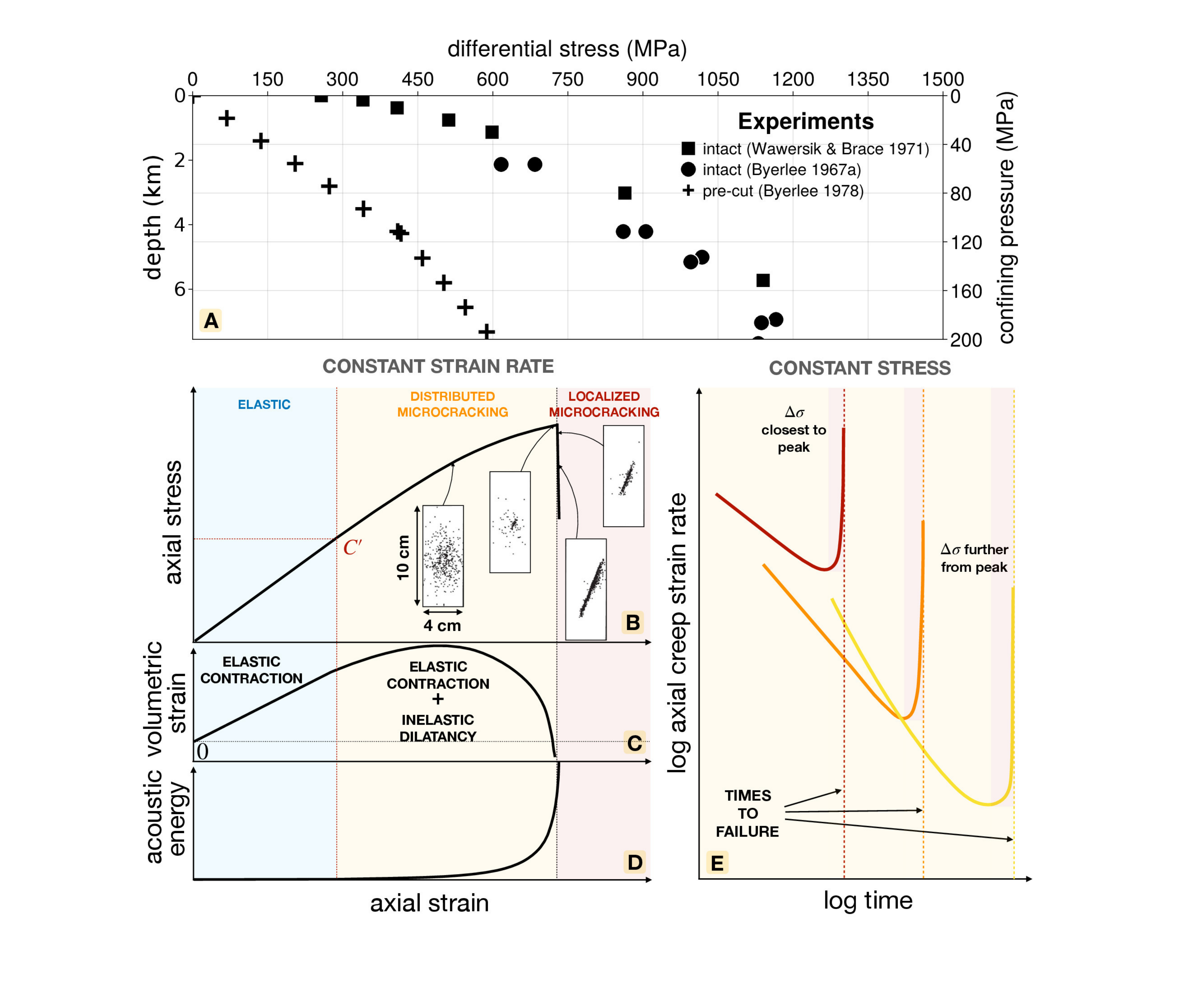}
\caption{\textbf{A.} Experimental constraints on brittle rock strength as a function of increasing pressure for intact (squares and circles) and pre-cut (crosses) samples. 
\textbf{B–D.}Schematic illustration of axial stress, volumetric strain and acoustic energy vs. axial strain, in triaxial experiments performed at a constant strain rate. Point $C'$ marks the onset of dilatancy. Sample cross sections showing the spatial pattern of acoustic emissions are reproduced from \protect\citeA{LocknerEtAl1991}. \textbf{E.} Typical pattern of axial strain vs. time in a brittle creep experiment in which the differential stress $\Delta \sigma$ is imposed and maintained. Colors correspond to values of $\Delta \sigma$ close to (red) or far from (yellow) the rock's peak strength. Here compressive axial strains, strain rates and stresses are plotted as positive numbers for clarity.
}
\label{exp}
\end{figure}

Laboratory experiments have long been used to learn about rock deformation mechanisms in the brittle regime \cite{PatersonWong2005}. The brittle behavior of low-porosity crustal rocks (Figure \ref{exp}) has some defining characteristics. First and foremost, the differential stress that must be applied to break a rock (the rock's strength) increases with pressure \cite{Byerlee1967} (Figure \ref{exp}A, squares and circles). The stress required to slip on a pre-existing discontinuity is also pressure dependent, and both stresses weakly depend on lithology \cite{Byerlee1978}. The contrast between these two stresses (intact vs. pre-cut) is typically on the order of hundreds of MPas (Figure \ref{exp}A). Experiments further reveal a number of phenomena that precede macroscopic failure of a rock sample, such as: a reduction of effective elastic moduli, volume expansion, and acoustic emissions (Figure \ref{exp}B-D). Failure is a catastrophic phenomenon that occurs when stresses reach a peak strength which is greater when the imposed strain rate is faster \cite<e.g.,>{Lockner1998,PatersonWong2005}. Failure manifests as a transition from distributed to localized strain along macroscopic fractures oriented in a systematic manner with respect to the stress field.  
It is also well documented that rocks can creep when subjected to a constant stress below their peak strength \cite<e.g.,>[Figure\ref{exp}D]{Kranz1979,CarterEtAl1981,BaudMeredith1997,HeapEtAl2009, BrantutEtAl2013}. Such brittle creep is typically described as involving three phases: A first phase (primary creep) where strain rate decelerates, a prolonged second phase (secondary creep), during which creep rate remains nearly steady, and a final stage (tertiary creep) when deformation accelerates until macroscopic failure (Figure \ref{exp}E, note that the log time representation does not convey the long duration of the secondary phase).

This seemingly complex phenomenology is reasonably well understood as the macroscopic manifestation of the growth and interaction of microcracks that nucleated on pre-existing defects \cite{TapponnierBrace1976}. Crack growth first occurs in a distributed fashion across the sample (Figure \ref{exp}B). Macroscopic failure then results from the sudden coalescence of interacting microcracks (Figure \ref{exp}B), whose growth is enabled by differential stress \cite<e.g.,>{LocknerEtAl1991,Lockner1998, McBeckEtAl2019}. Sample dilatancy points to the tensile nature (mode-I) of some of these cracks, which are susceptible to radiate acoustic energy as they grow (Figure \ref{exp}D). The time and strain rate dependence of these phenomena further suggests that the speed of crack propagation in the bulk rock depends on the forces acting at crack tips, which is typical of sub-critical crack growth processes. The main underlying mechanism in the brittle regime is known as stress corrosion \cite{Atkinson1984}. It refers to reactions occurring between a chemically active fluid and the strained atomic bonds at the tip of microcracks, which induce stress-dependent kinetics of bond breaking \cite{Eppes2017}. While other mechanisms such as pressure-solution \cite<e.g.,>{GratierEtAl2013}, can also contribute to rate-dependent deformation as pressure increases, sub-critical crack growth has been identified as a key contributor to the strain rate (i.e., time-) dependent behavior of brittle rocks in the brittle regime that is particularly well highlighted by brittle creep experiments \cite{BrantutEtAl2013}.

\begin{figure}[t!]
\noindent\includegraphics[width=\textwidth]{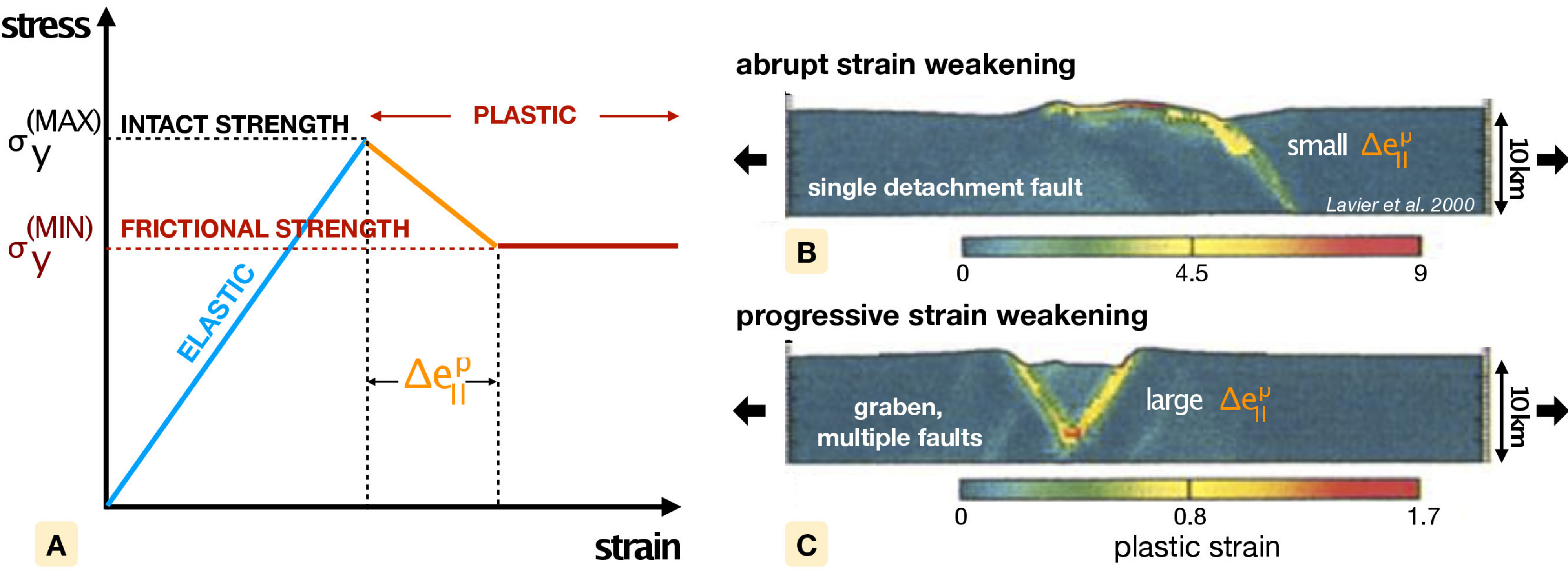} 
\caption{\textbf{A.} Schematic representation of an elastic-plastic rheology with strain weakening, under constant applied strain rate. The difference between initial ($\sigma_y^{(MAX)}$) and final ($\sigma_y^{(MIN)}$) yield stresses are caused by a prescribed decrease of frictional properties $\mu$ and $C$ over a specified amount of plastic strain $\Delta e^P_{II}$. \textbf{B}–\textbf{C.} Example simulation of extension in a 10 km-thick elastic-plastic upper crust overlying an inviscid medium \protect\cite{LavierEtAl2000}. The only difference between the two panels is the choice of $\Delta e^P_{II}$, which is small in \textbf{B}, producing a large-offset normal fault (detachment), and large in \textbf{C}, producing two conjugate faults that outline a graben structure. In this example the difference between $\sigma_y^{(MAX)}$ and $\sigma_y^{(MIN)}$ is caused by a drop in material cohesion while friction is kept constant.}
\label{lavier}
\end{figure}

Though the phenomenology of brittle failure was well known long before geodynamicists harnessed the power of microprocessors, most tectonic simulations currently rely on a simplified treatment that consists in capping stresses at a rate-independent Mohr-Coulomb yield envelope \cite<e.g.,>{PoliakovBuck1998,Gerya2010}. This has the advantage of being numerically efficient, adequately capturing the pressure-dependent frictional strength of pre-cut rocks, and spontaneously localizing plastic strain through the bifurcative properties of the Mohr-Coulomb plastic flow rule \cite<e.g.,>{RudnickiRice1975, VermeerDeBorst1984, LemialeEtAl2008, Kaus2010}. In this framework, strain localization is typically accompanied by a rotation of the principal stresses inside the incipient shear band, which leads to a reduction of the remote stresses \cite{lepourhiet2013}.
By itself, this rotation-induced "structural" softening does not account for the $100$s of MPas that separate the strength of intact rocks from their residual strength once faulted (Figure \ref{exp}A). An approach commonly used to promote sustained strain localization in tectonic simulations (Figure \ref{lavier}) is to weaken the material friction $\mu$ and cohesion $C$, from ${\{\mu_{max},C_{max}\}}$ to ${\{\mu_{min},C_{min}\}}$ over a certain amount of non-recoverable (plastic) strain $\Delta e_{II}^p$ \cite<e.g.,>[Figure \ref{lavier}A]{PoliakovBuck1998, LavierEtAl2000}. 
This amounts to enforcing a contrast between intact and broken rocks reminiscent of the strength contrast observed experimentally.

Strain-weakened Mohr-Coulomb plasticity however presents several drawbacks. This parameterization typically ignores the strain rate dependence of rocks' intact strength, and relies on a single value of intact friction and cohesion to determine the intact yield strength. Further, the critical plastic strain $\Delta e_{II}^p$ is meant to represent a wide range of possible weakening mechanisms, and is therefore not easily quantified through laboratory experiments. These limitations can be problematic since the choice of weakening parameters can have major consequences on the outcome of a tectonic simulation. \citeA{LavierEtAl2000} for example pointed out the spectacular effect of $\Delta e_{II}^p$ on tectonic styles produced in a rifting simulation (Figure \ref{lavier}B vs. \ref{lavier}C). While some recent studies have investigated the effects of various weakening parameterizations (e.g., \citeA{DuretzEtAl2021,Meyer2017,Naliboff2020,Pan2023}), it remains common practice to rely on ad-hoc softening rules in geodynamic simulations without assessing their impact on model behavior.

One path toward remedying this issue is to improve the way geodynamic simulations parameterize the transition from intact to broken rock, in a manner that allows more direct comparison with experimental data and can be interpreted in terms of underlying deformation mechanisms. An adequate parameterization of progressive brittle failure should indeed account for standard observations such as the pre-peak reduction in elastic moduli, the evolving spatial pattern of acoustic emissions, or sample dilatancy which ceases upon failure (Figure \ref{exp}B–D). It should also account for the strain rate dependence of brittle yielding and the occurrence of brittle creep. Finally, it should include a representation of the ever-evolving internal state of the rock to include a memory of past deformation events. A promising alternative is to turn to models that describe brittle yielding as the accumulation of damage which ultimately leads to macroscopic failure.

A first family of such models are Continuum Damage Mechanics models. They treat failure as a progressive phenomenon indexed on the alteration of a rock's internal state (damage), and can produce strain rate-dependent brittle strengths, as well as pre-peak softening. Some are built on thermodynamic descriptions of energy dissipation during inelastic deformation \cite<e.g.,>[]{LyakhovskyEtAl1997, HamielEtAl2004, KarrechEtAl2011a}, others simply index damage growth on excess stresses above a yield stress, and strain \cite<e.g.,>{ManakerEtAl2006}. They do not assume a specific microstructure, which makes them flexible but also not directly interpretable in terms of deformation processes.

In that regard, micromechanics-based models have been particularly successful at capturing the broad range of behaviors associated with brittle deformation \cite{PatersonWong2005}. In this family of damage models, assumptions about the distribution and geometry of pre-existing defects in the material allow the analytical determination of stress concentrations around them, using linear elastic fracture mechanics.
Motion along defects cause the stress intensity factors (i.e., a measure of the stress state at the edge of discontinuities) at their tips to increase up to the fracture toughness of the rock, allowing tensile crack propagation. Drivers of such stress heterogeneities can be planar flaws such as grain boundaries, pre-existing microcracks \cite <e.g.,>[]{Kachanov1982a, Kachanov1982b, Nemat-NasserHorii1982, AshbySammis1990}, pores \cite{SammisAshby1986}, moduli contrasts across grains in contact \cite{DeyWang1981}, or can even be left undetermined \cite<\eg, >{Costin1985}.
Tensile cracks, in turn, alter the effective elastic properties of the rock as they lengthen, in an anisotropic fashion \cite{Walsh1965a,Walsh1965ba,BudianskyOconnell1976, Kachanov1993, DeshpandeEvans2008}.
This framework has been used to model high strain rate deformation \cite<e.g., during seismic rupture, >{BhatEtAl2012,ThomasEtAl2017} assuming critical fracture propagation, as well as slow deformation assuming sub-critical crack growth \cite{Kachanov1982c}. The latter class of models has also been used to describe brittle creep, assuming pre-existing planar defects \cite{BrantutEtAl2012}, successfully accounting for the multi-phased dynamics of brittle creep (Figure \ref{exp}E).

One drawback of this approach is its computational cost, because it requires to accurately resolve the kinetics of fracture lengthening, which crack interactions ultimately render unstable close to macroscopic failure. This may explain why it has not yet been implemented in long-term, large scale tectonic simulations, even though the processes it describes are clearly central to the initiation and evolution of crustal faults. By representing specific deformation mechanisms that can be studied in the laboratory, these models can indeed be calibrated against experiments and need not resort to ad-hoc macroscopic parameters \cite<e.g.,>{Costin1983,Costin1985, BhatEtAl2011,BrantutEtAl2012}.

As a first step in this direction, this study aims at constructing a constitutive brittle rheology rooted in the subcritical growth of microcracks from pre-existing rock defects. We seek a formulation that (1) captures the essence of brittle rock behavior at the expense of a few simplifications, (2) has a straigthforward micromechanical interpretation, (3) can be calibrated against experimental data, and (4) is usable in standard 2-D plane strain numerical geodynamic models. We propose such a constitutive law in Section \ref{sec:scam}, and describe its fundamental behavior in terms of stress-strain curves in Section \ref{sec:triaxial_application}. This allows us to calibrate its parameters using experimental data from both constant strain rate and brittle creep tests. We then implement our constitutive law in 2-D plane strain numerical simulations that reproduce experimental conditions (Section \ref{sec:2-D_application}), and discuss the model's key features in Section \ref{sec:discussion}. Finally, we implement our constitutive law in a crustal-scale tectonic simulator and compare it to the standard elasto-plastic approach (Section \ref{sec:tecto}).

\section{A Sub-Critically Altered Maxwell (SCAM) constitutive law for brittle deformation}
\label{sec:scam}

\begin{notation}
    \notation{Mohr-Coulomb plasticity}
    \notation{$\mu$} friction coefficient
    \notation{$\phi$} ($=\arctan{\mu}$) friction angle on shear defects ($\phi_m$ at macroscopic scale)
    \notation{$C_m$} (macroscopic) cohesion 
    \notation{$\sigma_y$} yield stress
    \notation{$\sigma_y^{(max)}$} intact plastic yield stress (determined by $\mu_{max}$ and $C_{max}$)
    \notation{$\mu_{max}$} initial friction coefficient (in strain weakened Mohr-Coulomb plasticity)
    \notation{$C_{max}$} initial cohesion (in strain weakened Mohr-Coulomb plasticity)
    \notation{$\sigma_y^{(min)}$} fully weakened plastic yield stress (determined by $\mu_{min}$ and $C_{min}$)
    \notation{$\mu_{min}$} fully weakened friction coefficient
    \notation{$C_{min}$} fully weakened cohesion
    \notation{$\Delta e_{II}^p$} accumulated plastic strain needed to fully weaken the frictional properties
    
    \vspace{0.5cm}
    \notation{Damage mechanics}
    \notation{$D$} $\in [D_0,1]$ damage internal state variable
    \notation{$D_{0}$} $\in [0,1]$ damage value corresponding to no tensile defect in the rock
    \notation{$D_i$} initial damage
    \notation{$D_c$} critical damage at the transition between the isolated crack regime and the interacting crack regime
    \notation{$\gamma$} $=\text{f}(D=1)$ ratio of residual over reference shear modulus
    \notation{$N_v$} number of shear defects per unit volume
    \notation{$V_c$} characteristic volume per crack ($1/N_v$)
    \notation{$A_c$} characteristic area per crack
    \notation{$A_b$} average area that separates neighboring cracks (bulk area in the ($\sigma_1$, $\sigma_2$) plane)
    \notation{$\psi$} shear defect angle with respect to $\sigma_1$
    \notation{$\alpha$} $\cos{\psi}$
    \notation{$a$} shear defect radius
    \notation{$l$} tensile "wing" crack length
    \notation{$K_I$} mode I stress intensity factor
    \notation{$K_I^{(w)}$} mode I stress intensity factor due to the wedging force $F_w$
    \notation{$K_I^{(\sigma_3)}$} mode I stress intensity factor due to $\sigma_3$
    \notation{$K_I^{(i)}$} mode I stress intensity factor due to interactions between cracks
    \notation{$\sigma_3^{i}$} internal stress acting in the direction of $\sigma_3$ resulting from cracks interaction
    \notation{$K_{IC}$} mode I fracture toughness 
    \notation{$\dot{l}_0$} characteristic crack growth rate
    \notation{$n$} Charles law exponent (corrosion index)
    \notation{$\beta$} geometric regularization factor
    \notation{$A_1$, $A_3$} constants that depend on friction and the orientation of shear defects
    
    \vspace{0.5cm}
    \notation{Stresses and strains}
    \notation{$\varepsilon_{ij}$} strain tensor
    \notation{$\sigma_1$} most compressive principal stress
    \notation{$\sigma_3$} least compressive principal stress
    \notation{$\Delta\sigma$} $ = \sigma_3-\sigma_1$, differential stress
    \notation{$\Delta\sigma^c$} differential stress at $K_I=0$ and $D=D_0$
    \notation{$\Delta\sigma_{bc}$} Minimum brittle strength
    \notation{$P$} $= - \sigma_{kk}/3$ pressure
    \notation{$p_c$} $=-\sigma_3$ confining pressure in experiments
    \notation{$e_{ij}$} deviatoric strain tensor
    \notation{$s_{ij}$} deviatoric stress tensor
    \notation{$e_{ax}$} deviatoric axial strain
    \notation{$s_{ax}$} deviatoric axial stress
    \notation{$J_2(X)$} $=(\text{dev}({X}_{ij})\text{dev}({X}_{ij}))/2$ second invariant of the deviator of second order tensor $X$
    \notation{$s_{II}$} $=\sqrt{J_2(\sigma_{ij})}$ scalar shear stress magnitude
    \notation{$s_{ij}^y$} deviatoric stress tensor satisfying the Mohr-Coulomb yield criterion
    \notation{$e_{II}$} $=\sqrt{J_2(\varepsilon_{ij})}$ scalar shear strain magnitude
    \notation{$\varepsilon^p_{ij}$} plastic strain tensor
    
    \vspace{0.5cm}
    \notation{Additional notations}
    \notation{$G$} effective shear modulus
    \notation{$G_0$} $= G(D=D_0)$ reference shear modulus corresponding to the lowest damage state (no tensile defect)
    \notation{$\text{f}(D)$} weakening function
    \notation{$\nu$} Poisson's ratio
    \notation{$\eta_{D}$} damage viscosity
    \notation{$\eta_p$} plastic viscosity (2-D SCAM simulations)
    \notation{$\eta_{eff}$} effective viscosity (2-D SCAM simulations)
    \notation{$\eta_{min}$} minimum viscosity (2-D SCAM simulations)
    \notation{$\eta_{max}$} maximum viscosity (2-D SCAM simulations)
    \notation{$v_i$} Components of the velocity field
    \notation{$\rho$} density
    \notation{$g_i$} Components of the gravity field
    \notation{$\Delta t$} time step
    
\end{notation}

\subsection{Generic stress-strain relation}

Our constitutive model builds upon an isotropic, incompressible elastic stress-strain relationship : 

\begin{equation}
\label{elastic}   
e_{ij} = \frac{s_{ij}}{2G}\ ,
\end{equation} 

linking the deviatoric strain and stress tensors, $e_{ij}$ and $s_{ij}$, through shear modulus $G$. In the following, we adopt the convention of summed repeated indices. 
Our fundamental assumption is that the shear modulus is altered as a function of the internal state of the material, which leads to path-dependent behavior. Specifically, we assume that $G$ decreases as a function of a scalar state variable $D$, a measure of rock damage, to be defined in section \ref{sec:dam_def}:

\begin{equation}
\label{eq:G_general}
G = G_{0}\text{f}(D)\ .
\end{equation}

In equation \eqref{eq:G_general}, $G_0$ denotes the shear modulus of the material in its least damaged state, and $\text{f}(D)$ a decreasing scalar function of $D$, hereafter referred to as ``weakening function'', satisfying $\text{f}(D) \in \left]0, 1\right]$ and $f'(D) < 0$. The incompressible elastic relationship \eqref{elastic} can be recast as a damaged-elastic constitutive law

\begin{equation}
\label{damage-elastic}   
e_{ij} = \frac{s_{ij}}{2G_{0}\text{f}(D)}\ ,
\end{equation} 

which takes the form of a Maxwell visco-elastic constitutive law upon time differentiation:

\begin{equation}
\label{eq:damaged-elastic_constitutive_law}   
\dot{e}_{ij} = \frac{\dot{s}_{ij}}{2G_{0}\text{f}(D)} + \frac{s_{ij}}{2\eta_D}\ .
\end{equation}

In equation \eqref{eq:damaged-elastic_constitutive_law} $\eta_D$ is a viscosity associated with damage growth:

\begin{equation}
\label{eq:eta_dam}   
\eta_{D} = \frac{\text{f}^2(D)\ G_0}{|\text{f}'(D)| \ \dot{D}}\ .
\end{equation}

The damage state variable is related to the lengthening of mode-I microfractures, an intrinsically dilatant process. Throughout this study, inelastic dilatancy is neglected in favor of a purely deviatoric description of the damaged rheology, focusing on the role of microcracking on shear modulus alteration, and on fault nucleation. A strategy to account for damage-induced dilatancy within the SCAM framework will nonetheless be outlined in Section \ref{sec:conclusions_perspectives}. In the following, we detail the micromechanical interpretation of the damage variable, the model governing its growth rate, as well as the weakening function.

\subsection{Micromechanical representation of rock damage}
\label{sec:dam_def}

\begin{figure}[t!]
\noindent\includegraphics[width=\textwidth]{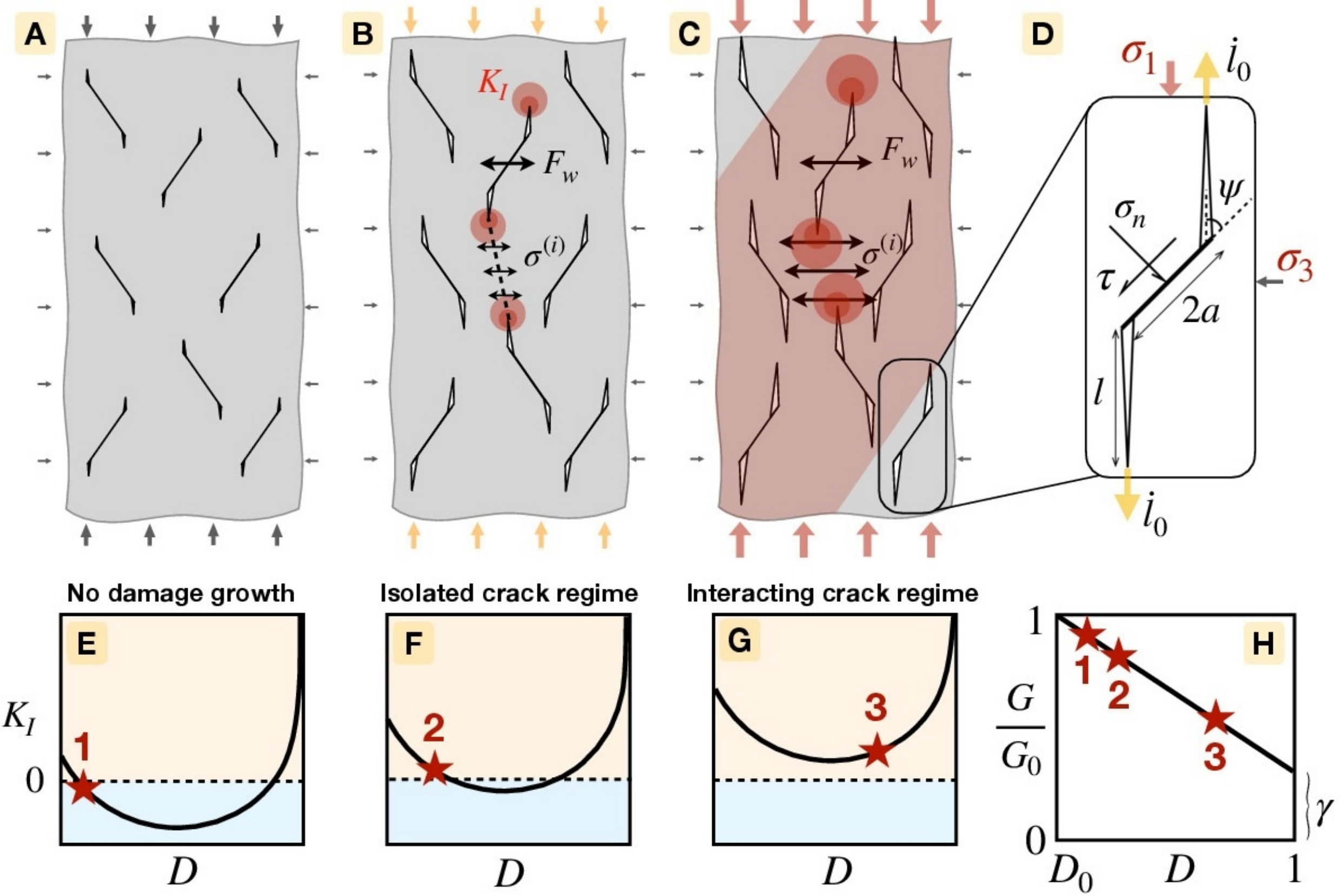} 
\caption{\textbf{A–C.} Cross-sections in the $(\sigma_1, \sigma_3)$ plane showing the array of cracks growing in a rock, in the model of \protect\citeA{AshbySammis1990}. These cracks initially grow from a shear defect of radius $a$, oriented at an angle $\psi$ to $\sigma_1$. At its tips, "wing" tensile cracks of length $l$ (\textbf{D}) may preexist. Under differential stresses unable to overcome the frictional resistance along shear defects (\textbf{A}), the stress concentration ($K_I$) at their edges is negative (\textbf{E}) and the material deforms elastically. Once stresses overcome the frictional resistance, the sliding shear defects exert a wedging force $F_w$ increasing $K_I$ (\textbf{F}). Positive $K_I$ promotes stable tensile cracks growth (\textbf{B}). As cracks lengthen, they begin to interact (\textbf{C}). At this point damage growth enters an unstable, accelerating regime (\textbf{G}). \textbf{H.} Alteration of the shear modulus as a function of damage.}
\label{model}
\end{figure}

Our goal is to model the accumulation of damage in the upper crust, which is primarily composed of low-porosity ($< 1\% $) magmatic and metamorphic silicate rocks. These units lie in an overall compressive stress state, with pressures up to hundreds of MPas. Yet, distributed brittle deformation typically involves the opening of mode-I microcracks (Figure \ref{exp}), which is made possible by stress concentrations around defects or grain boundaries. To describe these processes, we adopt the damage framework developed by \citeA{AshbySammis1990} which has been used successfully to predict the brittle strength of several rocks \cite<e.g.,>{BaudEtAl2000,WuEtAl2000,BhatEtAl2011} at low confining pressure, and the dynamics of fracturing during seismic ruptures \cite{BhatEtAl2012,ThomasEtAl2017}. 
This model considers the growth of tensile ``wing''-cracks from the tips of penny-shaped shear defects distributed within the rock (Figure \ref{model}A-D).

The damage variable represents the relative volume occupied by cracks as the wings lengthen in the direction of the most compressive stress (Figure \ref{model}D). It is defined as

\begin{equation}
\label{D}   
D = \frac{4}{3}\pi N_{v} (\alpha a + l)^{3}\ ,
\end{equation} 

where $N_v$ is the number of shear defects per unit volume, $a$ and $l$ are the radius of the shear defects and the length of the wing cracks, respectively. $\alpha=\cos{\psi}$ is the cosine of the angle $\psi$ between the shear defects and the most compressive stress (Figure \ref{model}D).
The least damaged state ($D=D_0$) corresponds to the state of a rock containing only shear defects, where no wing crack has nucleated (i.e., $l=0$).
The most damaged state occurs at $D=1$ when the volume of the spheres enclosing each wing crack has grown to match the characteristic volume defined by the spacing of defects ($V_c = 1/N_v$). This upper bound is the result of the formulation of the interaction between cracks, detailed in Section \ref{sec:damage_growth}. It represents a stage at which coalescence of cracks becomes unavoidable.
For simplicity, we only consider shear defects with normal vectors lying in the plane of the two extreme principal stresses $\sigma_1$ and $\sigma_3$. This allows us to index their activity on a 2-D Mohr-Coulomb yield criterion.

\subsection{Damage growth}
\label{sec:damage_growth}

We assume that under low strain rates and on long time scales, wing cracks lengthen in a sub-critical manner, i.e., with stress intensity factors ($K_I$) lower than the fracture toughness ($K_{IC}$) of the material \cite{Atkinson1984}.  To capture this process in our constitutive law, we adopt the stress corrosion law introduced by \citeA{Charles1958}, which has proven successful at explaining experimental data \cite<e.g.,>{Kachanov1982b,Atkinson1984,DeshpandeEvans2008,BrantutEtAl2012}. Specifically, the crack growth rate writes 

\begin{equation}
\label{eq:charles}
\dot{l} = \dot{l}_{0} \left(\frac{K_I}{K_{IC}}\right)^n\ ,
\end{equation}

where $\dot{l}_{0}$ is a characteristic crack growth rate and $n$ the Charles law exponent. The damage growth rate is then retrieved from the wing-crack tip speed

\begin{equation}
\label{eq:Ddot}   
\dot{D}  =  \frac{\partial D}{\partial l} \dot{l} =  \frac{3 D^{\frac{2}{3}} D_{0}^{\frac{1}{3}} \dot{l}_{0}}{\alpha a} \left(\frac{K_I}{K_{IC}}\right)^n\ .
\end{equation}

This equation applies only when $K_I>0$, otherwise $\dot{D}=0$.
Using equation \eqref{eq:charles} requires an expression for $K_I$, the stress intensity factor at the tip of the wing cracks. Following \citeA{AshbySammis1990}, we assemble $K_I$ as the sum of three terms : 

\begin{equation}
    \label{eq:KI_terms}
    K_I = K_{I}^{(w)} + K_{I}^{(\sigma_3)} + K_{I}^{(i)}\ .
\end{equation}

The first term ($K_{I}^{(w)}$) represents the stress concentration due to frictional slip on the shear defects wedging open the wing cracks. Following \citeA{TadaEtAl1973}, it can be expressed as the action of a tensile wedging force $F_w$ at the center of an equivalent penny-shaped crack. The radius of this circular crack is that of the sphere enclosing one entire wing crack (shear defect + tensile wings, Figure \ref{model}D). However, instead of writing it $l + \alpha a$, as in equation \eqref{D}, we write it $l + \beta a$, where $\beta$ is a regularization factor. This approach was adopted by \citeA{AshbySammis1990} to ensure that in the absence of wing cracks ($l=0, D=D_0$), $K_I$ matches the stress intensity factor at the tip of shear defects as derived by \citeA{AshbyHallam1986}. This yields $\beta = 1/\pi$ \cite{BhatEtAl2011} and the following expression for $K_{I}^{(w)}$:

\begin{equation}
    \label{eq:KI_wedging}
    K_{I}^{(w)}=\frac{F_{w}}{[\pi(l+\beta a)]^{3 / 2}}\ .
\end{equation}

The wedging force relates to the excess shear stress acting on the defects (of area $\pi a^2$) relative to their frictional resistance. Following \citeA{AshbySammis1990}, we write :

\begin{equation}
    \label{eq:Fw_principals}
    F_{w} = \left(\sigma_{3} A_{3}-\sigma_{1} A_{1}\right)a^2\ .
\end{equation}

$A_1$ and $A_3$ are constants that depend on the friction and orientation of the shear defects. In the following, we assume $\psi = 45^\circ$ as \citeA{AshbyHallam1986} showed that this orientation maximizes the wedging force over a wide range of wing crack lengths. This yields :

\begin{eqnarray}
\label{A1A3}   
A_1 & = & \pi \sqrt{\frac{\beta}{3}}\left[\sqrt{1+\mu^{2}}-\mu\right] \\
A_3 & = & A_{1}\left[\frac{\sqrt{1+\mu^{2}}+\mu}{\sqrt{1+\mu^{2}}-\mu}\right]\ .
\end{eqnarray} 
Overall, $K_I^{(w)}$ strongly depends on the differential stress $\Delta\sigma = \sigma_3 - \sigma_1$ that develops in the rock, as it allows frictional slip on the defects and wedging of the wings.
By contrast, the second term in equation \ref{eq:KI_terms} ($K_I^{(\sigma_3)}$) represents remote wing-normal compression $\sigma_3$ acting to close tensile cracks. \citeA{BhatEtAl2011} estimated it based on results from \citeA{TadaEtAl1973} as:

\begin{equation}
    \label{eq:KI_sigma3}
    K_{I}^{(\sigma_3)}=\frac{2}{\pi}\left(\sigma_{3}\right) \sqrt{\pi l}\ .
\end{equation}

Finally, the third term ($K_{I}^{(i)}$) serves to describe the interaction between cracks as they lengthen, and is a core feature of this micromechanical
model. \citeA{AshbySammis1990} required that the wedging forces applied to cracks be compensated by an internal stress ($\sigma^{(i)}$ in Figure \ref{model}) to satisfy mechanical equilibrium. The internal stress is applied on an effective area perpendicular to $\sigma_3$ that separates neighboring cracks ($A_b$). The sum of this area with the characteristic area of each wing crack ($\pi(l + \alpha a)^2$) amounts to the area $A_c$ that is obtained by projecting the spherical volume $V_c = 1/N_v$ along $\sigma_3$. Therefore, $A_b = A_c - \pi(l + \alpha a)^2$, with 

\begin{equation}
    \label{eq:area_per_crack}
    A_c = \pi^{1/3}\left(\frac{3}{4N_v} \right)^{2/3}\ .
\end{equation}

This leads to the following expression for the internal stress acting in the direction of least compression, $\sigma_3^i$ :

\begin{equation}
    \label{eq:internal_stress}
    \sigma_3^i = \frac{F_w}{A_b}\ .
\end{equation}

Internal stress $\sigma_3^i$ increases dramatically as wings lengthen ($D$ approaches $1$) and the areas between fractures ($A_b$) shrink. This is when crack interactions become dominant.
$K_{I}^{(i)}$ is readily obtained from $\sigma_3^i$ by analogy with \eqref{eq:KI_sigma3} :

\begin{equation}
    \label{eq:KI_interactions}
    K_{I}^{(i)}=\frac{2}{\pi}\left(\sigma_{3}^i\right) \sqrt{\pi l}\ .
\end{equation}

The full expression of $K_I$ then reads

\begin{equation}
    \label{eq:KI_full}
    K_{I}=\frac{F_{w}}{[\pi(l+\beta a)]^{3 / 2}}+\frac{2}{\pi}\left(\sigma_{3}+\sigma_{3}^{i}\right) \sqrt{\pi l}\ .
\end{equation}

It can be recast as a function of damage rather than crack length, following \citeA{BhatEtAl2011}, yielding

\begin{equation}
\label{eq:KI}   
K_I = \sqrt{\pi a} \left[ \left(\sigma_{3} A_{3}-\sigma_{1} A_{1}\right)\left(c_{1}+c_{2}\right)+\sigma_{3} c_{3} \right]\ ,
\end{equation} 
where $c_1$, $c_2$ and $c_3$ are functions of the damage state that write

\begin{eqnarray}
    c_1 & = & \frac{1}{\pi^2 \alpha^{3 / 2}\left[\left(D / D_{0}\right)^{1 / 3}-1+\beta / \alpha\right]^{3 / 2}} \label{eq:c1} \\
    c_2 & = & \frac{2}{\pi^2 \alpha^{3 / 2}} \left[ \left(D / D_{0}\right)^{1 / 3}-1 \right]^{1/2} \left[\frac{D_{0}^{2 / 3}}{1-D^{2 / 3}}\right] \label{eq:c2} \\
    c_3 & = & \frac{2 \sqrt{\alpha}}{\pi}\left[\left(D / D_{0}\right)^{1 / 3}-1\right]^{1 / 2}\ . \label{eq:c3}
\end{eqnarray} 

\subsection{Weakening function}
\label{sec:weakening_function}

We next turn to the formulation of the function $\text{f}(D)$ used to weaken the shear modulus as damage accumulates. The simplest effective medium representation of a cracked isotropic material assumes non-interacting cracks \cite{Kachanov1993}. Within this approximation, the change in elastic strain energy due to a population of cracks can be inferred by summing their individual contribution. This amounts to elastic compliances scaling linearly with damage. Elastic stiffnesses  therefore scale as $(1 + CD)^{-1}$, where C is a constant that depends on the orientation distribution and geometry of cracks. Linearization of this form provides a reasonable estimate of elastic stiffnesses at low damage values. Because the damage framework of \citeA{AshbySammis1990} sets an upper bound on damage at $1$, we use this approximation and postulate a linear weakening of $G$ with respect to damage $D$ :  
 
\begin{equation}
\label{Geff}
\text{f}(D) = \frac{\gamma-1}{1 - D_0} D + \frac{1-\gamma D_0}{1-D_0}\ ,
\end{equation}

such that $G(D_0) = G_0$ and $G(D=1) = \gamma G_0$. The weakening parameter $\gamma \in \left]0,1\right] $ can be thought of as a property of the material representing the stiffness of a fully damaged rock (e.g., a fault zone) normalized by its maximum possible stiffness in a low-damage state. The derivative of our weakening function with respect to $D$ is :

\begin{equation}
\label{f'}   
f' = \frac{\gamma - 1}{1 - D_0}\ .
\end{equation}

It should be noted that for simplicity our model weakens the shear modulus isotropically, even though damage grows in a highly anisotropic fashion. 

To recap, equations \ref{eq:damaged-elastic_constitutive_law} and \ref{eq:eta_dam}, combined with equations \ref{eq:Ddot}, \ref{eq:KI} and \ref{Geff} make up the complete SCAM constitutive law, which is akin to Maxwell visco-elasticity with a strongly non linear dependence of viscosity on stress, and progressive alteration of the elastic modulus with increasing damage.
These equations are reiterated below:

\begin{eqnarray}
\dot{e}_{ij} & = & \frac{\dot{s}_{ij}}{2G_{0}\text{f}(D)} + \frac{s_{ij}}{2\eta_D} \nonumber\\ 
\eta_{D} & = & \frac{\text{f}^2(D)\ G_0}{|\text{f}'(D)| \ \dot{D}} \nonumber\\
\dot{D} & = & \frac{3 D^{\frac{2}{3}} D_{0}^{\frac{1}{3}} \dot{l}_{0}}{\alpha a} \left(\frac{K_I}{K_{IC}}\right)^n \nonumber\\
K_I & = & \sqrt{\pi a} \left[ \left(\sigma_{3} A_{3}-\sigma_{1} A_{1}\right)\left(c_{1}+c_{2}\right)+\sigma_{3} c_{3} \right] \nonumber\\
\text{f}(D) & = & \frac{\gamma-1}{1 - D_0} D + \frac{1-\gamma D_0}{1-D_0} \nonumber
\end{eqnarray}.

\section{Application to a 0-D triaxial loading setup}
\label{sec:triaxial_application}

\subsection{Constitutive SCAM equations in a triaxial setup}
\label{Pointwise numerical integration}

To illustrate the behavior of the SCAM flow law, we implement it in a geometry typical of rock deformation experiments (Figure \ref{0D_damaged_vs_plast}A) : Compression along the axis of a cylindrical sample ($\sigma_1$, along direction $x_1$) subjected to axially symmetric confining stress ($\sigma_3$, along directions $x_2$ and $x_3$). The corresponding stress tensor writes 

\begin{equation}
    \label{eq:stress_triaxial}
    \underline{\underline{\mathbf{\sigma}}} = \begin{bmatrix}
        \sigma_{ax} & 0 & 0\\
        0 & -p_c & 0\\
        0 & 0 & -p_c
    \end{bmatrix},
\end{equation}
where $\sigma_{ax}$ is the axial stress and $p_c$ the confining pressure surrounding the curved surface of the sample. 
We use a simplified point-wise formulation of our differential constitutive relationship  \eqref{eq:damaged-elastic_constitutive_law} assuming homogeneous deformation within the sample.
As stated previously, we ignore volumetric strain and focus solely on the relationship between the deviatoric axial strain rate $\dot{e}_{ax}$ and the deviatoric axial stress $s_{ax}$.
The constitutive equations reduce to the following ordinary differential equation (ODE) :

\begin{equation}
\label{eq:axial_constitutive_law}
\dot{e}_{ax} = \frac{\dot{s}_{ax}}{2G_{0}\text{f}(D)} + \frac{s_{ax}}{2\eta_{D}},
\end{equation} 

to be solved jointly with the damage evolution equation (\ref{eq:Ddot}).

A first type of experiment consists of applying a constant axial strain rate and measuring the axial stress. In our framework, $s_{ax}$ verifies :

\begin{equation}
\label{eq:csr_ode}
\dot{s}_{ax} = 2G_{0}\text{f}(D)\left(\dot{e}_{ax} - \frac{s_{ax}}{2\eta_{D}}\right),
\end{equation}

with $s_{ax}(t=0) = 0$ and $D(t=0) = D_i \geq D_0$.

Another class of experiments (brittle creep tests) consists of applying a constant axial stress and measuring the axial strain. In our model, the latter is given by

\begin{equation}
\label{eq:cs_ode}
\dot{e}_{ax} = \frac{s_{ax}}{2\eta_{D}},
\end{equation} 

In this case, the initial value of $D$ cannot be chosen arbitrarily and must be consistent with the imposed stress. To ensure this, we first integrate the constant strain rate and damage growth ODEs (equations \ref{eq:csr_ode} and \ref{eq:Ddot}) up to the desired value of axial deviatoric stress $s_{ax}$ assuming a known strain rate. The damage value reached at the end of this preliminary step is used as initial condition for equations \ref{eq:cs_ode} and \ref{eq:Ddot}, along with $e_{ax}(t=0)=0$. These equations are integrated up to $D$ close -- but not equal -- to $1$, due to the singular behavior at this limit, coming from the $c_2$ term in equation \ref{eq:c2}.

The above ODEs are integrated numerically using a $5/4$th order Runge-Kutta method \cite{TsitourasEtAl2009}. This is done within the DifferentialEquations.jl Julia package \cite{RackauckasNie2017} using adaptive time-stepping with absolute and relative tolerance of $10^{-6}$ and $10^{-4}$ respectively.

\subsection{Stress-strain curves and creep regimes}
\label{sec:0D_description}

We illustrate the fundamental behavior of the SCAM model in triaxial experiments using reference micromechanical parameters appropriate for Westerly granite (Table \ref{tab:parameters}), which will be rigorously determined in Section \ref{sec:model_calibration}. Constitutive equations are integrated up to $D=0.95$.

\begin{figure}[t!]
\noindent\includegraphics[width=\textwidth]{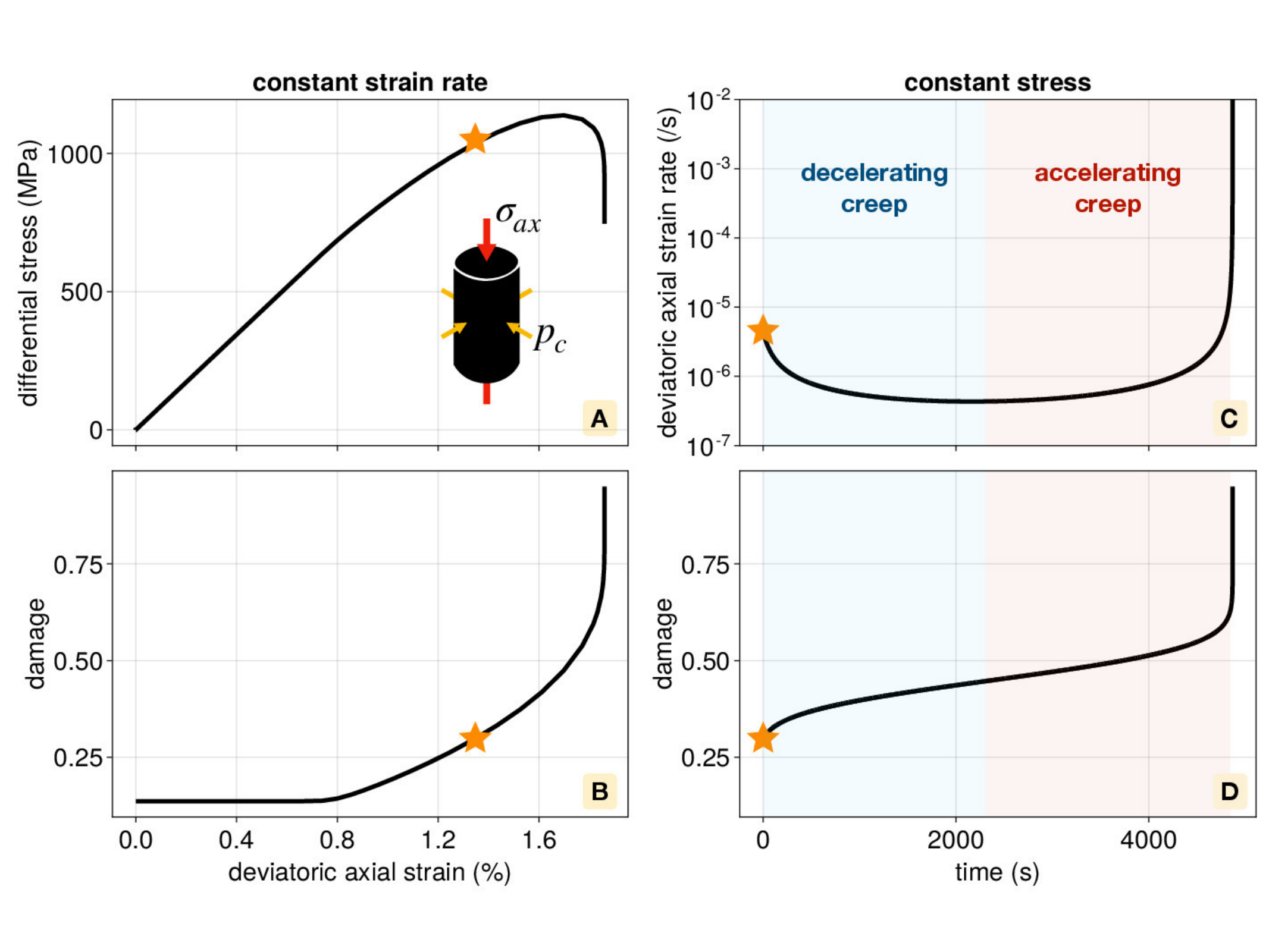} 
\caption{0-D simulations based on the SCAM model in a typical triaxial setup with a confining pressure of $150$ MPa (stress state shown in Panel \textbf{A}). \textbf{A}-\textbf{B}. Differential stress and damage vs. strain in a constant strain rate experiment ($10^{-5}\ s^{-1}$). \textbf{C}-\textbf{D.} Deviatoric axial strain rate and damage vs. time in an experiment where stress is kept constant after reaching the stress and damage state pictured by the stars in panels \textbf{A} and \textbf{B}. Here strains and stresses are represented positive in compression for clarity.}
\label{0D_damaged_vs_plast}
\end{figure}

Figure \ref{0D_damaged_vs_plast}A, B correspond to a constant strain rate setup at $10^{-5}\ s^{-1}$, under $150$ MPa of confining pressure. The axial stress-strain curve displays an initial elastic phase followed by visible weakening of the effective modulus when differential stress exceeds $\sim 700$ MPa. This is accompanied by damage growth (Figure \ref{0D_damaged_vs_plast}B) which accelerates catastrophically as the sample reaches its peak stress. The post-peak stress drop is similarly abrupt as damage approaches $1$. 

Figure \ref{0D_damaged_vs_plast}C, D correspond to a constant stress simulation starting at the yellow star shown in panels A and B. Strain rate first decelerates, then remains steady for hours, and ultimately accelerates up to the macroscopic failure of the material, consistent with the subsequent phases of brittle creep observed experimentally (Figure \ref{exp}E). What is usually referred to as secondary creep would here be associated to the transition between decelerating and accelerating creep, and was not depicted in Figure \ref{0D_damaged_vs_plast}C, D because of the clear bimodal dynamics of brittle creep expressed by the SCAM model. This strain rate behavior is associated with dynamics similar to those of the damage growth rate, visible through the slope of damage evolution with respect to time in Figure \ref{0D_damaged_vs_plast}D.

\begin{figure}[t!]
\noindent\includegraphics[width=\textwidth]{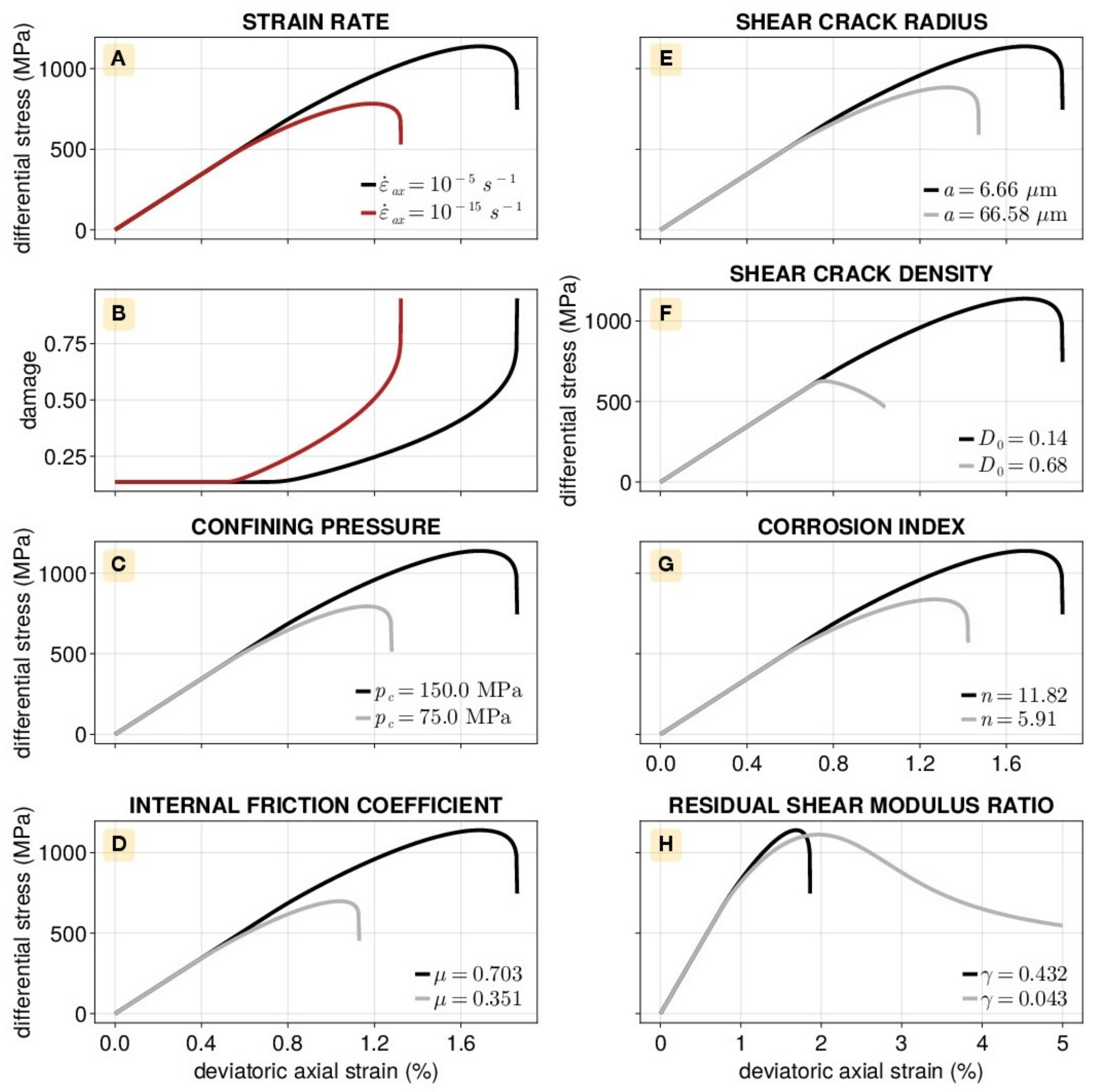} 
\caption{Effect of strain rate (\textbf{A},\textbf{B}), confining pressure $p_c$ (\textbf{C}), friction coefficient $\mu$ (\textbf{D}), shear defect radius $a$ (\textbf{E}), shear crack density $D_0$ (\textbf{F}), corrosion index $n$ (\textbf{G}) and residual shear modulus $\gamma$ (\textbf{H}) on differential stress with respect to deviatoric axial strain for 0-D SCAM simulations. Black lines correspond to the best fitting parameters for Westerly granite detailed in Section (\ref{sec:model_calibration}). Here strains and stresses are represented positive in compression for clarity.}
\label{0D_params_effect}
\end{figure}

The effect of various model parameters and experimental conditions on the behavior of the SCAM model under constant strain rate is shown in Figure \ref{0D_params_effect}. The black curves correspond to a strain rate of $10^{-5}\ s^{-1}$, a confining pressure of $150$ MPa and the reference set of micromechanical parameters for Westerly granite (Table \ref{tab:parameters}). Figure \ref{0D_params_effect}A,B shows that a reduction of the imposed axial strain rate leads to a lower peak stress due to damage having more time to accumulate under lower axial stress, precipitating failure (Figure \ref{0D_params_effect}B). Increasing the radius of the shear defects (Panel E) while keeping $D_0$ constant leads to a decrease of the peak stress. This is because the stress intensity factor at the wing crack tips increases with increasing shear defect size, prompting faster crack growth. Thus, significant damage can build under lower stresses, and the peak stress is reached sooner.
Increasing $D_0$ while keeping the shear defect size constant (Panel F) also leads to a lower peak stress, but limits the amount of softening that takes place pre-peak. This is because cracks arranged in a denser array will interact and coalesce sooner. The stress decrease additionally does not display the abrupt drop seen with the reference case, which we attribute to the larger reduction of shear modulus per damage increment.
A decrease of the Charles law exponent $n$ (Panel G) or friction coefficient (Panel D) similarly lowers the peak stress, by enabling damage build-up under lower stress intensity factors (\ref{eq:charles}) and under lower differential stress, respectively.
Finally, a greater degree of modulus weakening (via a reduced $\gamma$ parameter) leads to more pre-peak softening but has a limited impact on the peak stress (Panel H). It also stabilizes the stress drop by limiting the unstable growth of damage as it gets close to $1$. In this case of extreme loss of elastic stiffness, the larger negative stress increment associated with damage increment post-peak tends to reduce the catastrophic increase in damage growth.

\begin{figure}[t!]
\noindent\includegraphics[width=\textwidth]{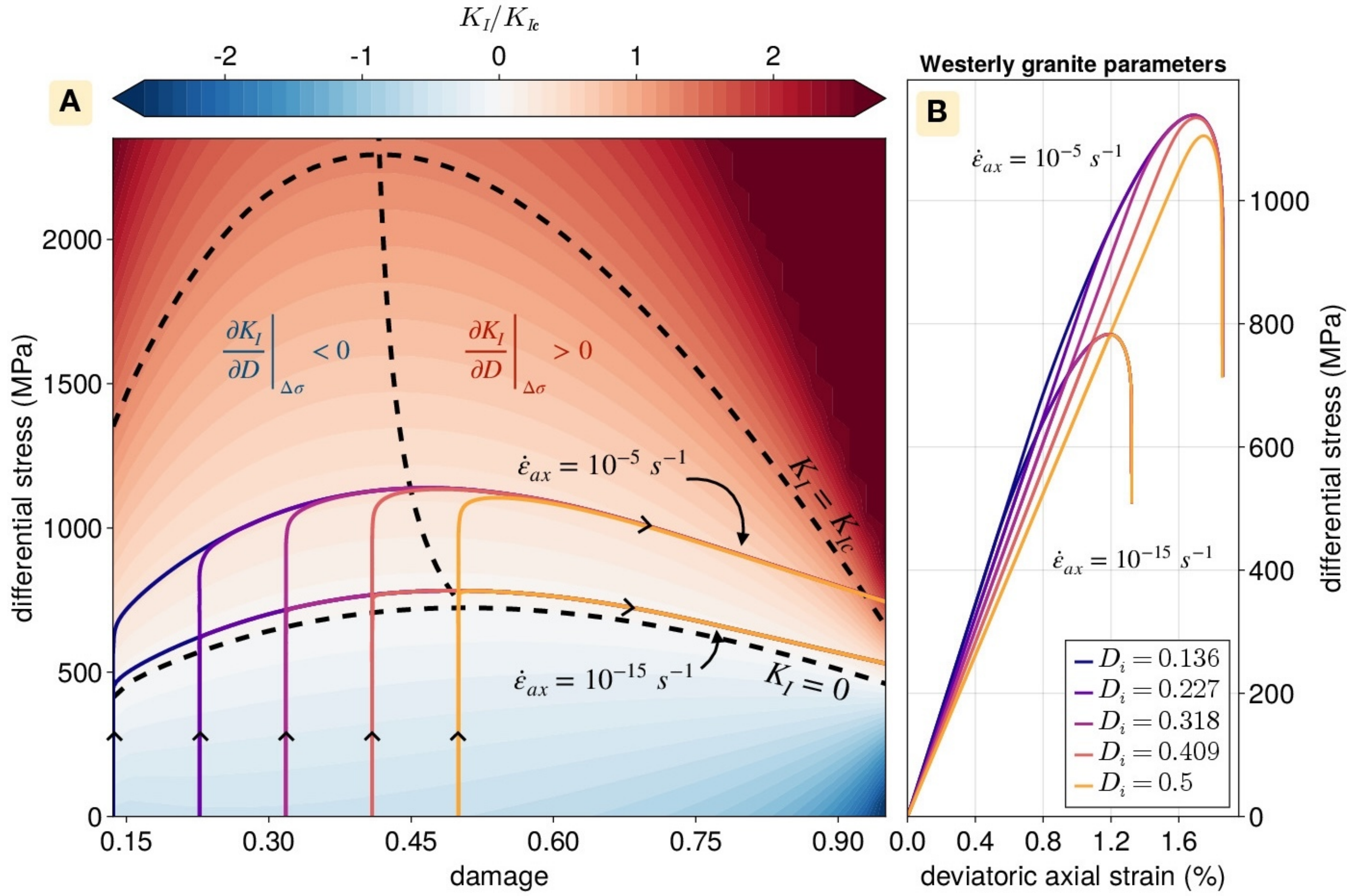} 
\caption{\textbf{A.} Trajectories of differential stress with respect to damage at $\dot{\varepsilon}_{ax}=-10^{-5}\ s^{-1}$ and $\dot{\varepsilon}_{ax}=-10^{-15}\ s^{-1}$ for various initial damages (color code) and at constant confining pressure $p_c = 150 MPa$. The background is colored according to $K_{I}/K_{IC}$, with the two iso-values pictured with black dashed contours representing $K_I = 0$ and $K_I = K_{IC}$ respectively. 
The near-vertical dashed black line highlights damage values where $\partial{K_I}/\partial{D} = 0$ under constant $\Delta\sigma$. \textbf{B.} Same trajectories plotted as standard deviatoric axial strain vs. differential stress, along with the color code for initial damage state. Strains and stresses are represented positive in compression for clarity.}
\label{KI_map}
\end{figure}

To better visualize the dynamics of damage growth in the SCAM model, we represent constant strain rate experiments in a plot of differential stress vs. damage (Figure \ref{KI_map}A). This representation allows us to map the stress intensity factor at the wing-crack tip (colors and contours in panel A), which gives us a proxy for damage growth rate. We specifically highlight two sets of experiments. The first set is performed at a laboratory strain rate $\dot{\varepsilon}_{ax} = 10^{-5} \ s^{-1}$, and the second at a tectonic strain rate of $10^{-15} \ s^{-1}$, both under a confining pressure $p_c = 150$ MPa. In each set, we vary the initial damage $D_i$, using values of $0.136$ ($D_0$), $0.227$, $0.318$, $0.409$ and $0.5$. 

Each experiment follows a specific trajectory in differential stress vs. damage space. For example, in the case of no initial tensile cracks ($D_i=D_0$), differential stress first increases while damage remains constant. This is because in the initial elastic regime, $K_I <= 0$ and wing cracks cannot grow. Once the system reaches the domain of positive $K_I$, damage can start growing, and increases with stress. The system appears to follow a contour of constant $K_I$ up to the peak differential stress ($\sim 1140$ MPa). Past this point, the differential stress starts to decrease while damage keeps increasing at an accelerating pace. This is due to the fact that $K_I$, which sets the rate of damage growth, now increases with increasing damage. This final phase of rapid failure manifests as an abrupt post-peak stress drop in the stress-strain curve (Panel B).

These three regimes, characterized by the absence of growth, the stable and then the unstable growth of damage is illustrated in Figure \ref{model} with the three numbered stars respectively.
Simulations carried out under the same strain rate, but with greater initial damage show the same behavior, and their trajectories tend to align along the same iso-$K_I$ ($\sim 0.3 K_{IC}$) path as followed by the $D_i = D_0$ case. This forms an envelope that materializes an upper bound of the differential stress value with respect to damage. This envelope corresponds to ($K_I \sim 0.05 K_{IC}$) for tectonic strain rates, and therefore lies at lower stress values.  
If, however, a simulation is initiated with damage in excess of $\sim0.45$ (e.g., orange paths in panel A), damage will immediately start growing in the unstable regime, where $\partial K_I/\partial D > 0$. In this case, the system reaches a peak stress which is lower than that of the other simulations.
\subsection{Calibration of SCAM parameters with laboratory experiments}
\label{sec:model_calibration}

The ability of the SCAM model to reproduce both constant strain rate and constant stress experiments suggests that laboratory data can be used to constrain its micromechanical parameters (Table \ref{tab:parameters}). Specifically, stress-strain curves from constant strain rate experiments under various confining pressures can help constrain elastic and frictional properties, while strain rates and time to failure in brittle creep tests contain information about the kinetics of damage build-up. 

To leverage this information, we use the 0-D "forward" models presented in the previous section in a Bayesian inversion framework \cite[see \ref{ap:inversion} for details]{tarantola2005}. We expect 0-D models to be representative of the homogeneous deformation stage up to the peak stress (prior to localization), as micro-cracking is known to first develop in a distributed fashion (Figure \ref{exp}).

\begin{table}
 \caption{Inverted parameters}
 \centering
 \begin{tabular}{l l l l}
 \hline
  symbol  & description & Westerly granite & Darley Dale sandstone \\
 \hline 
   $G_0$        & shear modulus at $D = D_0$ (GPa)          & $28.72 \pm 0.02$             & $5.202 \pm 3$       \\
   $\gamma$     & residual ratio $G/G_0$ at $D = 1$         & $0.432 \pm 0.004$            & $0.281 \pm 0.003$   \\
   $\mu$        & friction coefficient of the shear flaws   & $0.703 \pm 0.001$            & $0.5093 \pm 0.0003$ \\
   $a$          & shear flaws radius ($\mu$m)               & $6.66 \pm 0.16$              & $656.8 \pm 4.3$     \\
   $n$          & Charles law exponent                    & $11.82 \pm 0.03$             & $24.96 \pm 0.07$    \\
   $\dot{l}_0$  & Charles law reference crack growth rate (mm $s^{-1}$) & $16.44 \pm 0.19$ & $0.0029 \pm 0.0001$ \\
   $K_{IC}$     & fracture toughness (MPa $m^{1/2}$)        & $1.29 \pm 0.01$              & $1.412 \pm 0.004$   \\
   $D_0$        & $D$ associated to the shear flaws only    & $0.1358 \pm 0.0003$          & $0.3724 \pm 0.0008$ \\
   $D_i$        & initial value of $D$                      & $0.1361 \pm 0.0003$          & $0.27 \pm 0.05$     \\
 \hline
 \end{tabular}
 \label{tab:parameters}
 \end{table}

\subsubsection{Experimental data}

We apply the Bayesian inversion method to experimental data corresponding to two lithologies. The first is Westerly granite, a rock type widely used in experiments that is representative of the continental upper crust in term of mineralogy and low porosity. This rock has been shown to experience the type of diffuse cracking and catastrophic fracture coalescence that our model seeks to capture \cite<e.g.,>[]{TapponnierBrace1976,LocknerEtAl1991}. We specifically use constant strain rate ($\dot{\varepsilon}_{ax} = 10^{-5} s^{-1}$) experiments under confining pressures of $20$, $30$, $80$ and $150$ MPa in dry conditions from \citeA{WawersikBrace1971} (Figure \ref{wg_fit}A). We complement these data with minimum brittle (secondary) creep strain rates measured under seven imposed differential stresses ranging from $77\%$ to $93\%$ of the short-term strength (meaning the peak strength at a laboratory strain rate) of the rock subjected to an effective confining pressure of $30$ MPa in water-saturated samples by \citeA{BrantutEtAl2012} (Figure \ref{wg_fit}B).

The second rock type we consider is Darley Dale feldspar-rich sandstone, another widely studied lithology. While its properties are likely less representative of the upper crust than that of Westerly granite, inferring its micro-mechanical parameters can provide helpful comparisons to assess the validity of our model. One caveat of this choice is that porous sandstone may deform according to mechanisms other than the growth of tensile cracks from shear defects, such as Hertz-contact driven microfracturing \cite<e.g.,>[]{ZhangEtAl1990}, tensile cracks nucleating from pores \cite<e.g.,>[]{SammisAshby1986}, or distributed cataclastic flow where microcracking grows along grain boundaries \cite<e.g.,>[]{MenendezEtAl1996}.
\citeA{HeapEtAl2009} however report that for confining pressures up to their maximum of $50$ MPa, stress-induced damage grows predominantly in a direction subparallel to the axis of compression. Additionally, dilatancy patterns at constant strain rate are very similar to what is observed in low-porosity rocks such as Westerly granite \cite<e.g.,>{ZobackByerlee1975,WuEtAl2000}. These observations suggest that the wing-crack model remains a relevant conceptual framework for the brittle deformation of Darley Dale sandstone, at least below 50 MPa of effective confining pressure.

Friction coefficient $\mu$, fracture toughness $K_{IC}$ and pre-existing crack length $a$ were previously estimated for Darley-Dale  by \citeA{WuEtAl2000} within the \citeA{AshbySammis1990} wing-crack framework, but assuming critical crack growth (i.e., at $K_I=K_{IC}$). Here, we additionally use entire time series of brittle creep strain rates from \citeA{HeapEtAl2009} to provide strong constraints on the kinetics of crack growth. We specifically use experimental results performed under constant stresses of $80$, $85$ and $90\%$ of the short-term strength from \citeA{HeapEtAl2009} (Figure \ref{sandstone_fit}B). We combine these time series with stress-strain curves determined under a constant strain rate of $10^{-5} s^{-1}$ and confining pressures of $10$ and $50$ MPa in water-saturated samples from the same authors (Figure \ref{sandstone_fit}A). While data was also available for a confining pressure of $30$ MPa, it displayed a significantly different shear modulus compared to the other two. We thus decided to exclude it from the joint inversion procedure. 

It should be noted that because our constitutive law is based on incompressible elasticity, we remove the volumetric component of elastic deformation (i.e., the Poisson effect) from constant strain rate experimental data. In practice, this means that the reported strain rate is converted to a deviatoric axial strain rate $\dot{\varepsilon}'_{ax} = \frac{2}{3}(1 + \nu) \dot{\varepsilon}_{ax}$, with $\nu$ the Poisson's ratio, in order to compare our simulations with laboratory results.

\subsubsection{A-priori parameters}

The a-priori value of $G$ is chosen based on the initial slope of the elastic (linear) portion of available stress-strain curves. The initial value of the friction coefficient $\mu$ is set to the standard value of $0.65$ \cite{Byerlee1978}. We initially assume $K_{IC}$ to fall between values determined experimentally in quartz (1 $\mathrm{MPa}\cdot m^{1/2}$) and wet Westerly granite (1.74 $\mathrm{MPa}\cdot m^{1/2}$) from the compilation of \citeA{Atkinson1984}. We initialize $\gamma$ at an intermediate value of $0.5$, and $a$ at the mean grain size of Westerly granite: $0.5$ mm. We also set $D_0 = D_i$ at $0.2$, $n=12$ \cite{atkinson1979}, and $\dot{l}_0 = 10^{-2}\ m\ s^{-1} $.
This set of a-priori guesses on the parameter values is first used to invert only the data from constant strain rate experiments. The results of this step are used to construct new priors on the model parameters, shown as gray shadings in the right columns of Figures \ref{wg_fit} and \ref{sandstone_fit}. These priors are then used for a combined inversion of constant strain rate and brittle creep experiments, using a step multiplier $\kappa = 0.1$ (See \ref{ap:inversion} for details). A hundred steps were typically sufficient to reach convergence, yielding the posterior model parameter distributions shown in black in the right column of Figures \ref{wg_fit} and \ref{sandstone_fit}.

\subsubsection{Results}

\begin{figure}[t!]
\noindent\includegraphics[width=\textwidth]{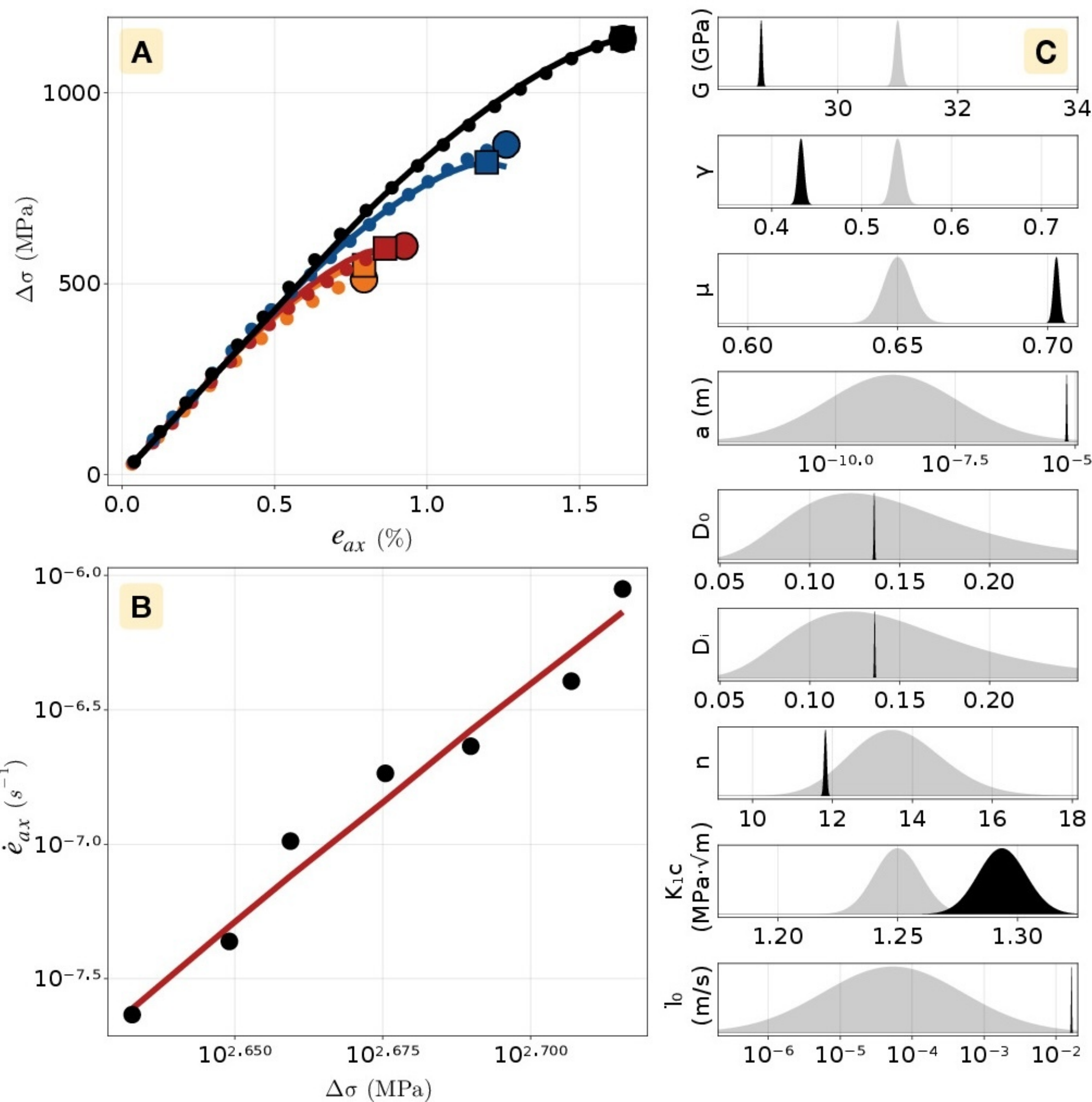} 
\caption{SCAM model fit to experimental data on Westerly granite. \textbf{A.} Constant strain rate experiments with $\dot{\varepsilon}_{ax}=-10^{-5}\ s^{-1}$ and confining pressures of 20 (orange), 30 (red), 80 (blue) and 150 MPa (black). Dots show the data and lines the best fitting models. \textbf{B.} Dataset of minimum brittle creep strain rate as a function of imposed differential stress (black dots), with relationship derived from best fitting model (red line). \textbf{C.} Prior (grey) and posterior (black) distributions of inverted micromechanical parameters.}
\label{wg_fit}
\end{figure}

Figure \ref{wg_fit} shows the results of our joint inversion of constant strain rate and brittle creep data in Westerly granite. Panel A compares the SCAM-simulated stress-strain curves (plain lines) and the experimental data points. The agreement is good at confining pressures of 30, 80 and 150 MPa. At 20 MPa, however, the model slightly over-estimates the peak stress. The \citeA{WawersikBrace1971} study also contains data at atmospheric, $3.5$ and $10$ MPa confining pressures, but under these conditions our model was not able to accurately represent the pressure dependence of the peak stress. Figure \ref{wg_fit}B also shows secondary creep strain rates as a function of imposed differential stress from our simulations (red line), which are in good agreement with experimental data (black dots). The relationship between brittle creep strain rate and stress is effectively a power law with a stress exponent of $\sim18$.

\begin{figure}[t!]
\noindent\includegraphics[width=\textwidth]{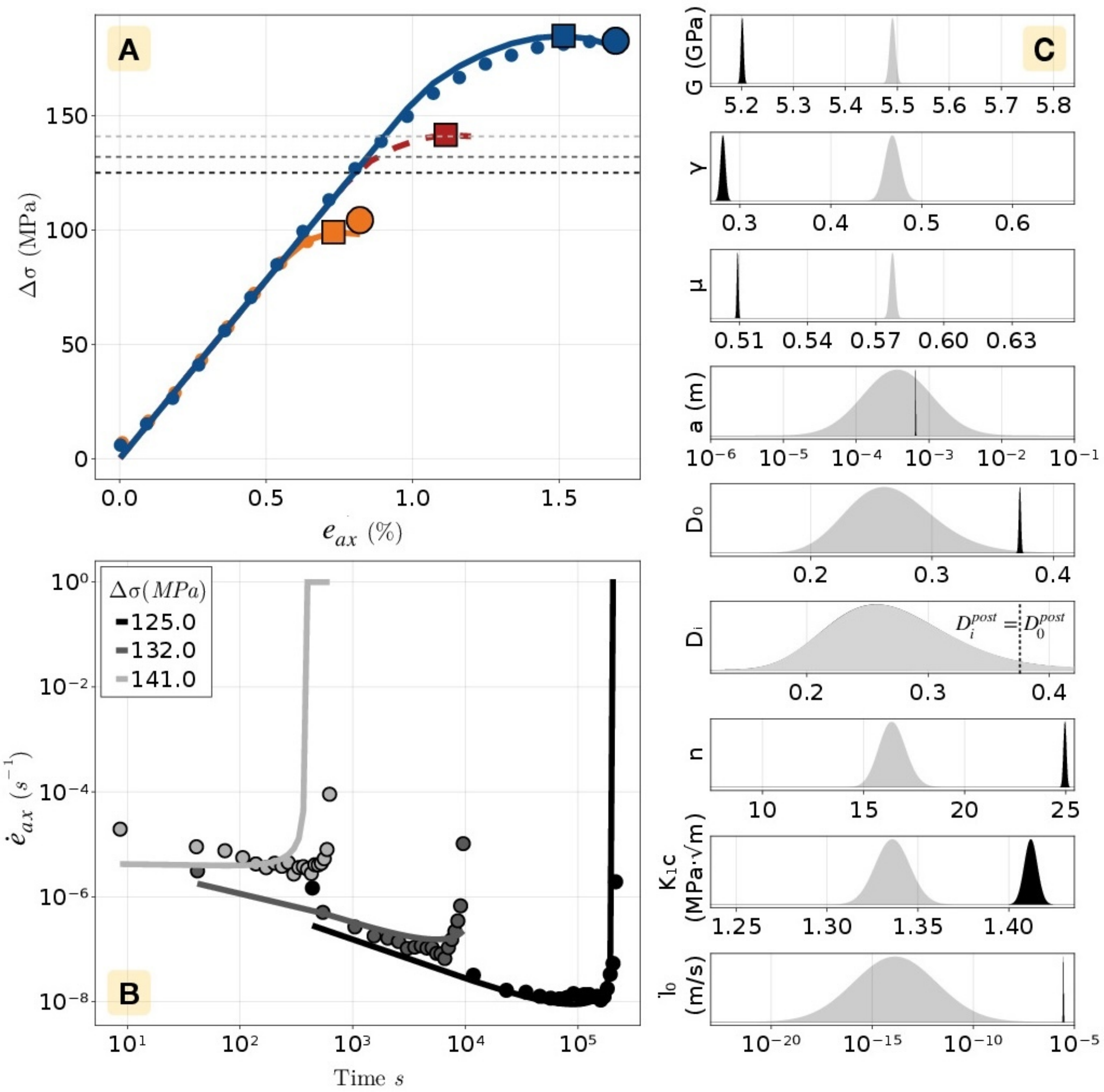} 
\caption{SCAM model fit to experimental data on Darley Dale sandstone. \textbf{A.} Constant strain rate experiments with $\dot{\varepsilon}_{ax}=-10^{-5}\ s^{-1}$ and confining pressures of 10 (orange) and 50 MPa (blue). Dots show the data and lines the best fitting models. Larger circles and squares indicate experimentally determined and modeled peak stress, respectively. Horizontal dashed lines show the imposed stress levels for the experiments shown in Panel B. \textbf{B.} Brittle creep tests shown as time series of axial strain rate under imposed differential stress and confining pressure $p_c=30$ MPa (dots and lines correspond to data and models, respectively). A simulated constant strain rate curve at this confining pressure is included in Panel \textbf{A} (red dashed curve with red square marking the peak stress). \textbf{C.} Prior (grey) and posterior (black) distributions of inverted parameters. The posterior distribution of the initial damage state $D_i$ is the same as its prior because the inversion kept lowering the $D_i$ value to below $D_0$, which was not permitted. This resulted in a null gradient of the log-likelihood with respect to $D_i$ and thus no change of the parameter nor of its posterior distribution relative to the prior.}
\label{sandstone_fit}
\end{figure}
The best fitting parameter values as well as their log-normal standard deviations are listed in Table \ref{tab:parameters}, and shown as probability distributions (in black) in Figure \ref{wg_fit}. Because the prior distributions (in gray) were determined by fitting only constant strain rate data, the differences between prior and posterior distributions highlight the information provided by brittle creep data. This information specifically constrains the initial damage state, as well as parameters related to the kinetics of damage growth such as $\dot{l}_0$ or $n$. It also strongly constrains the size of shear defects (to $\sim7 \mu$m), which influences $K_I$ and therefore the damage growth rate. Parameters such as $K_{IC}$, $\mu$ and $\gamma$ are also slightly re-evaluated.

Best-fitting stress-strain curves for Darley Dale sandstone are shown in Figure \ref{sandstone_fit}A for $10$ and $50$ MPa of confining pressure as plain lines, along with experimental data (points). Larger markers mark the peak stress of simulations (squares) and experiments (circles). The red dashed line shows an additional simulated curve at an intermediate pressure of $30$ MPa. This pressure corresponds to that of the brittle creep tests (Figure \ref{sandstone_fit}B), which were conducted under three axial stresses indicated as dashed lines in Figure \ref{sandstone_fit}A. It can be seen that the greatest applied differential stress ($141$ MPa) is very close to the inferred peak stress at 30 MPa of confining pressure. Figure \ref{sandstone_fit}B compares simulated and measured strain rates in the brittle creep experiments. Our best fitting parameters do a good job at reproducing the shape of the strain rate curves as well as the time to macroscopic failure (the final, near-vertical portion of the curves). 

Similarly to our results in Westerly granite, joint inversions of brittle creep tests and constant strain rate experiments provide strong constraints on parameters such as shear defect size, initial damage, Charles law exponent, and $\dot{l}_0$. Our inversions yield a significantly greater defect size ($\sim700\mu$m vs. $\sim7\mu$m) and Charles law exponent ($\sim25$ vs. $\sim12$) in sandstone compared to granite, as well as a lower shear modulus and greater degree of elastic weakening (lower $\gamma$). The initial damage state of sandstone also appears greater. We however find comparable fracture toughness in both lithologies, and a slightly greater coefficient of (defect-scale) friction in granite ($0.7$ vs. $0.5$).

\section{Application to a 2-D plane-strain numerical press}
\label{sec:2-D_application}
\subsection{Conservation equations and numerical methodology}

In order to perform 2-D simulations of material deformation governed by the SCAM constitutive equations, we adapt the long-term tectonic modeling code \textit{SiStER} (Simple Stokes solver with Exotic Rheologies, \citeA{OliveEtAl2016}), which solves for conservation of mass, momentum (and energy if needed), in a 2-D continuum assuming elastic incompressibility and planar deformation. Conservation of mass and momentum write:

\begin{equation}
    \label{eq:mass_conservation}
    \frac{\partial v_{i}}{\partial x_{i}}=0,
\end{equation}
and
\begin{equation}
    \label{eq:momentum_conservation}
    \frac{\partial s_{i j}}{\partial x_{j}}-\frac{\partial P}{\partial x_{i}}+\rho g_{i}=0,
\end{equation}

where $v_i$ are velocities, $P = -1/3\ \sigma_{kk}$ is pressure, $\rho$ is density and $g_{i}$ the gravitational acceleration.
Deviatoric stresses $s_{ij}$ are related to velocities in equation \eqref{eq:momentum_conservation} using a Maxwell visco-elastic constitutive relationship between deviatoric stresses and strain rates \cite<e.g.,>[]{Gerya2010, MoresiEtAl2003} :

\begin{equation}
    \label{eq:maxwell}
\dot{e}_{ij}=\frac{1}{2 G} \dot{s}_{ij}+\frac{1}{2 \eta} s_{ij}
\end{equation}

where $\dot{s}_{ij}$ is discretized using a first-order backward finite difference scheme with time step $\Delta t$, so that the deviatoric stress at time $t$ becomes

\begin{equation}
    \label{eq:s(t)}
    s_{ij}^{t}= 2 Z \eta \dot{e}_{ij}^{t} + (1-Z) s_{ij}^{t-\Delta t},
\end{equation}

with $\dot{e}_{ij} = (\partial v_{i}/\partial x_j + \partial v_{j}/\partial x_i)/2$ and

\begin{equation}
    \label{eq:Z}
    Z=\frac{G \Delta t}{\eta+G \Delta t}.
\end{equation}

The effective viscosity in equation \ref{eq:maxwell} can represent a range of rheologies. A very high value sets a very long Maxwell time, which effectively renders the material elastic. In the viscous regime, $\eta$ can represent brittle plasticity (e.g., as detailed in Section \ref{long_term_rheology}), or a specific creep mechanism of known flow law. In practice, $\eta$ is constructed as the harmonic average of several viscosities, each representing individual flow mechanisms.

The mass \eqref{eq:mass_conservation} and momentum \eqref{eq:momentum_conservation} conservation equations, expressed in terms of velocities \eqref{eq:s(t)} and pressure, are discretized with a conservative finite difference scheme formulated on a staggered grid \cite<e.g.,>[]{GeryaYuen2003}. This leads to a linear system that is solved for velocities and pressure over the entire domain using a direct solver. Retroactions between the viscosity and velocity fields require the use of non-linear iterations (here approximate-Newton, described as Algorithm 2 in \citeA{SpiegelmanEtAl2016}) to reach convergence, which is assessed by comparing the L2 norm of the residual vector to a specified tolerance (relative tolerance between $10^{-7}$ and $10^{-2}$, see \textit{readme} documentation in the code repository linked in the Acknowledgements section).

Once a reasonably converged solution is found, the time evolution is performed explicitly using the time step $\Delta t$ introduced in equation \eqref{eq:s(t)}. This is specifically done by advecting Lagrangian markers which carry material properties such as density, viscosity and friction. Markers are advected within the velocity field interpolated from the nodes. Marker properties are then passed back to the nodes to prepare the next solve of the conservation equations at the next time step. Markers also carry material stresses in order to solve equation \eqref{eq:s(t)}. In addition to being advected, these stress components are also rotated according to the local rotation rate determined from the velocity field at each timestep \cite{Gerya2010}.

\subsection{Numerical implementation of the SCAM rheology}
\label{sec:num}

The implementation of the SCAM model in the 2-D code was performed as described in the following subsections.
First, a damage property and its evolution rules are implemented, along with the shear modulus dependence on damage. Then, a smooth transition to long term plastic behavior in fully damaged parts of the material is introduced. 

\subsubsection{Damage growth and viscosity}

The damage state $D$ is added as an additional variable discretized on both markers and nodes. Its evolution equation (\ref{eq:Ddot}) is solved with a finite difference method.
The damage rate $\dot{D}$ and its associated viscosity $\eta_D$ (equation \ref{eq:eta_dam}) are evaluated on nodes at each non-linear iteration using previous stresses and interpolated damage values from markers. The shear modulus is also altered according to the damage state.

When stepping through time, marker damage is incremented by interpolating the damage rate from nodes to markers. Due to the non-linearity of equation \ref{eq:Ddot}, damage is prone to catastrophic growth, which can be challenging for a numerical solver. We therefore adapt the time step to the dynamics of damage growth by limiting the maximum increment of damage on a node at each time iteration by an amount $\Delta D_{max}$.

\subsubsection{Switching from damaged to plastic rheology after crack coalescence}
\label{long_term_rheology}

$D$ values approaching $1$ can be thought of as a state when the rock looses its macroscopic cohesion through crack coalescence. Our damage rheology is not well suited to represent large strains that may develop beyond this point, for example within localized fault zones. Mohr-Coulomb plasticity, on the other hand, is perfectly relevant to model the frictional rheology of such fault gouges. Crack coalescence is however a necessary condition to the formation of macroscopic fault zones, such that damage growth up to $1$ has to precede Mohr-Coulomb plastic deformation. Our damage model being formulated as an effective Maxwell rheology, we choose to retain this framework in our implementation of plasticity.
We therefore implement a continuous effective viscosity that smoothly switches from our damage viscosity $\eta_D$ to the standard plastic viscosity $\eta_p$ \cite{DuretzEtAl2021} as $D$ approaches $1$.

Because in our micromechanical model crack normals lie in the $\{\sigma_1,\sigma_3\}$ plane, plastic deformation beyond coalescence should be confined to that same plane, and can therefore be modeled with a Mohr-Coulomb yield criterion ensuring that

\begin{equation}
\label{eq:MC_inequality}
s_{II} \leq \sigma_y\ ,
\end{equation}

where $s_{II} = \sqrt{J_2(\mathbf{\sigma})}$, with $J_2(\mathbf{\sigma}) = 1/2\ s_{ij}s_{ij}$ the second invariant of the deviatoric stress tensor. $s_{II}$ is also the radius of Mohr's circle in 2-D incompressible plane strain :

\begin{equation}
\label{eq:MC_radius}  
s_{II} = \frac{1}{2}(\sigma_{3} - \sigma_{1})\ .
\end{equation} 

In equation \eqref{eq:MC_inequality}, the plastic yield stress $\sigma_y$ writes :

\begin{equation}
\label{eq:MC_yield_stress}   
\sigma_y = \sin{\phi_m}\ P + \cos{\phi_m}\ C_m\ .
\end{equation} 

The yield stress is a function of the macroscopic friction angle $\phi_m=\arctan{\mu_m}$ and cohesion $C_m$, as well as of the in-plane pressure $P$ : 

\begin{equation}
\label{eq:inplane_pressure}   
P = -\frac{1}{2}(\sigma_{1} + \sigma_{3})\ ,
\end{equation} 

which in elastically incompressible 2-D plane strain is also equal to total pressure. Satisfying the Mohr-Coulomb yield criterion within a Maxwell visco-elastic framework can be done through an effective ``plastic viscosity'' approach (e.g., \citeA{Gerya2010}; \citeA{DuretzEtAl2021}). As fully-damaged areas become incompressible elastic-plastic zones, equation \eqref{eq:damaged-elastic_constitutive_law} becomes :

\begin{equation}
\label{eq:plastic-elastic_constitutive_law}   
\dot{e}_{ij} = \frac{\dot{s}_{ij}}{2G_{0}\text{f}(D)} + \frac{s_{ij}}{2\eta_p}\ .
\end{equation}

The plastic viscosity $\eta_p$ is set to guarantee that stresses satisfy the Mohr-Coulomb criterion \eqref{eq:MC_inequality} once "broken" material starts behaving as a plastic fault zone. If $s_{II}$ lies below $\sigma_{y}$, $\eta_p$ is effectively infinite (plasticity is not activate), otherwise $\eta_p$ reads (see \ref{ap:eta_plas}) :
\begin{equation}
\label{eq:plastic_viscosity_full}
\eta_p = \frac{s_{II}}{2\left(  \dot{e}_{II} - \frac{\sigma_{y} - s_{II}}{2G_{0}\text{f}(D) \Delta t})\right)\ },
\end{equation}

where $\dot{e}_{II} = \sqrt{J_2(\dot{\varepsilon})}$ is the second invariant of the strain rate tensor.
A smooth transition from damage to plastic viscosity is implemented using a hyperbolic tangent function S$(x)$ that goes from $0$ to $1$ as its argument goes from negative to positive.  We set the lowest viscosity to the plastic viscosity and thus write the continuously differentiable effective viscosity : 

\begin{equation}
\label{eq:effective_viscosity}
\eta_{eff} = \text{S}(\eta_D - \eta_p)\eta_D + \left(1 - \text{S}(\eta_D - \eta_p)\right)\eta_p\ .
\end{equation}

To ensure that the effective viscosity remains plastic when material is fully damaged, we set $\eta_D(D=1)$ to the smallest viscosity that can be resolved by our numerical solver (see Section \ref{sec:2-D_setup}). 

Including the transition to large-strain plasticity, the complete set of differential equations that constitute the SCAM model can be summarized as

\begin{equation}
\label{eq:damaged_elasto_plastic_constitutive_law}
\dot{e}_{ij} = \frac{\dot{s}_{ij}}{2G_{0}\text{f}(D)} + \frac{s_{ij}}{2\eta_{eff}}\ .
\end{equation}

When subjected to loading, the material first responds elastically with $D=D_i$ (its initial damage state $\geq D_0$), until $K_I$ becomes positive and damage starts growing (Figure \ref{model}). At that moment damage-driven alteration of the shear modulus generates an effective damage viscosity which affects the material behavior. Up to peak stress, the damage viscosity is greater than the plastic viscosity since it allows stress build-up, therefore $\eta_{eff}\sim\eta_D$. During the post-peak stress drop, $\eta_D$ quickly drops below the plastic viscosity which then becomes the effective viscosity. This allows the accumulation of large strains under stresses capped by the Mohr-Coulomb yield stress. Said yield stress is computed according to equation \eqref{eq:MC_yield_stress}. In the following, we adopt a macroscopic friction angle that matches the frictional properties of the shear defects, i.e., $\phi_m = \tan^{-1}{\mu}$. We also assume that the fully damaged material is cohesionless, i.e., $C_m=0$. It should be noted that a fully damaged material may also return to an elastic behavior if $s_{II}$ happens to drop below $\sigma_y$. However, its shear modulus will have been permanently reduced by damage ($G = \gamma G_0$).

\subsection{2-D setup: the numerical press}
\label{sec:2-D_setup}

\begin{figure}[t!]
\begin{center}
\noindent\includegraphics[width=0.5\textwidth]{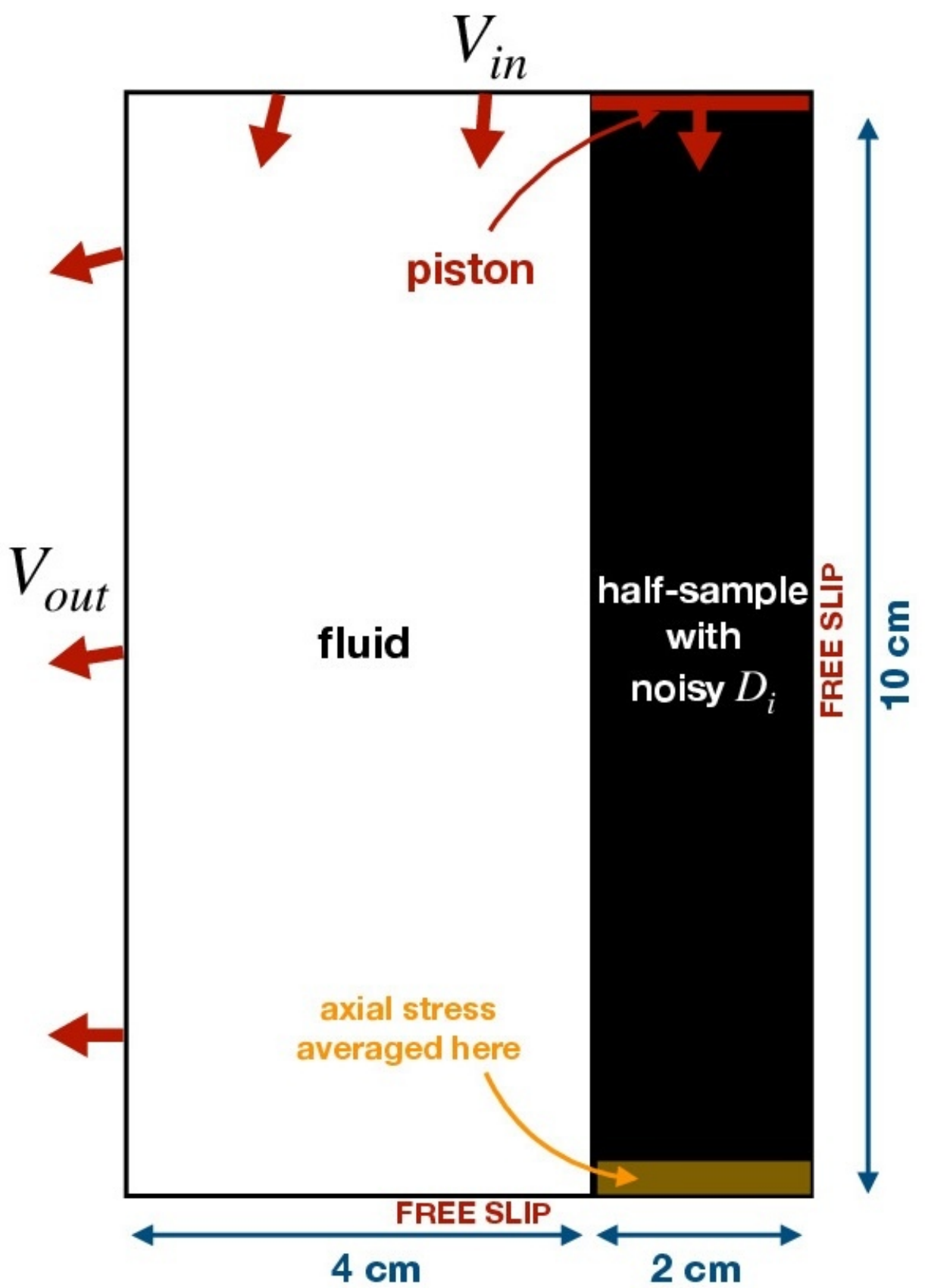} 
\end{center}
\caption{Geometric setup used to simulate experimental deformation in 2D plane-strain. Exploiting the axial symmetry of triaxial experiments, we only model the left-half of a cross section containing the sample axis. The half-sample has a length of $10$ cm and a width of $2$ cm. It is initially seeded with a noisy damage field. A small piston one cell tall (with the same mechanical properties as the sample) is pushed in above the sample. Left of this assemblage lies the confining fluid kept at a constant pressure. Top and left borders are associated with Dirichlet boundary conditions on the velocity component normal to the borders, while the tangential components are left free. The right and bottom borders are free slip. Axial stress is evaluated by averaging vertical stresses in a $0.5$ cm  thick slice at the bottom of the sample.}
\label{setup_2D}
\end{figure}

We construct a 2-D plane-strain analog to the triaxial experimental setup described in Section \ref{sec:triaxial_application}, following the geometry shown in Figure \ref{setup_2D}. This allows us to simulate constant strain rate deformation of a Westerly granite sample, with micromechanical properties determined in Section \ref{sec:model_calibration} (Table \ref{tab:parameters}), up to large strains and including localization.
The axial symmetry of triaxial tests allows us to only consider half of the sample's cross-section. Our geometry thus consists of a half-sample $10$ cm tall and $2$ cm wide on the right side of a wider box ($10.5\times 6$ cm) that includes confining fluid left of the sample, and a $0.5$ mm-thick "piston" above the sample (Figure \ref{setup_2D}). 
Constant strain rate conditions are enforced by pushing material inward from the top of the domain at a constant velocity. The piston is here to ensure that new material flowing in during deformation is not of sample type. Outward velocities are prescribed along the left boundary to preserve a constant volume in the computational domain. The confining fluid is modeled as a low-viscosity Newtonian medium, with pressure imposed at the lower left corner of the domain. Gravity is ignored.
The initial sample damage fluctuates spatially between $D_i = 0.136$ and $0.236$  with an isotropic Perlin noise structure that represents material heterogeneities. 
The spatial domain is discretized using cell sizes of $2 \times 0.5$ mm within $3$ cm of the left wall, and $0.5 \times 0.5$ mm within $3$ cm of the right wall, i.e., the part of the domain containing the sample.

The convergence of Stokes solvers being very sensitive to viscosity contrasts, we restrict the maximum variation in viscosity across the domain to five orders of magnitude. We do so by setting an upper bound on viscosity $\eta_{max}$ such that its associated Maxwell time $\eta_{max}/G$ is $50$ times longer than the time required to elastically reach the peak stress at the imposed axial strain rate. This large viscosity is initially assigned to the sample, rendering it effectively elastic at the beginning of the simulation. 
A lower bound on viscosity in the numerical domain is obtained through $\eta_{min} = 10^{-5} \times \eta_{max}$. This low viscosity is assigned to the confining fluid, ensuring that it behaves viscously throughout the simulation. 
Because the damage viscosity $\eta_D$ drops significantly as damage accumulates, the effective viscosity of the sample will decrease as it begins to fail. As $\eta_D$ approaches $\eta_p$, a smooth transition towards plastic viscosity is 
performed over a viscosity range $|\eta_D - \eta_p| \approx \eta_{min}/50$. Regardless of the viscosity transition, damage keeps increasing until reaching $1$. At this point, it stops evolving and $\eta_D$ is fixed at $\eta_{min}$.
Once parts of the sample are fully damaged, they effectively behave as a Mohr-Coulomb plastic solid with no cohesion and the same (macroscopic) friction coefficient as that determined to act on the microscopic shear defects ($0.7$).

Finally, the comparisons of simulated constant strain rate experiments with laboratory data requires the evaluation of macroscopic axial strains and deviatoric stresses. The axial strains are measured by tracking the displacement of the top boundary of the sample through time, and normalizing it by the initial size of the sample. 
Axial deviatoric stresses are obtained by averaging the vertical deviatoric stress $s_{ax}$ in a horizontal ``stress gauge'', i.e., a $0.5$ cm thick band at the bottom of the numerical sample (Figure \ref{setup_2D}), excluding a cell size length near the left boundary, to avoid any influence from interpolations at the interface between sample and fluid.

\subsection{Results}
\label{sec:numerical_results}

\begin{figure}[t!]
\noindent\includegraphics[width=\textwidth]{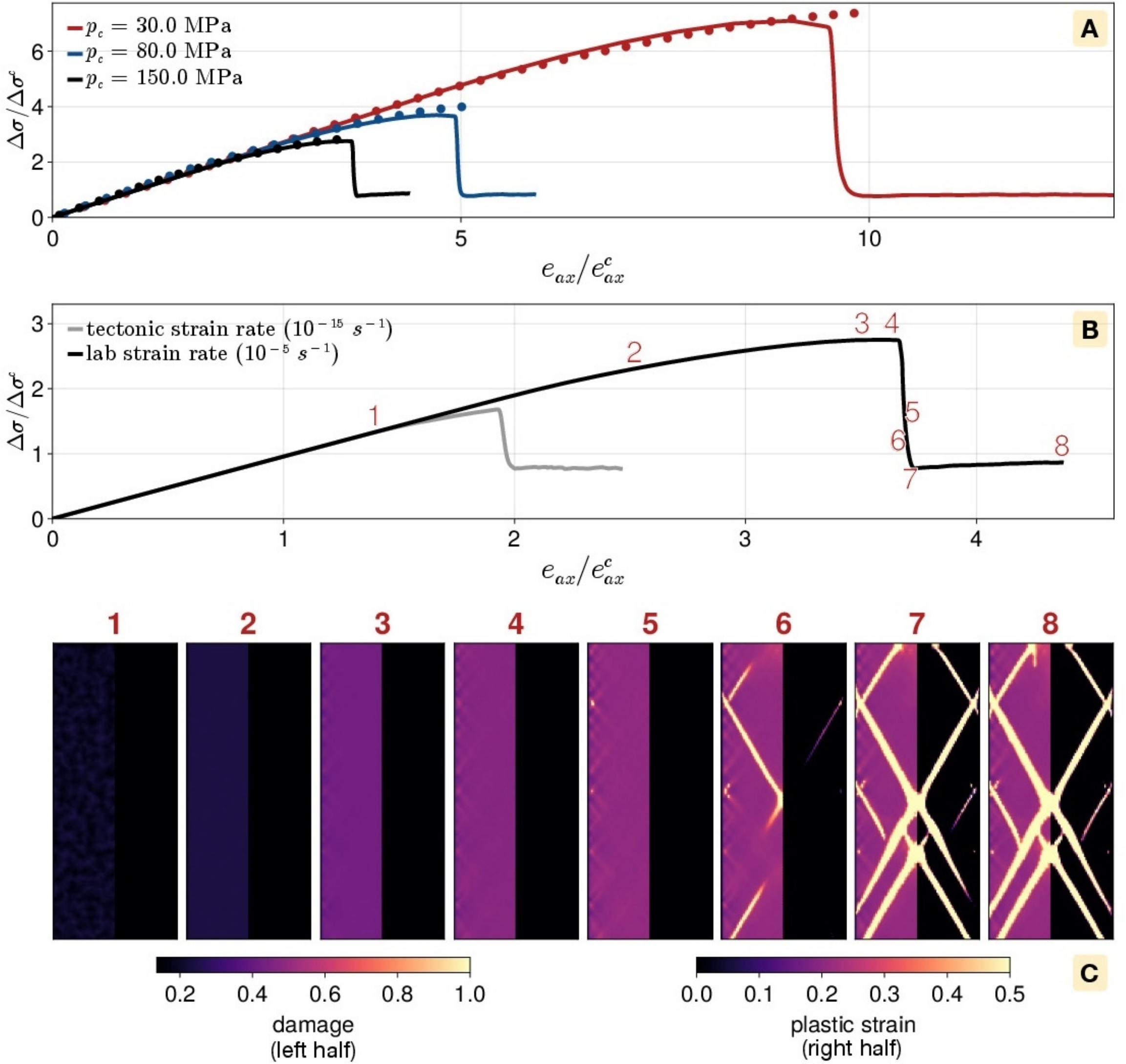} 
\caption{Results of the SCAM model simulations in a 2-D plane-strain setup using Westerly granite parameters under constant strain rate. The numerical samples are initialized with a noisy $D_i$ field. \textbf{A.} Simulated stress-strain curves at $\dot{\varepsilon}_{ax}=-10^{-5}\ s^{-1}$ and confining pressures of 30 (red), 80 (blue) and 150 MPa (black curve), to be compared with the corresponding experimental data (dots). \textbf{B.} Simulation performed at $p_c = 150$ MPa described above compared to a simulation at a ``tectonic'' strain rate of $10^{-15} \ s^{-1}$. Differential stresses and deviatoric axial strains are normalized by their value when $K_I=0$ at $D=D_0$ to allow comparison between triaxial experimental data and 2-D plane strain deformation (see Text). The red numbers in Panel \textbf{B} correspond to the damage (left) and plastic strain (right) snapshots shown in Panel \textbf{C}.}
\label{2D_noseed}
\end{figure}

Figure \ref{2D_noseed} shows results from the numerical press performed under a constant axial strain rate of $10^{-5}\ s^{-1}$ (Panels A and B) and $10^{-15}\ s^{-1}$ (Panel B), and confining pressures of $30$, $80$ and $150$ MPa. Figure \ref{2D_noseed}C illustrates the patterns of damage growth and plastic strain for the simulation performed under $150$ MPa of confining pressure (black lines in Panels A and B), on the left and right halves of each snapshot, respectively. The timing of each snapshot is indicated by the numbers on the stress-strain curve in Panel B. Up to snapshot 3, damage grows in a distributed fashion, which smoothes the initial heterogeneities. Damage increases homogeneously up to $\sim 0.45$ during that stage. Just prior to the peak stress, damage growth starts to localize close to the sample border, forming fast-growing damage bands at angles of $\sim 30^{\circ}$ with respect to the compression direction. They develop within a strain range of less than 0.01 \% (snapshots 5,6,7). Plastic strain is estimated by integrating the second invariant of the inelastic deviatoric strain rate tensor through time (once Mohr-Coulomb plasticity has been activated), and accumulates within fully damaged bands. Plastic shear banding first lags behind damage banding. Once a sample-scale damage band has grown, it effectively becomes a plastic shear band. This process begins as the axial stress drops abruptly (snapshots 6-7). Interestingly, off-band distributed damage does not evolve significantly during shear band development.

Figure \ref{2D_noseed}A compares the stress-strain curves produced in our 2-D simulations to the experiments of \citeA{WawersikBrace1971} described in Section \ref{sec:model_calibration}, performed under the same conditions. To facilitate comparisons between a 2-D plane strain and an axisymmetric setup, we normalize the differential stress $\Delta\sigma = \sigma_3 - \sigma_1$ by its value at $K_I=0$ and $D=D_0$, which is the criterion for the onset of tensile crack growth (even though the crack growth rate is infinitely slow at $K_I = O^+$) :

\begin{equation}
    \label{eq:damage_initiation_stress}
    \Delta\sigma^c = p_c(1-\frac{\sqrt{1 + \mu^2} + \mu}{\sqrt{1 + \mu^2} - \mu})
\end{equation}
The above expression is obtained by applying the Mohr-Coulomb criterion to optimally-oriented planes in a principal stress field. The axial deviatoric strain is then normalized by the deviatoric strain needed to reach $\Delta\sigma^c$  elastically with the reference shear modulus $G_0$ that corresponds to $D=D_0$ :

\begin{equation}
    \label{eq:damage_initiation_strain}
    e_{ax}^{c} = \frac{1}{2G_0} s_{ax}^c.
\end{equation}

In equation \ref{eq:damage_initiation_strain}, $s_{ax}^c$ is the deviatoric axial stress at the onset of microcraking, and is equal to $\Delta\sigma^c/2$ in a 2-D plane strain configuration, and to $(2/3)\Delta\sigma^c$ for a triaxial configuration.
This non-dimensionalization of stresses and strains accounts for the fact that the mean stress, which impacts damage growth and the position of the peak stress, has a different expression in a triaxial vs. plane-strain geometry. It ensures that experiments conducted with the same parameters in either geometry will show the same non-dimensional tangent modulus and peak differential stress.

Our 2-D simulations are in good agreement with the experimental data from \citeA{WawersikBrace1971} (Figure \ref{2D_noseed}A) at all three confining pressures. This shows that the parameters determined by fitting 0-D (point-wise) simulations to triaxial data produce sensible behavior when implemented in a 2-D ``spatialized'' geometry.
Figure \ref{2D_noseed}B shows our reference simulation at $\dot{\varepsilon} = 10^{-5}\ s^{-1}$ and $p_c = 150$ MPa in black, compared to a simulation performed at a ``tectonic'' strain rate of $10^{-15}\ s^{-1}$. The peak stress of the slower simulation is significantly lower than that of the reference simulation, with a loss of strength during macroscopic failure that is approximately divided by 2.

\section{Discussion: A brittle constitutive law rooted in micromechanics}
\label{sec:discussion}

\subsection{Features of brittle deformation captured by the SCAM model}

As illustrated in Sections \ref{sec:triaxial_application} and \ref{sec:2-D_application}, the SCAM model captures a range of features typical of brittle deformation revealed by laboratory experiments (Figure \ref{exp}). These include: (1) the co-existence of several measures of rock strength, such as the intact strength and the residual (i.e., ``pre-cut'') frictional strength, all of which depend on confining pressure (e.g., \citeA{Byerlee1978}); (2) the permanent weakening of elastic properties occurring prior to the peak stress; (3) the strain rate-dependence of brittle strength, which enables (4) the occurrence of brittle creep under constant imposed stress. Here we discuss the parameters of the SCAM framework that control these various macroscopic properties.

\subsubsection{Microscopic vs. macroscopic strength, elastic weakening, and strain rate dependence}
\label{sec:micro_macro}

The SCAM framework involves several thresholds of inelastic deformation. The first is when slip on pre-existing, small-scale shear defects becomes able to wedge open tensile wing cracks. It corresponds to $K_I = 0$ (Figures \ref{model} and \ref{KI_map}), and is closely related to the Mohr-Coulomb criterion, in the sense that opening wing cracks requires a greater differential stress under greater confining pressure \cite{Costin1985}. The second threshold is when cracks have sufficiently lengthened to transition from a non-interacting to an interacting regime. This aspect will be further detailed in Section \ref{sec:retroactions}. The third threshold is when $D$ reaches its maximum value $\sim 1$, at which point cracks coalesce into a macroscopic fault, which is modeled as a shear band with a macroscopic ``bulk'' friction equal to that acting on the microscopic shear defects, and no cohesion. In practice, the second and third thresholds occur in very close succession because the damage growth rate accelerates catastrophically as soon as cracks enter the interacting regime, especially under constant axial strain rate. Some amount of deformation is still required for the shear band to reach its steady-state stress after the third threshold (e.g., after snapshot 7 in Figure \ref{2D_noseed}). After that, the macroscopically broken rock has a ``residual'' strength that is entirely set by its friction coefficient. 

The material's elastic properties are altered by damage, causing pre-peak softening of the rock and permanent weakening of the shear modulus. The ratio of the fully-damaged ($V_s^D = V_s(D=1)$) over the reference ($V_s^0 = V_s(D=D_0)$) shear-wave velocity can be related to the shear modulus weakening ratio $\gamma$, assuming small density variations, as: $V_s^D/V_s^0 \approx \sqrt{\gamma}$. Using $\gamma$ values inverted from Westerly granite and Darley Dale sandstone (Table \ref{tab:parameters}) we obtain shear-wave velocity reductions of $34\%$ and $47\%$, consistent with values measured in the damage zone of natural faults, which range from $20$ to $50\%$ \cite<e.g.,>[]{KarabulutBouchon2007, WuEtAl2009}, as well as laboratory tomography on granite showing a reduction in P-wave velocity of around $50\%$ \cite{AbenEtAl2019}.

The SCAM model accounts for the temporal dependence of brittle deformation via a sub-critical crack growth law, which allows cracks to grow below the fracture toughness of the material \cite{Costin1983,Costin1985}. This assumption introduces a characteristic crack growth time $ a/\dot{l}_0$ that is modulated by the stress intensity factor ($K_I$) and the Charles law exponent $n$ (equation \ref{eq:charles}).
These parameters themselves depend on ambient conditions such as moisture levels \cite{atkinson1979,Eppes2017} or temperature \cite{HeapEtAl2009}. To first order, the strain rate dependence of the SCAM flow law reflects the ratio of the characteristic duration of the deformation of interest to the characteristic crack growth time. A very slow (``tectonic'') experiment will for example leave ample time for cracks to grow, weaken the material and cause macroscopic failure, preventing the build-up of very large stresses (Figure 
\ref{2D_noseed}B). Conversely, experiments conducted under laboratory strain rates will reveal greater peak stress (e.g., Figure 7 of \citeA{Costin1983}). Interestingly, the strain rate dependence of brittle deformation implies that wide portions of the upper crust should behave in an effectively viscous fashion (with viscosity $\eta_D$) when undergoing progressive failure (i.e., prior to localization). This behavior must however be inherently transient because the amount of damage a rock can withstand before macroscopic failure is necessarily finite. A competition between crack growth and crack healing processes may prolong this distributed viscous deformation phase, but is beyond the scope of the present study.

\subsubsection{Retroactions between damage and damage growth rate}
\label{sec:retroactions}

The transition between a regime of lengthening but non-interacting wing cracks, and one of catastrophically-interacting long cracks (Figure \ref{model}) is at the heart of many macroscopic behaviors manifested by the SCAM model. As detailed in Section \ref{sec:damage_growth}, the stress intensity factor ($K_I$) at the tip of wing cracks is constructed as the sum of three terms (equation \ref{eq:KI_terms}). The first two terms lead to a decrease in $K_I$ as wing cracks lengthen, i.e., as $D$ increases. This corresponds to the isolated crack regime. The third term has the opposite effect: increasing $D$ increases $K_I$, and thus the damage growth rate through Charles' law (equation \ref{eq:charles}). This last term becomes dominant at larger values of $D$ and describes the interacting crack regime. The transition between successive regimes is closely related to the convexity of $K_I$ as a function of $D$, and its dependence on the evolving differential stress, as illustrated in Figure \ref{model}E–G.

In constant strain rate experiments, damage starts to grow when $K_I$ becomes positive. This is made possible by elastic loading raising the differential stress at constant initial damage state $D_i$ (vertical trajectories in Figure \ref{KI_map}). When an experiment is started with a low $D_i$, damage growth first occurs in the non-interacting regime, where $\partial K_I / \partial D < 0$ (e.g., purple trajectory in Figure \ref{KI_map}). Damage growth rates are initially very slow, because $K_I / K_{IC}$ raised to a large exponent ($n$ in equation \ref{eq:charles}) gives an extremely slow crack growth speed when $K_I$ barely exceeds $0$. The damage viscosity is initially very high, and the material continues to behave elastically. As damage increases, both the shear modulus and the damage viscosity decrease because of the decreasing $\text{f}(D)$ and $f^2(D)$ terms in equations \ref{eq:G_general} and \ref{eq:eta_dam}. This leads to pre-peak softening of the stress-strain curve. 
The crack growth rate –strongly controlled by $K_I$– is the sole mechanism that can lead to a stress rate decrease. It thus competes with the elastic stress rate increase imparted by far-field loading, controlling the stress level at which material softening occurs. In Figure \ref{KI_map}, this manifests as trajectories aligning on a contour of constant $K_I$, which is greater for a greater imposed strain rate.

Then, the system transitions to the interacting regime where $\partial K_I / \partial D > 0$. The regime transition as illustrated in Figure \ref{KI_map}A connects all the differential stress maxima spanning all values of $K_I$ between $0$ and $K_{IC}$. We write $D_c$ the ``critical'' damage value that marks this regime transition. $D_c$ decreases with increasing differential stress (vertical black dashed curve in Figure \ref{KI_map}A), and verifies:
\begin{equation}
    \label{eq:Dc}
    \left . \frac{\partial K_I(D)}{\partial D}\right |_{D = D_c(\Delta\sigma)} = 0\ .
\end{equation}
Because prior to crossing the regime transition stress trajectories align close to an iso-$K_I$ contour (which depends on strain rate), the value of $D_c$ can be thought of as a decreasing function of strain rate. $D_c$ is bounded by the value of damage that maximizes the differential stress at $K_I = K_{IC}$ (here $\sim 0.42$), and the value that maximizes stress at $K_I = 0$ (here $\sim 0.5$). These end-member cases respectively represent a very fast strain rate experiment, in which cracks would grow critically (at elastic wave speeds), and an extremely long and slow experiment in which cracks can grow sub-critically at $K_I\sim0$.
The upper and lower bound on $D_c$ happen to be close to each other, yielding a narrow range of critical damage values ($\sim 0.42-0.5$ in Figure \ref{KI_map}A). When damage exceeds $D_c$, the system enters the interacting regime, in which an increase in $D$ increases $\dot{D}$ at constant stress, thereby accelerating the reduction of the shear modulus (equation \ref{eq:G_general}) and damage viscosity $\eta_D$ (equation \ref{eq:eta_dam}). The material can no longer accumulate stress, and stresses decrease below their peak value. 
At this point the stress trajectories in Figure \ref{KI_map}A begin to deviate from an iso-$K_I$. For our best-fitting set of parameters, $K_I$ increases drastically as $D$ exceeds $D_c$, which manifests as a sharp stress drop as $D$ approches $1$ (Figure \ref{KI_map}).

An interesting consequence of the fact that pre-peak stress trajectories tend to first align on the same iso-$K_I$ contour regardless of initial damage state is that they all experience a regime transition at the same $D_c$ and at the same peak differential stress (for a given imposed strain rate). In Figure \ref{KI_map}B, this manifests as peak stress magnitudes that are largely insensitive to any value of initial damage lower than $D_c$. This feature of the model is consistent with experimental results from \citeA{WangEtAl2013}, who found similar peak strength in samples initially subjected to varying degrees of thermal cracking, which we interpret to represent varying $D_i$ (i.e., varying wing crack lengths at fixed shear defect size and density).
On the other hand, if an experiment is started with a damage state that exceeds $D_c$, the system will entirely bypass the non-interacting regime and will display very little post-peak softening (e.g., light orange trajectories in Figure \ref{KI_map}A). In this case, the degree of initial damage affects the position of the peak stress.

We note that in a few instances, large values of damage do not lead to a catastrophic stress drop. This occurs for example for low values of $\gamma$ (Figure \ref{0D_params_effect}H), or a high value of $D_0$ (Figure \ref{0D_params_effect}F), where stresses slowly decay over a few percent of axial strain.
The only way to prevent a catastrophic stress decrease is for $K_I$ to decrease as $D$ approaches 1. In Figure \ref{KI_map}A, this would manifest as a steeply decreasing stress trajectory that crosses iso-$K_I$ contours for $D>D_c$. The slope of a stress trajectory in ($D$, $\Delta \sigma$) space is equal to $\frac{\partial \Delta \sigma}{\partial D} = \frac{3}{2}\frac{\dot{s}_{ax}}{\dot{D}}$. Post peak, the deviatoric axial stress rate (equation \ref{eq:csr_ode}) is increasingly dominated by the viscous term, as $\dot{D}$ accelerates. Neglecting the elastic term in equation \ref{eq:csr_ode} yields :

\begin{equation}
    \label{dsdD}
    \frac{\partial \Delta \sigma}{\partial D} = - \frac{3}{2} \frac{|f^{\prime}(D)|}{\text{f}(D)} s_{ax}\ .
\end{equation}

Using the equations for $f^{\prime}(D)$ \eqref{f'} and $\text{f}(D)$ \eqref{Geff}, it can be seen that a low value of $\gamma$ leads to a very steep $\frac{\partial \Delta \sigma}{\partial D}$. This likely explains the gentler stress drops shown in Figure \ref{0D_params_effect}H. We suspect that a large $D_0$ leads to a similar effect on $\frac{\partial \Delta \sigma}{\partial D}$ and accounts for the progressive stress drop in Figure \ref{0D_params_effect}F.
It is noteworthy that our inversion for Westerly granite predicts a sharp stress drop, even though it relies on data that does not span the interacting crack regime ($D > D_c$, i.e., only pre-peak data is used in Figure \ref{wg_fit}A, and minimum creep strain rates in Figure \ref{wg_fit}B).

Our 2-D simulations help us assess how the change of crack growth regime affects the spatial pattern of damage as it transitions from distributed to localized (Figure \ref{2D_noseed}C). In particular, the initial phase of distributed damage growth appears to coincide with the non-interacting regime. Damage increases uniformly, smoothing any pre-existing initial damage heterogeneity, and reaches a near constant value ($\sim 0.45$) in the bulk rock when damage bands begin forming. We interpret this uniform value as related to $D_c$. Specifically, the distributed build-up of damage (snapshots 1, 2 and 3 in Figure \ref{2D_noseed}C) proceeds in the non-interacting regime, in which damage is uniformly capped at $\sim D_c$. Stress concentrations due to numerical noise or prescribed heterogeneities can however trigger the switch to the interacting regime in some portions of the sample (along the sides in snapshot 5 of Figure \ref{2D_noseed}), leading to the localization of damage bands. Interestingly, the stable, uniform growth of damage when $D<D_c$ is probably the reason for the good agreement between our 2-D simulations and our 0-D models, which are by definition ``homogeneous'' (Figure \ref{2D_noseed}A). The post-peak behavior predicted by the SCAM model is however significantly different in 2-D vs. 0-D simulations. This is because it is driven by retro-actions between damage localization within a band and the stress field of the surrounding rock that cannot be captured in a pointwise model.

The distinct regimes of crack growth are also responsible for the two stages of creep observed in our constant stress simulations (Figure \ref{0D_damaged_vs_plast}C, D, see Section \ref{sec:triaxial_application}), as well as in brittle creep experiments (Figure \ref{exp}).
In the representation of Figure \ref{KI_map}A, a constant stress experiment simply maps as a horizontal line starting from any point of the constant strain rate trajectories prior to the peak stress. In order to break the material under constant stress, the $K_I$ trajectory has to remain in the domain $K_I > 0$ up to $D=1$. There thus exists a threshold in differential stress $\Delta\sigma_{bc}$ that must be met for the sample to fail macroscopically. Otherwise, the accumulation of damage under constant stress will decrease $K_I$ all the way to negative values, inhibiting the growth of further damage. This threshold corresponds to the largest differential stress able to produce a stress concentration factor equal to zero. It is represented in Figure \ref{KI_map}A by the summit of the dashed contour of $K_I = 0$, and corresponds to a value of around $720$ MPa for a confining pressure of $150$ MPa with our inverted Westerly granite parameters. Overall, $\Delta \sigma_{bc}$ can be thought of as a theoretical minimum strength of the rock, which is a function of confining pressure only (continuous blue line in Figure \ref{byerlee_results}). 
If a brittle creep test is carried out under a constant differential stress above $\Delta \sigma_{bc}$, the damage state will eventually reach $D_c$ in a finite amount of time, then transition to the interacting regime that allows failure. In this case, the creep test will begin by a decrease in $K_I$ that manifests as a decrease in the macroscopic strain rate referred to as decelerating or primary creep (Figure \ref{0D_damaged_vs_plast}C). 
We note that the minimum brittle creep strain rate should be captured accurately in 0-D simulations, since it corresponds to the strain rate at $D\sim D_c$, the extreme value of damage at which damage can grow in a distributed fashion. Finally, as the system switches to the interacting regime, $K_I$ and the macroscopic strain rate both increase, first slowly then catastrophically, accounting for tertiary creep and failure of the sample.
The idea that the transition to an accelerating regime of crack interaction up to failure corresponds to crossing a threshold in damage $D_c$ explains the observations of \citeA{BaudMeredith1997}, who noted that the transition to tertiary brittle creep coincided with a critical extent of microcracking.

\subsection{Revisiting the Byerlee limit}
\label{sec:new_approach}

\begin{figure}[t!]
\noindent\includegraphics[width=\textwidth]{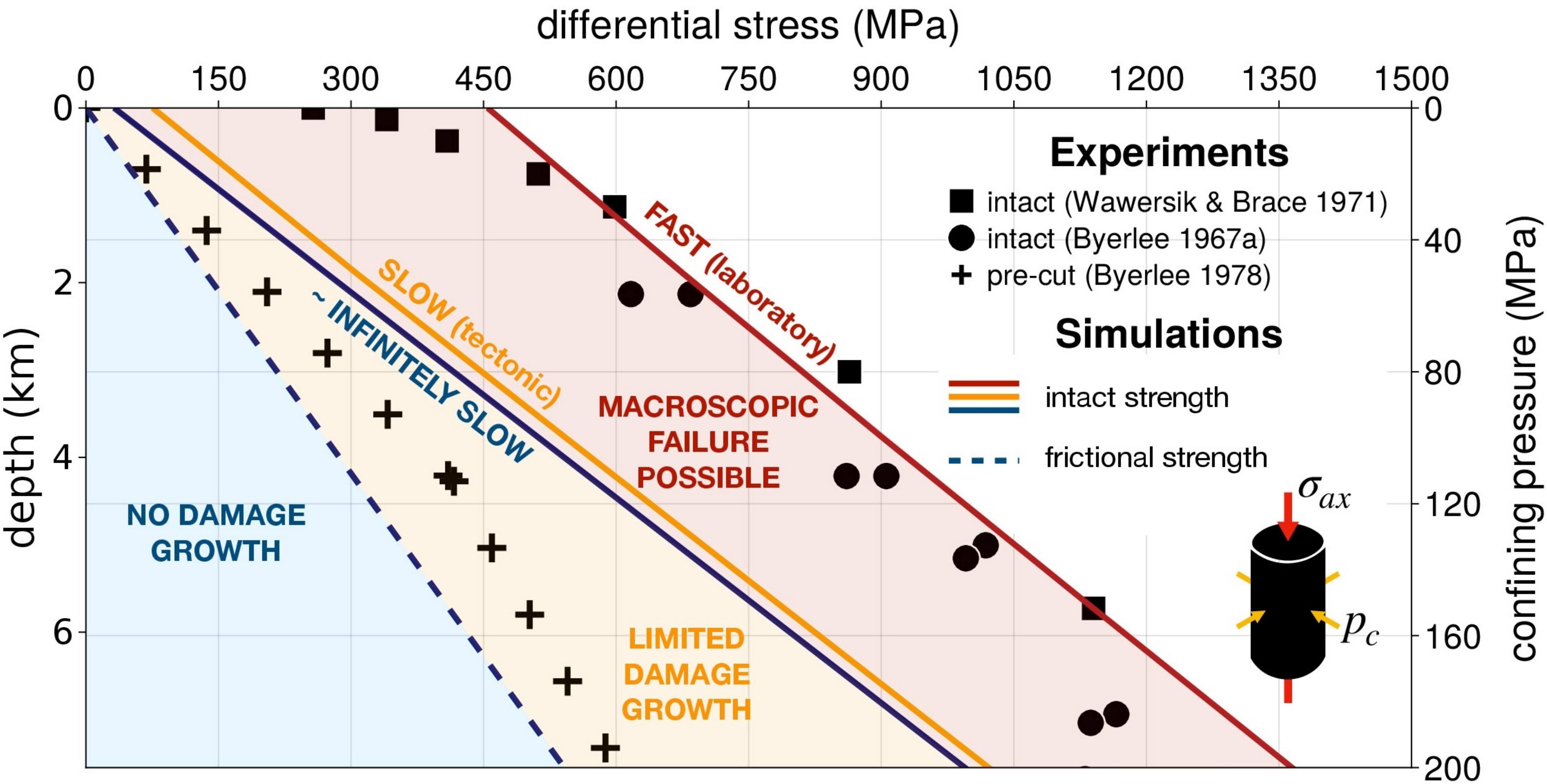} 
\caption{Brittle yield envelopes of the SCAM model calibrated against Westerly granite data (lines), compared to the experimentally-determined strength of intact samples  \protect\cite{WawersikBrace1971,Byerlee1967} (squares and circles) and pre-cut samples \protect\cite{Byerlee1978} (crosses). Red and orange plain lines correspond to peak strengths at $\dot{\varepsilon}_{ax}=-10^{-5}\ s^{-1}$ and $-10^{-15}\ s^{-1}$, respectively. The plain blue line corresponds to the minimum failure strength near the limit of infinitely slow strain rates. Modeled failure envelopes exhibit a constant effective friction coefficient of $\sim0.96$ and strain rate-dependent cohesion. The dashed blue line represents the Mohr-Coulomb yield envelope associated with no cohesion and a friction coefficient $\mu=0.7$, which corresponds to the friction used on shear defects and on macroscopic shear bands in the SCAM simulations.}
\label{byerlee_results}
\end{figure}

Figure \ref{byerlee_results} shows failure envelopes for Westerly granite as determined with intact samples (peak stresses at laboratory strain rate, \citeA<e.g.,>{Byerlee1967}, \citeA{WawersikBrace1971}, black circles and squares), as well as pre-cut samples (the ``maximum friction'' point from \citeA{Byerlee1978}, black crosses).
Both are on the order of hundreds of MPa, and increase linearly with confining pressures in the $20$ to $200$ MPa range. The intact envelope has a steeper slope and lies $\sim 400-500$ MPa above the pre-cut strength.
Both our 0-D and 2-D models reproduce the intact envelope under the same laboratory strain rate of $10^{-5}~s^{-1}$ (red line). 
The simulated envelopes are linear in confining pressure and display an effective cohesion of $94$ MPa (inferred by linear regression of the red curve).
On the other hand, the standard Mohr-Coulomb plasticity framework, with no cohesion and a friction coefficient of $0.7$ provides a good fit to the strength of pre-cut samples (dashed blue line). 

If one was to model the transition from intact to broken through strain-softened plasticity (Figure \ref{lavier}A), the friction should drop from $\sim 0.96$ to $\sim 0.7$, and the cohesion from $94$ to $0$ MPa. This should occur over a very small amount of plastic strain $\Delta e_{II}^p$ to produce a sharp stress drop. It should be noted that a high "intact" friction coefficient, such as $\sim 0.96$, would produce unrealistic shear band orientations (e.g., Coulomb angles of $\sim 23^{\circ}$ between the band and $\sigma_1$).
Within this model, friction is a property of the bulk material that must evolve as deformation accrues. By contrast, within the SCAM framework, friction is an intrinsic property of planar discontinuities in the rock that manifests at two scales. Friction first conditions slip on small-scale discontinuities (shear defects) whose interaction leads to the formation of larger-scale frictional interfaces (macroscopic shear bands). Those two frictional scales are characterized by the same friction coefficient and no cohesion. Until cracks coalesce, frictional sliding only occurs at the scale of shear defects, and its effect on the material is resolved through its induced stress concentration leading to tensile cracking. After coalescence, the broken material acts as a new frictional zone that generates its own stress perturbations on the surrounding ``unbroken'' material. It leads, in 2-D setups, to a shear band growing along a direction in which macroscopic stress concentrations amplify damage growth, yielding a large differential stress drop of hundreds of MPas , which takes place over a very small range of axial strain (Figure \ref{2D_noseed}A, B).

It is remarkable that our best fitting coefficient of friction for Westerly granite data ($0.7$) --which is constrained by data up to the peak strength-- also fits the strength of pre-cut samples (dashed blue line in Figure \ref{byerlee_results}). This supports our approach of switching from a damage model to cohesionless Mohr-Coulomb plasticity while retaining a constant coefficient of friction. This approach has the advantage of producing consistent shear band angles of around $30^{\circ}$ to the most compressive stress (Figure \ref{2D_noseed}C). 

The SCAM framework also allows us to investigate failure at much slower deformation rates because brittle creep data contributes strong constraints on the rate dependence of the pre-peak behavior (Figures \ref{0D_params_effect}A and \ref{2D_noseed}B), which is rooted in sub-critical crack growth. 
As an example, the 0-D failure envelope at a tectonic strain rate of $10^{-15}\ s^{-1}$ is shown by the orange line in Figure \ref{byerlee_results}, and represents a constant strength contrast of $\sim 300$ MPa relative to laboratory strain rates (red line). Compared to the laboratory strain rate simulations, the tectonic strain rate envelope amounts to lower effective cohesion ($\sim 15~$MPa) and a similar effective friction. 
To investigate the model's behavior in the limit of extremely slow strain rates, we construct an estimate of minimum intact strength using the 0-D SCAM model (solid blue line). We calculate it as the differential stress value at $D=D_c$ that would drive a damage growth rate arbitrarily set to $0.1$ per billion year. This yields a line with an intercept that is still significantly greater than zero ($C \sim 6.4$ MPa). Overall, the effective cohesion of the brittle failure envelope can be thought of as a strain rate dependent term that does not entirely vanish in the limit of long loading times (e.g., planetary lifetime). The effective friction coefficient however remains invariant with respect to strain rate.

In summary, while standard strain-softened plasticity treats the brittle limit as an envelope that can move with accumulated plastic strain, the SCAM model treats it as a tenuous failure domain (red area in Figure \ref{byerlee_results}) whose upper boundary depends on strain rate. Specifically, the maximum stress that must be attained for the rock to fail macroscopically can vary by $\sim 400$ MPa between tectonic and laboratory conditions (ten orders of magnitude in strain rate). This effect is, by definition, not captured by rate-independent elasto-plastic models. 
On the other hand, the onset of damage growth is not sufficient to define the lower boundary of the failure domain.
In order to activate inelastic strain, stresses must exceed a threshold that corresponds to $K_I(D_0)=0$, i.e., the activation of tensile cracking through frictional sliding on small defects. Because frictional sliding is indexed on meeting a cohesionless Mohr-Coulomb criterion with a friction of $0.7$, this threshold is closely related to the dashed blue line in Figure \ref{byerlee_results}.
Exceeding this threshold however does not guarantee macroscopic failure. If differential stress remains between the dashed and solid blue lines (orange area in Figure \ref{byerlee_results}), damage can be generated in the isolated crack regime, but will never reach $D_c$ in a reasonable amount of time. The rock will thus never fail.

\subsection{On the meaning of the SCAM micromechanical parameters}
\label{sec:parameters_meaning}

Modeling progressive brittle failure with the SCAM model involves a number of parameters that lend themselves to micro-mechanical interpretations. However, the numerous simplifications made by the wing crack model \cite{AshbySammis1990}, and the use of empirical rules for shear modulus weakening and sub-critical crack growth warrant some caution in doing so. Here we discuss the extent to which parameter values determined by calibrating SCAM against laboratory data provides meaningful information on a rock's microstructure.

Previous studies applying the wing crack micromechanical framework \cite{AshbySammis1990}, such as \citeA{BhatEtAl2011} and \citeA{BrantutEtAl2012}, made the assumption that wing crack initiation happens at $K_I=K_{IC}$. In that case, the analytical expression for $K_I$ (equation (3) in \citeA{AshbyHallam1986}) can be used to constrain $\mu$, $a$ and $K_{IC}$ given prior knowledge of $a$ or $K_{IC}$, and of the axial stress of a sample at the onset of microcracking $\sigma_{1c}$, for various confining pressures. The onset of microcracking is typically indexed on the onset of dilatancy or acoustic emissions \cite{BraceEtAl1966a,AshbySammis1990} (Figure \ref{exp}). Applying this method to a linear fit of the \citeA{BraceEtAl1966a} data ($\sigma_{1c}=3.6 \sigma_{3}+100\ \mathrm{MPa}$) yields $\mu = 0.69$, $a=0.58~$mm, assuming the fracture toughness $K_{IC} = 1.29$ MPa$\cdot$m$^{1/2}$ determined by our joint inversions in Section \ref{sec:model_calibration}. This coefficient of friction is very close to our estimate, but the crack radius is two orders of magnitude larger than our value of $6.66\ \mu$m, closer to the rock's grain size: the length scale considered susceptible to drive wing-cracking. 
For Darley Dale sandstone, \citeA{WuEtAl2000} used the same methodology and obtained $\mu=0.69$ as well as $K_{IC} \leq 0.1$ MPa$\cdot$m$^{1/2}$, assuming $a=0.11~$mm. These values differ from our inversion results of $\mu=0.51$, $K_{IC}=1.4$ MPa$\cdot m^{1/2}$ and $a=0.65$ mm, especially for $K_{IC}$. These differences may be attributed to our use of a sub-critical crack growth law as opposed to $K_I = K_{IC}$. We also note that because $K_{I}$ scales as $\sqrt{a}$, our inversions do a better job at constraining a ratio of $K_{IC} / \sqrt{a}$ rather than $K_{I}$ or $a$ individually. Some trade-off is therefore expected between these values.
It is noteworthy that the data against which the SCAM model parameters were inverted are much more complete than the data used in the studies discussed above. Specifically, our data includes stress-strain curves up to the peak stress (as opposed to a single point: the onset of dilatancy). We however acknowledge that at the peak stress, the discrepancy between the wing-crack model and the real rock microstructure may become large, introducing some bias in our parameter estimate.

Inspecting Figure 12 of \citeA{AshbySammis1990} also reveals that wing-crack based damage mechanics can fit the very first portion of the experimental failure envelope of Westerly granite (i.e, at confining pressures between $0$ and a few $100$s of MPas), but fail to predict peak stresses at greater pressures due to the non-linear dependence of intact rock strength with respect to confining pressure (visible in Figure \ref{byerlee_results}: squares and circles). 
\citeA{AshbySammis1990} suggested that the curvature of the experimental failure envelope of Westerly granite might result from low temperature ductile flow occurring within weak granite minerals. This hypothesis was tested by \citeA{BhatEtAl2011} by modeling a bi-mineralic quartz-feldspar assemblage with a dislocation glide flow law. It was found consistent with the experimental failure envelope at greater confining pressures. 
Broadly speaking, inverting experimental data over a large range of pressures using a model based only on fracture growth likely neglects processes that can significantly affect the mechanical response, such as intra-grain dislocation glide. This leads to inverted fracture parameters whose value can deviate from their expected range, since they are forced to explain behaviors caused by other deformation mechanisms. The exact value of the parameters should therefore not be over-interpreted. The value of the SCAM model lies more in its ability to extrapolate micromechanics-based rock behavior to larger scales, rather than in its informative power about rocks' intrinsic parameters.

\section{Towards an application to tectonic problems}
\label{sec:tecto}

In Section \ref{sec:triaxial_application}, we introduced and calibrated the 0-D SCAM micro-mechanical model against experimental data to capture the mechanical behaviors of Westerly granite and Darley Dale sandstone in the brittle regime. Section \ref{sec:2-D_application} was dedicated to the implementation of this model in a 2-D plane strain tectonic solver to investigate the behavior of the calibrated model when strain localization is made possible. We validated the model by showing that it still accurately predicts experimental data in 2-D. Next, we turn to the end-goal of the SCAM framework by showcasing initial attempts at using it to model tectonic deformation.

\subsection{Crustal-scale numerical setup}
\label{sec:tectonic_setup}

\begin{figure}[t!]
\noindent\includegraphics[width=\textwidth]{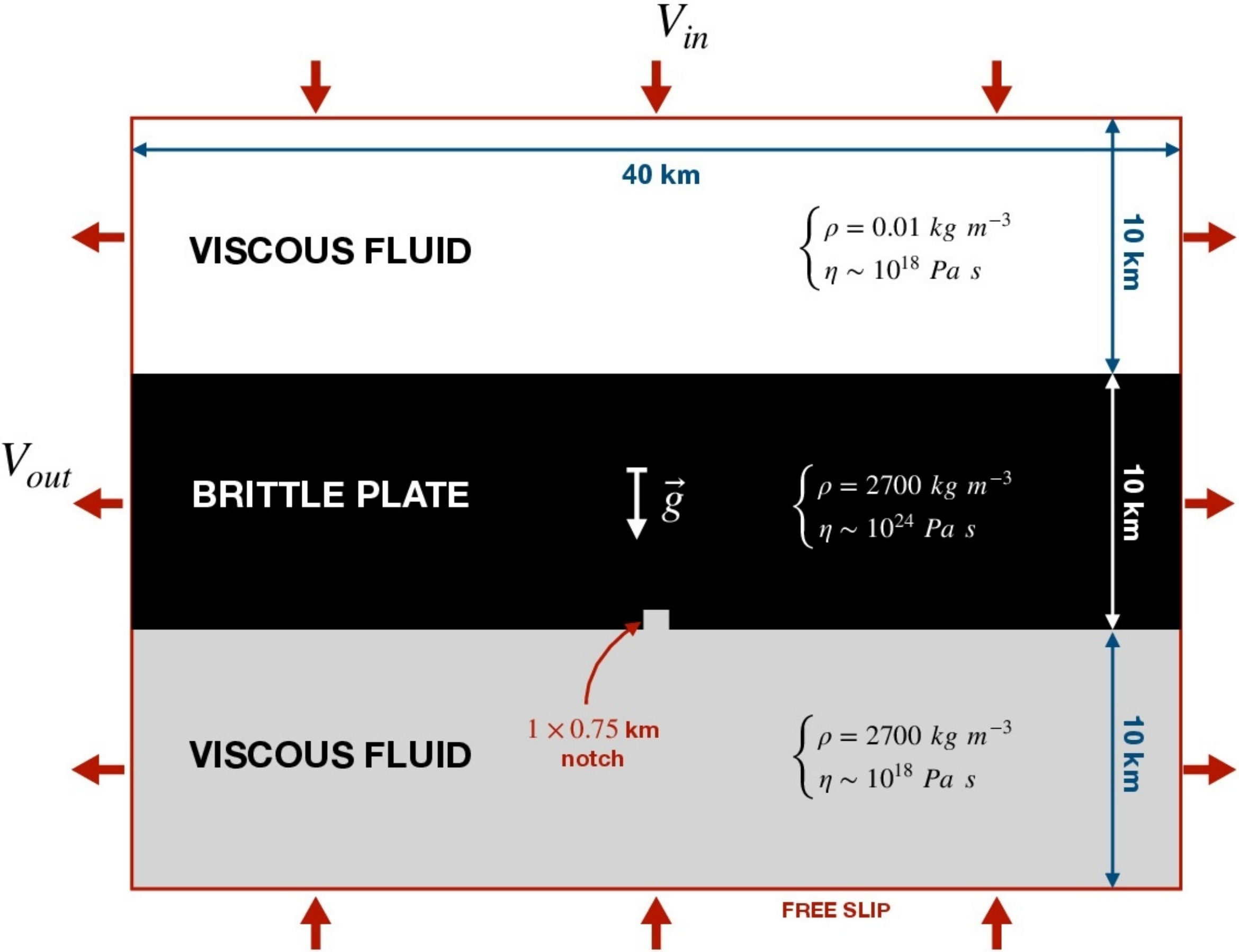} 
\caption{Numerical setup used to simulate the stretching of a rigid crustal unit in 2-D plane strain. The domain is $40~$km long and $30~$km thick and contains two $10$-km thick low-viscosity layers at the top and bottom of a brittle plate of the same thickness. The brittle plate is notched by a 1 $\times$ 0.75 km protrusion of the underlying viscous fluid. Constant outward velocities at the left and right boundaries apply a constant stretching rate on the brittle plate, while the top and bottom boundaries are also assigned constant inward velocities to satisfy volume conservation. All boundaries are free of shear tractions.}
\label{fig:tecto_setup}
\end{figure}

As a first step toward using the SCAM model for long-term tectonic problems, we focus on the initial stages of faulting of a $10$-km thick brittle plate overlying a $40$-km wide and $10$-km thick low-viscosity Newtonian medium subjected to gravity ($9.8~$m$~s^{-2}$) and to a constant horizontal extension rate (Figure \ref{fig:tecto_setup}). An additional upper layer of low-viscosity ``sticky air'' ensures the brittle plate has a traction-free top boundary \cite{Gerya2010}.
The plate is notched at the center of its lower edge by a $1~$km long and $0.75~$km thick protrusion of the underlying viscous layer to promote strain localization in the middle of the domain. The latter is discretized using a cell size of 100 $\times$ 250 m within 8 km of the top and bottom walls, and an greater resolution, with a cell size of 100 $\times$ 100 m in the remainder of the domain containing the brittle plate.
The plate and the fluid layer underneath it are assigned a density of $2700~$kg~m$^{-3}$, while the uppermost fluid layer's density is $0.01~$kg~m$^{-3}$, to ensure negligible pressures at the top of the plate. The left and right sides are prescribed a fixed outward horizontal velocity amounting to a constant extension rate of $10~$cm/yr (strain rate of $\sim 10^{-13}\persec$). The velocities of the top and bottom boundaries are set to satisfy volume conservation within the domain and to preserve the height of the brittle plate's top surface. All boundaries are free slip. 
As in simulated experiments, the maximum viscosity $\eta_{max}$ allowed in the domain is chosen such that its Maxwell time ($\eta/G$) is $50$ times longer than the longest simulation time. This guarantees that the brittle plate --whose viscosity is set to $\eta_{max}$-- retains an elastic response throughout the entire simulation. The exact value of $\eta_{max}$ will therefore not matter as long as it is large enough. To avoid large, computationally challenging viscosity contrasts within the model domain, the lower bound on viscosity is set to 6 orders of magnitude below the brittle plate's viscosity, and is assigned to the air and viscous lower layer. This guarantees that they behave as low-viscosity fluids during each simulation.
This setup resembles that used by \citeA{LavierEtAl2000} and \citeA{OliveEtAl2016} to investigate the effect of brittle strain softening and elasticity on extensional tectonic styles, within the standard elasto-plastic framework. One notable difference is that here the base of the brittle plate is a lithological boundary that gets advected as the plate thins (unlike, e.g., a thermal boundary that may experience diffusion).

We run two suites of simulations, one with the SCAM parameterization of brittle failure in Westerly granite (Table \ref{tab:parameters}), and one with the standard strain weakened elasto-plastic (EP) approach, similar to the runs of \citeA{OliveEtAl2016}.
Tectonic simulations using the SCAM model, like those presented in Section \ref{sec:2-D_application}, include a switch to standard elasto-plasticity in fully damaged areas. Specifically, the damage viscosity is set to smoothly transition to a plastic viscosity as $\eta_D$ approaches $\eta_p$ over a viscosity range of $|\eta_D - \eta_p|$ equal to $\eta_{min}/50$. The locally broken material (\ie, $D=1$) behaves as a Mohr-Coulomb plastic solid, with no cohesion and the same friction coefficient as that of the shear defects.

Standard elasto-plastic simulations, on the other hand, are parameterized to match the intact strength of the SCAM model calibrated on Westerly granite (Figure \ref{byerlee_results}) under a laboratory strain rate ($10^{-5}\persec$, red curve) and under a tectonic strain rate ($10^{-15}\persec$, orange curve). These envelopes correspond to a similar friction coefficient $\mu=0.96$ and cohesions $C$ of $94\MPa$ and $15\MPa$, respectively (see Section \ref{sec:new_approach}). For concision, these simulations will be referred to as \emph{EP-lab} and \emph{EP-tecto}.
In both cases the frictional parameters are linearly weakened over a critical amount of accumulated plastic strain $\Delta\varepsilon^p_{II}=0.1$ down to $\mu=0.7$ and $C=0\MPa$, the frictional properties of the shear defects. 

\subsection{Development of fault networks}
\label{sec:pattern}

\begin{figure}[t!]
\noindent\includegraphics[width=\textwidth]{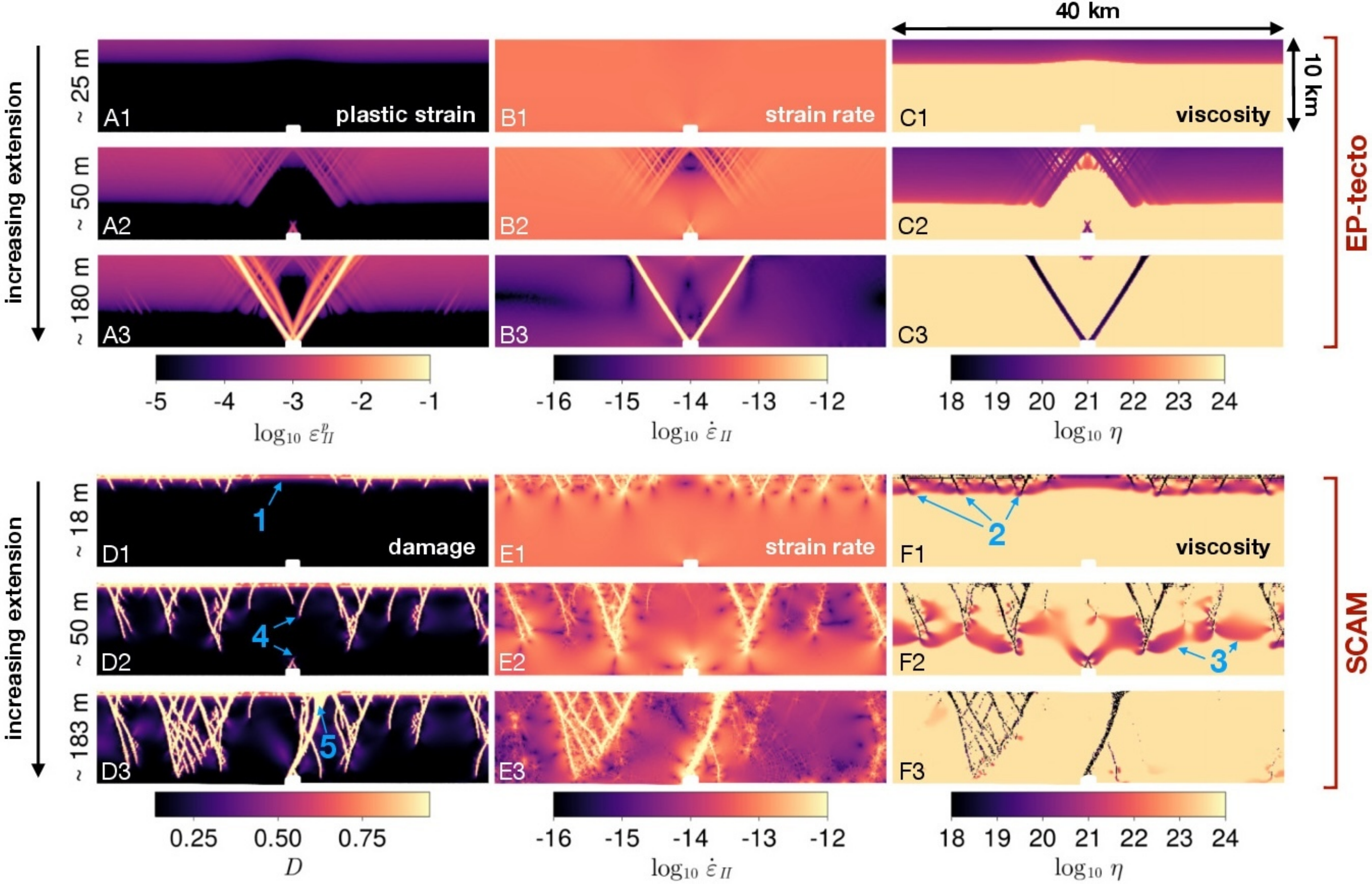} 
\caption{Snapshots of an early rifting simulation ($10~$cm/yr) in a $10$-km thick brittle plate using the elasto-plastic EP-tecto parameterization (\textbf{A–C.}) vs. the SCAM model (\textbf{D–F.})
Each row corresponds to a specific time (i.e., amount of finite extension) in the simulation. The first column shows the accumulated plastic strain for EP-tecto, and the damage field for the SCAM model. The second column shows the second invariant of the deviatoric strain rate tensor, and the third column displays the viscosity field within the brittle plate.}
\label{fig:tecto_snapshots}
\end{figure}

We first compare the faulting patterns produced by an elasto-plastic rheology (EP-tecto) vs. the SCAM rheology during early rifting (up to $\sim 180~$m of total extension). Figure \ref{fig:tecto_snapshots} shows successive snapshots of accumulated plastic strain $e^p_{II}$ (A1 to A3), second invariant of strain rate $\dot{e}_{II}$ (B1 to B3), and viscosity (C1 to C3), as extension of the EP-tecto plate progresses. Plastic yielding starts from the surface of the plate, where pressure, and therefore yield stress, is lowest. It progressively deepens as the yield criterion is met deeper and deeper due to elastic loading of the plate (Panels A1 and A2). Plastic yielding initially mostly develops in a distributed fashion (second row), with plastic strain localization occurring once almost half of the plate has reached yielding. The zone of distributed yielding closer to the viscous protrusion, then transforms into an area densely populated with shear bands, of dip angle near $55$\degree, each accommodating a very small fraction of the total extension rate (second row). Stress concentrations around the basal notch eventually lead to the formation of a pair of shear bands symmetrically cutting across the brittle plate, following the path of pre-existing superficial shear bands. Plastic strain accordingly localizes along two major antithetic shear bands (third row). The formation of these faults relax the elastic stresses within the plate and inhibit plastic yielding in the remainder of the plate (Panels B3 and C3).

The same setup using the SCAM flow law (Figure \ref{fig:tecto_snapshots}D–F) shows a different story. Damage first increases uniformly within $\sim 1~$km below the surface of the brittle plate (Panel D1, blue arrow 1). After shallow damage exceeds values of 0.4-0.5 ($\sim D_c$), damage localization proceeds through the downward propagation of damage bands which promptly turn into plastic shear bands (first row). This occurs when the isolated crack regime transitions to the interacting cracks regime (see Section \ref{sec:retroactions}). The thickness of the shallow distributed damage zone ($\sim 1~$km) is set by a competition between damage growth being activated deeper and deeper as $K_I$ turns positive with loading, and the unstable growth of damage bands (which soon turns plastic) once $D$ exceeds $D_c$. Damage banding unloads the surrounding material by releasing elastic stress through inelastic flow. The growth of these shear bands is self-promoted as the result of stress concentrations at their lower tips, which locally accelerate crack growth and coalescence. This dynamic is clearly visible on the viscosity snapshots, for example Panel F1, where each fault has its own associated lobe of low viscosity (blue arrow 2), driven by damage growth. These lobes are accompanied by a front of low (damage) viscosity which deepens through time (Panel F2, blue arrow 3). This reflects the fact that with continued far-field loading, the depth at which $K_I$ turns positive increases.

Growing faults tend to shield each other and alter the stress field in their vicinity, which leads to the development of complex networks shaped in a tree-like fashion (\eg, the imbricated fault structure that forms on the left side of the plate in Panels D3–F3). 
After about $50~$m of stretching, the concentration of tensile stresses around the viscous protrusion promotes the growth of a major fault, which first grows upward with a downward convex shape (Figure \ref{fig:tecto_snapshots}D2–F2), connecting with one of the deepening faults (Panel D2, blue arrow 4), before a more favorably oriented branch eventually grows and bypasses the less favorably oriented pre-existing segment of the fault. This master-fault's dip angle increases from $\sim 60\degree$ at the bottom of the plate, to almost $90\degree$ close to the surface. As extension progresses, the uppermost part of the fault (Panel D3, blue arrow 5) rotates towards gentler dips, which progressively widens the thickness of the damage band near the surface, as shown in the last row of snapshots (blue arrow 5 in Panel D3). The dip angle of secondary faults is variable and ranges from $60\degree$ to $75\degree$ with segments locally close to vertical, reflecting the spatially variable distribution of stresses induced by the complex array of faults.

The fault angles with respect to $\sigma_1$ (here vertical) are overall smaller in the SCAM tectonic simulations (15-30\degree) than what was observed in our 2-D simulations of triaxial experiments ($\sim 30\degree$, see Section \ref{sec:numerical_results}). These faults are closer to the Coulomb angle (23\degree) associated with the effective intact strength coefficient of internal friction of Westerly Granite ($\mu=0.96$), than with the coefficient of friction at the micro-crack scale ($\mu=0.7$). Some deviation from the theoretical value likely stems from heterogeneities in the stress field that develop through stress concentrations and complex interactions between growing shear bands. In simulation EP-tecto, the master fault orientations of around 33\degree, correspond to the Arthur angle $\theta_A = 45\degree - (\phi + \psi)/4$ \cite{ArthurDunstan1977} associated with the prescribed initial internal friction coefficient of $0.96$ without any plastic dilatancy ($\psi=0$), which is a typical shear band angle found (along with the Coulomb angle) in numerical elasto-plastic simulations \cite{Kaus2010}. 

\begin{figure}[t!]
\noindent\includegraphics[width=\textwidth]{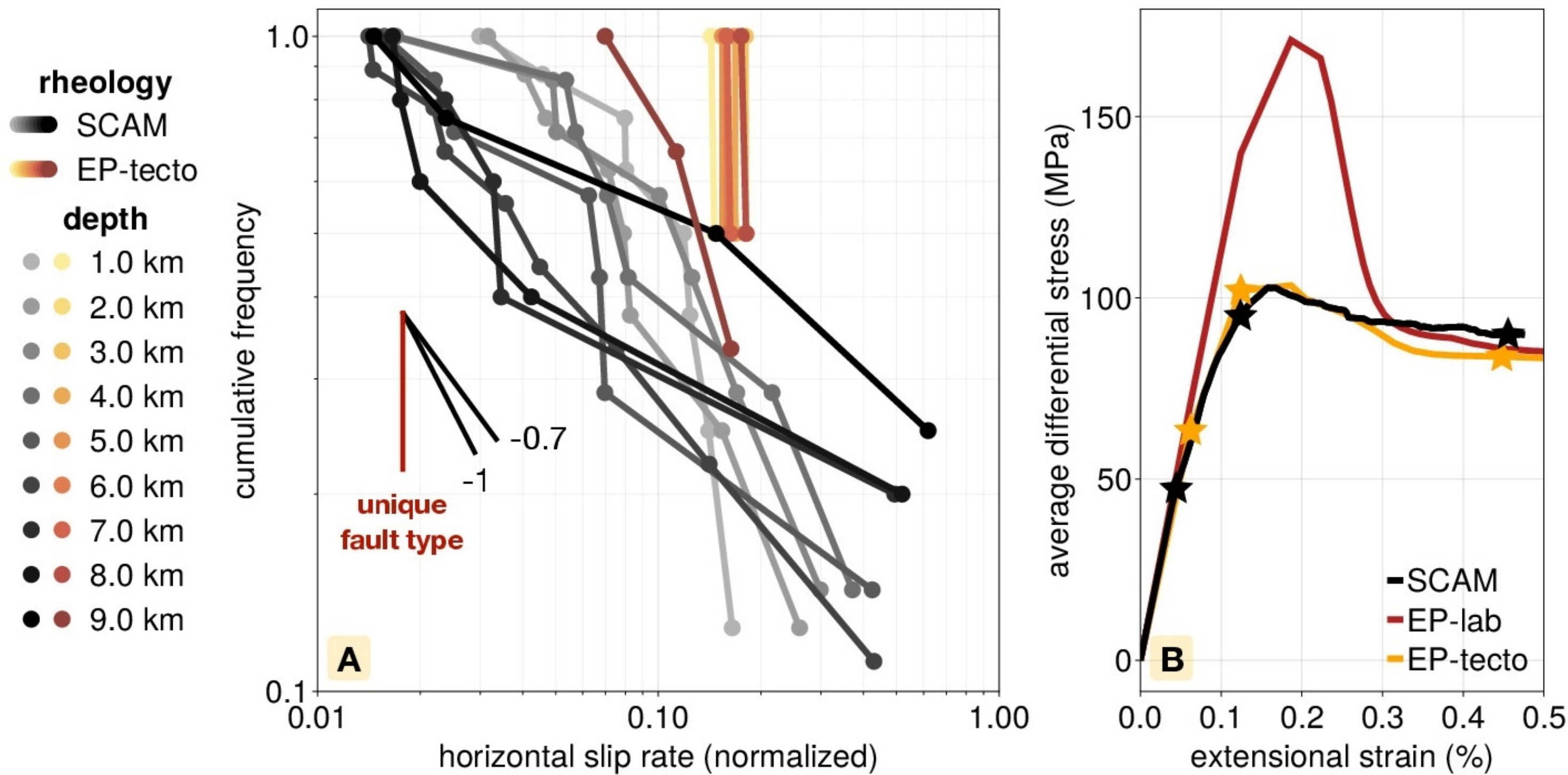} 
\caption{\textbf{A.} Distribution of slip rates on fault populations sampled along different depths in a SCAM-based simulation (grey lines) vs. simulation EP-tecto (yellow-to-red lines). Faults are measured after $180$ m of total stretching, corresponding to Panels A3–C3 and D3–F3 of Figure \ref{fig:tecto_snapshots}. \textbf{B.} Differential stress averaged across a $1~$km-wide vertical band along the left side of the extended brittle plate, plotted as a function of total extensional strain. The black curve corresponds to the SCAM simulation, whereas the red and orange curves are associated to elasto-plastic simulations parameterized with short (EP-lab) and long-term (EP-tecto) intact strengths. Stars mark the timing of the snapshots displayed in Figure \ref{fig:tecto_snapshots}.}
\label{fig:power_law_tecto}
\end{figure}

Contrary to the EP-tecto simulation, the initiation of a fault cutting across the plate does not inhibit the activity on secondary faults, which continue to accommodate a significant amount of extension in the SCAM run (Figure \ref{fig:tecto_snapshots}, Panel B3 vs. E3). We note that EP simulations can produce such behavior, but it generally requires a small strength contrast between the fault zone and surrounding lithosphere \cite{LavierEtAl2000}.
To further characterize the partitioning of strain in the SCAM runs, Figure \ref{fig:power_law_tecto}A represents the distribution of slip rates on the population of faults, sampled at different depths in the plate after $\sim180~$m of extension (Figure \ref{fig:tecto_snapshots}A3–C3 and D3–F3). We identify normal faults by locating positive peaks along a profile of horizontal strain rate $\dot{e}_{xx}$ along a line of constant depth. To mitigate resolution issues when two neighboring peaks are found, we only retain the larger of any two peaks distant by less than 4 cell sizes. We focus on peaks where strain rate exceeds $4 \cdot 10^{-13}\persec$, which amounts to five times the background horizontal strain rate imposed by boundary conditions. The slip rate on each fault is then calculated assuming that the thickness of the faults is one cell length ($100~$m in these simulations). 

Lastly, we find that the slip rates of the incipient fault population produced by the SCAM model follows a power law distribution of exponent close to $1$ (grey lines in Figure \ref{fig:power_law_tecto}A). This distribution likely originates in the tree-like, near-fractal nature of the fault network that emerges within the brittle plate. Simulation EP-tecto, on the other hand,  rapidly produces two dominant faults with very similar slip rates, and virtually no smaller fault with slower slip rates. The distribution of fault slip rates produced by the SCAM run evokes the power law distributions exhibited by natural populations of normal faults. For example, the lengths and offsets of Basin and Range normal faults have been shown to follow a power law distribution with exponent close to unity \cite<\eg, >{ScholzEtAl1993,MarrettEtAl1999}. Natural rift systems are also known to partition extension onto multiple faults of varying sizes, such that all minor intrabasinal faults typically accommodate as much strain as a single half-graben border fault \cite{Morley1996}.
Standard elasto-plastic models tend to localize deformation onto a small number of lithosphere-scale faults \cite{OliveEtAl2016}, but can also produce a wide range of fault sizes in the early stages of rifting, provided sufficient resolution is used \cite{Naliboff2020,Pan2022}. In this case, the distribution of fault sizes and offsets appears to depend on the amount and rate of strain softening, as well as on the overall extension rate \cite{Naliboff2020}. The SCAM model provides a novel framework to further explore this dependence without resorting to ad-hoc softening rules.

\subsection{Strength of the brittle lithosphere}

Figure \ref{fig:power_law_tecto}B shows the differential stress averaged along the left side of the model domain throughout our early rifting simulations. All show an initial elastic increase, followed by a peak and a softening phase. The average peak stress reached in the SCAM simulation is $\sim 100\MPa$. The softening phase is much less acute than in our simulations of triaxial experiments (Figure \ref{2D_noseed}). This possibly reflects the progressive nature of the development of the fault network (Figure \ref{fig:tecto_snapshots}). The peak stress roughly coincides with the moment when a fault first connects the top to the bottom of the plate, shortly after snapshots D2–F2 in Figure \ref{fig:tecto_snapshots}. The steady, long-term stress ($\sim 90\MPa$ in Figure \ref{fig:power_law_tecto}B) is attained when one of the throughcutting faults has clearly developed into the dominant one accommodating the largest extension rate (Figure \ref{fig:tecto_snapshots}E3).

On the other hand, simulation EP-lab illustrates what happens when one models rifting with strain-softened elasto-plasticity calibrated on experiments conducted at laboratory strain rates (Figure \ref{byerlee_results}). The plate reaches a significantly greater peak stress of $\sim 160\MPa$ (Figure \ref{fig:power_law_tecto}B), and experiences more drastic weakening down to its long-term frictional strength, which is inferred from pre-cut samples ($\mu = 0.7$ and no cohesion). By contrast, simulation EP-tecto, which is calibrated to reproduce the intact strength of samples predicted by the SCAM model at tectonic strain rates (yellow line in Figure \ref{byerlee_results}), unsurprisingly produces a stress-strain curve similar to the SCAM simulation (Figure \ref{fig:power_law_tecto}B).

Through its built-in dependence on strain rate, the SCAM rheology can extrapolate crustal strength under tectonic conditions, even though it is entirely calibrated on laboratory data. It thus constitutes a promising alternative to the standard approach in tectonic modeling, which consists of using ``Byerlee's law'', i.e., assigning the pre-cut sample strength to intact brittle lithosphere (\cite{BraceKohlstedt1980}). The SCAM flow law has the advantage of using a single value of friction and a handful of micro-mechanical parameters to predict the lithosphere's intact strength across ten orders of magnitude of strain rate. It self-consistently handles the transition from intact to broken (pre-cut), and does not require any empirical assumption on weakening strain. Alternatively, SCAM can be used to prescribe an intact strength that is appropriate at tectonic strain rates, for use in standard elasto-plastic models (e.g., simulation EP-tecto). This approach, however, does not solve the problem of the ad-hoc weakening strain.

\subsection{Growth of geological structures}

In order to compare the tectonic structures formed after greater amounts of finite extension by the SCAM and elasto-plastic (EP-tecto) rheologies, we performed three additional simulations under strain rates of $1~$mm/yr, $1~$cm/yr and $10~$mm/yr for each rheology. Due to the computationally demanding nature of the SCAM model, we decreased the resolution of the spatial domain using a cell size of 250 $\times$ 500 m within 8 km of the top and bottom walls, and a greater resolution, with a cell size of 250 $\times$ 250 m in the rest of the domain containing the brittle plate. 

\begin{figure}[t!]
\noindent\includegraphics[width=\textwidth]{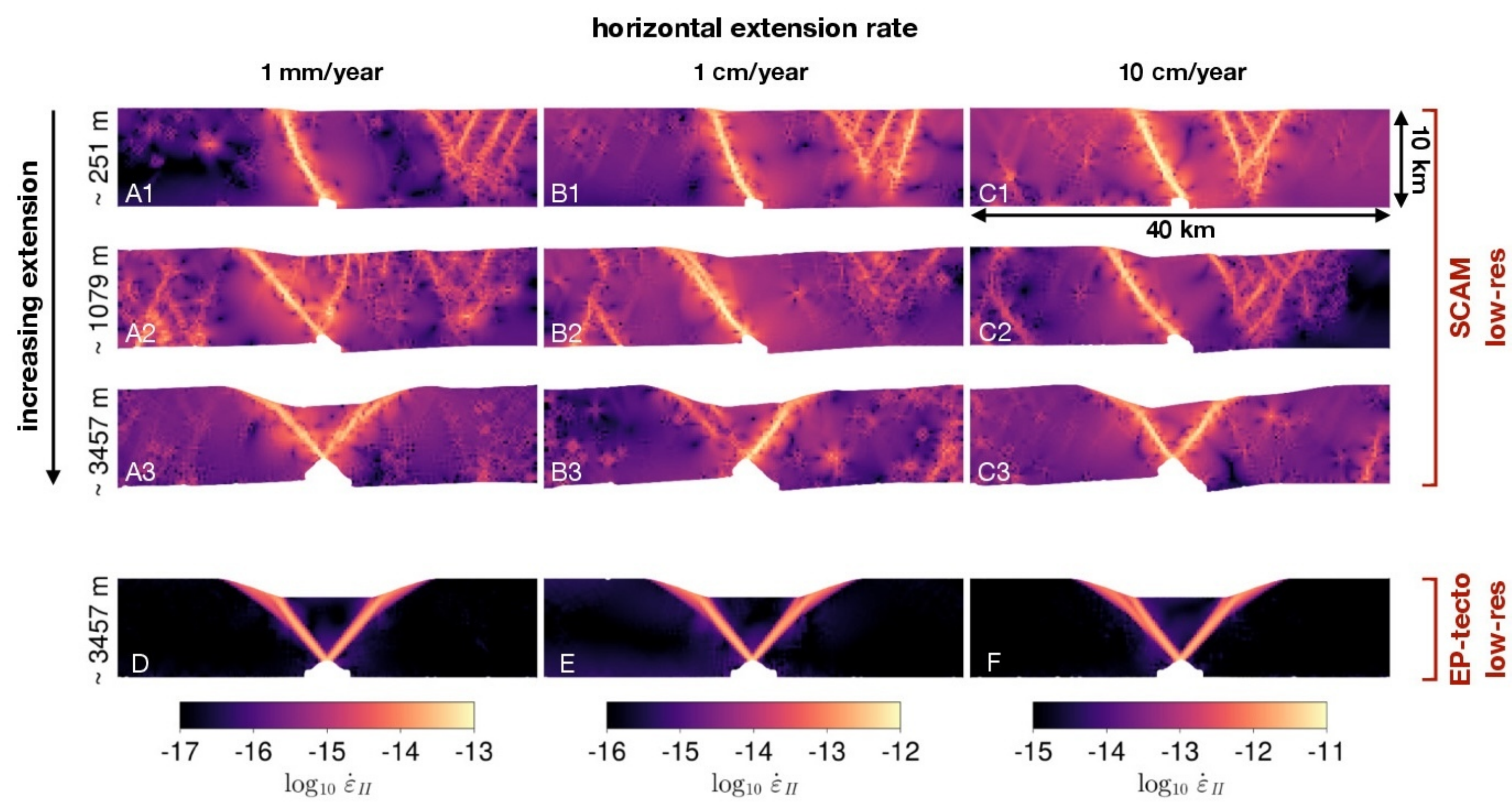} 
\caption{Numerical simulation of about $3.5$ km of extension of a $10$-km thick brittle plate, at varying rates (columns). Snapshots of the second invariant of strain rate at different amounts of finite extension (shown on the left), for a brittle plate governed by \textbf{A–C.} the SCAM flow law, and \textbf{D–F.} the EP-tecto rheology. The model resolution has been reduced compared to Figure \ref{fig:tecto_snapshots} to reach greater amounts of finite extension in reasonable time.}
\label{fig:tecto_strain_rate}
\end{figure}

Snapshots of the second invariant of strain rate ($\dot{\epsilon}_{II}$) fields at different times are displayed in Figure \ref{fig:tecto_strain_rate}. The first three rows correspond to successive times throughout the SCAM simulations after $\sim 251$, $\sim 1079$ and $\sim 3457~$m of horizontal stretching, respectively. The fourth row corresponds to snapshots of elasto-plastic simulations performed under the same extension rates, after $\sim 3457~$m of extension. 
The elasto-plastic simulations remain unsurprinsingly insensitive to a change of strain rate \cite<\eg, >{OliveEtAl2016}, as they all develop the same central graben structure (Figure \ref{fig:tecto_strain_rate}D–F)

The SCAM simulations, on the other hand, generate features that vary with extension rate. The first one is the location of the cluster of faults in the hanging-wall ((Figure \ref{fig:tecto_strain_rate}A1–C1). The faster the extension rate, the closer this cluster of secondary faults lies to the master fault, with distances ranging from $15$ to $7~$km. The second feature is the geometry of the rift after $3457~$ of extension (Panels A3–C3). The rift structure shows a greater degree of asymmetry with increasing extension rate. This is particularly well expressed in topography, which corresponds to a symmetric graben in Panel A3 but is closer to a half-graben in Panels B3 and C3, as
more extension is accommodated of the central right-dipping fault. We acknowledge that some of this variability could be attributed to stochasticity in marker positions, coupled with strong non-linearities of the SCAM flow law. We however ran the SCAM simulations a second time using different randomly assigned marker positions, and otherwise identical parameters. This second set yielded the same pattern of strain rate dependence as illustrated in Panels A3–C3.

Extension rate is known to influence rifting styles, primarily by modulating the thermal structure \cite{Buck91,LavierBuck2002}. Our results suggest that it may play an additional, more subtle role by modulating the very processes of fault development. Specifically, the SCAM model introduces a strain rate dependence of brittle deformation by transiently activating a moderate to low viscosity in portions of the upper crust where damage is actively growing (e.g., Panels F1 and F2 of Figure \ref{fig:tecto_snapshots}). An analogous strain rate dependence of deformation was previously noted by \citeA{OliveEtAl2016} in rifting simulations that treat the upper crust as a visco-plastic medium rather than an elasto-plastic medium. In the present case, however, the viscosity of the upper crust has a physical meaning ($\eta_D$, related to the damage growth rate), as opposed to an arbitrarily high value meant to simulate a stiff visco-plastic upper crust.
\subsection{Upscaling the SCAM parameters}

The results presented thus far show that a flow law indexed on the activity of sub-millimetric rock defects, and calibrated on decimeter-sized samples can produce reasonable deformation patterns when applied to tectonic problems at scales $> 10$ km. This fact does not negate the implication of larger defects ($> 10$ cm) in the nucleation of crustal faults, but highlights that they need not be explicitly described in our constitutive relation. These intermediate-scale defects can be thought of as emerging through the formation of distributed damage bands, albeit only at scales greater than the numerical grid size, before dominant crustal-scale bands fully develop (e.g., Figure \ref{fig:tecto_snapshots}D1–F1). In other words, the coalescence of sub-millimetric shear cracks produce diffuse proto-faults, and stress concentrations at their tips generate further localization at larger and larger scales.

Nonetheless, it is legitimate to wonder how differently our model would behave if it was based on larger (e.g., metric) shear defects. To address this question, we assess the effect of initial defect size $a$ on the two extreme measures of rock strength in the SCAM framework: the minimum strength that characterizes extremely slow deformation, and the maximum strength at very high strain rates.
The absolute lower bound on SCAM strength is given by the maximum differential stress satisfying $K_I=0$ (Figure \ref{KI_map}, end of Section \ref{sec:retroactions}). It corresponds to the differential stress that must be reached for macroscopic failure to be achievable, albeit after an infinitely long time (plain blue line in Figure \ref{byerlee_results}).
The expression of $K_I$ \eqref{eq:KI} is a function of $\sigma_1$, $\sigma_3$, $\mu$, $a$, and $N_v$ (through $D_0$), but in the special case where $K_I=0$, the explicit dependence on $a$ drops, such that the maximum differential stress at $K_I=0$ only depends on $\sigma_3$, $D_0$ and $\mu$. Assuming $\mu$ is constant across scales, we end up with a lower bound on strength that only depends on $D_0=\text{f}(a,N_v)$. 

Interestingly, this property can be used to constrain a plausible range of $D_0$ values to be used in tectonic simulations. The minimum strength (continuous blue line in Figure \ref{byerlee_results}) must exceed the stress required to slip on a favorably oriented pre-existing frictional surfaces (e.g., $\sim 500\MPa$ for $\sigma_3=150\MPa$, the value used to construct Figure \ref{KI_map}; dashed blue line in Figure \ref{byerlee_results}). The minimum strength must also be lower than the strength measured in the laboratory at the slowest possible strain rate (somewhere within the red area in Figure \ref{byerlee_results}). Taking $90\%$ of the peak value from \citeA{WawersikBrace1971} (e.g., $1150\MPa$ for $\sigma_3=150\MPa$) as a rather conservative estimate, considering that \citeA{BrantutEtAl2012} observed brittle creep failure at $\sim 77\%$ of estimated peak strength under $50\MPa$ of confining pressure, brackets the minimum strength for $\sigma_3=150\MPa$ between $\sim 500$ and $\sim 1150$ MPa. This corresponds to a range of $D_0$ between $0.06$ and $0.4$. 

On the other hand, the maximum strength of the material at very fast strain rates can be assessed by equating equation \eqref{eq:KI} to $K_{IC}$, and rewriting it as a function of $\sigma_1$. This gives an equation whose maximum with respect to $D$ yields the maximum differential stress a rock can withstand (Figure \ref{KI_map}), as cracks grow critically at seismic wave speeds \cite{BhatEtAl2012}. This maximum stress roughly scales as $K_{IC}/\sqrt{a}$.
It follows that the difference between the maximum differential stress at $K_I=K_{IC}$ and at $K_I=0$, a measure of the overall strain rate dependence of the SCAM model, approximately scales as $a^{-1/2}$. If fracture toughness is a scale-invariant, intrinsic property of the material, increasing the size of the initial shear cracks brings the maximum and minimum strengths closer and closer to each other, severely suppressing the strain-rate dependence of the material's intact strength. This result is a fundamental property of the wing-crack model of \citeA{AshbySammis1990}, and does not depend on the specifics of the SCAM model. Whether it applies to natural systems remains an open question, as it is well known that fracture toughness can vary significantly across scales.

\section{Conclusions and perspectives}
\label{sec:conclusions_perspectives}

In this paper, we introduced a Sub-Critically Altered Maxwell (SCAM) framework to describe brittle deformation in long-term tectonic models. It is a set of constitutive equations that capture experimentally-described behaviors of rocks at upper crustal pressures and temperatures. It is based on the evolution of an internal damage state and on its interactions with the elastic properties of the material. The model also allows large deformations by branching to plastic behavior after microcracks interact and coalesce.
The SCAM model has several notable properties that make it a promising alternative to standard elasto-plastic models, or a way to improve their parameterizations of brittle yielding. Elastic properties are permanently altered due to their indexation on damage. Damage growth is a time-dependent process that is activated at stresses far below failure strength, through frictional sliding on shear defects distributed throughout the rock. It promotes strain softening pre-peak and self-consistently generates the successive stages of brittle creep. Creep behavior results from the transition from negative to positive retro-action between damage and damage growth, which represents increasing interactions between lengthening cracks. 
The SCAM model also produces two yield strength using a single friction coefficient and no cohesion: a peak failure strength with high effective friction and strain rate dependent cohesion, which ultimately transitions to a rate-independent residual strength that obeys Byerlee's law. Despite the high effective friction ($\sim 1$), shear band orientations in 2-D plane strain simulations remain consistent with the true friction coefficient that describes the strength of shear defects ($\mu \sim 0.7$).
The SCAM model can be calibrated against experimental data using prior knowledge on well-constrained rock properties to model the deformation of a specific lithology. Here Bayesian inversions of experimental data on Westerly granite and Darley Dale sandstone led to a set of reasonable micromechanical parameters whose impact on the macroscopic behavior of the rock can be straightforwardly interpreted.

Preliminary results of rifting simulations in a $10$-km thick brittle plate subjected to gravity show that the SCAM model generates a population of faults with power-law distributed slip rates, akin to the distribution of natural fracture sizes and fault offsets, and likely introduces a strain rate dependence of the geological structures that develop at large strains. These features make it a good candidate to further investigate the complexity of brittle behavior across scales.

In designing the SCAM framework, we have strived to capture key micromechanical processes while keeping the model as simple as possible. This came at the cost of a few simplifications that may be relaxed in future work.
A first strong assumption is that weakening of the shear modulus occurs isotropically, even though the orientation of the shear defects is strongly anisotropic. This assumption has the advantage of reducing the complexity of the constitutive law and simplifying its interpretation, but lacks consistency when relating crack geometry to macroscopic behavior. 
Within this assumption, however, a more consistent approach to damage-induced elastic weakening could be to assume isotropic weakening in the plane containing the crack normals (the ($\sigma_1$, $\sigma_3$) plane), and no weakening in the perpendicular direction. Such anisotropy however implies that the 2-D plane strain condition can no longer be satisfied for arbitrary extremum stresses, i.e., enforcing $\varepsilon_2 = 0$ could lead to $\sigma_2$ not being an intermediate stress between $\sigma_1$ and $\sigma_3$. 

In addition to neglecting material anisotropy, the evaluation of $K_I$ (Section \ref{sec:damage_growth}) assumes that all cracks remain oriented at $45^{\circ}$ with respect to $\sigma_1$ regardless of the stress and material rotations that may occur during large deformations. This assumption can lead to significant errors on the stresses resolved on shear defects. We note that the strains associated with shear band localization are on the order of a few percent, and thus should not be accompanied by significant finite rotation. On the other hand, large strains could conceivably rotate shear defects out of an optimal orientation in a long-term tectonic simulation. From a theoretical and computational point of view, keeping track of multiple shear defect orientations in an evolving stress state is challenging, in part because evaluating stress intensity factors and interaction terms would become very complex. The very definition of damage should also be revised to account for curving wing-cracks that grow along a changing $\sigma_1$ direction, and/or different sets of tensile wings associated with different sets of shear defects.

Another possible improvement for long-term tectonic modeling would be to allow damage to partially or completely heal over long time scales. This would promote the progressive recovery of intact elastic properties within abandoned shear bands, and require a re-mobilisation of micro-scale frictional processes in order to re-activate a de-activated fault.
To this end, a sub-critical crack growth law similar to that of \citeA{DarotGueguen1986} seems adequate. Crack growth is formulated as a thermally activated process where the growth or healing of a crack is indexed on the sign of the energy balance associated to the incremental advance of the crack front. This formulation has a more robust thermodynamical foundation than \citeauthor{Charles1958}' \citeyear{Charles1958} law and also accounts for the effect of temperature. It however involves more parameters, which is why we restricted ourselves to a simpler version of \citeauthor{Charles1958}' law in this work

Finally, and importantly, the present study neglected the intrinsically dilatant effect of mode-I crack growth. Incompressibility is a common simplification in long-term tectonic models \cite{Gerya2010}, which amounts to setting the Poisson's ratio to $0.5$ and the dilatancy angle to zero when modeling brittle failure with Mohr-Coulomb plasticity. Recent studies have shown that accounting for elastic and plastic compressibility noticeably impacts the outcome of tectonic simulations. Specifically, \citeA{DuretzEtAl2021} found that compressibility hinders strain localization (it produces broader shear zones), and can facilitate the convergence of numerical solvers. A mechanically consistent way of introducing inelastic dilatancy within the elastically incompressible ($\nu = 0.5$) SCAM model would be to use the constitutive law developed by \citeA{BhatEtAl2012}. Their approach consisted of assessing the Gibbs free energy of a damaged solid (assuming the \cite{AshbySammis1990} micro-mechanical model), and deriving it with respect to stress to yield effective elastic compliances. This approach outlines two sources of volumetric strain during damage accumulation: one directly results from the elastic compressibility of the material, which is altered by damage, the other represents a coupling between shear and volumetric deformation (a direct consequence of mode-I cracks being wedged open by shear defects). An elastically incompressible, yet inelastically dilatant SCAM model could focus on capturing this shear-volumetric coupling through an effective dilatancy angle that can be directly related to damage and stress \citeA{BhatEtAl2012}. Dilatancy should cease as soon as cracks coalesce ($D \sim 1$), and give way to standard incompressible Mohr-Coulomb plasticity.

In its current state, the SCAM framework already unveils novel prospects for tectonic modelers. Keeping track of a damage field enables new connections between crustal scale simulations and key observables. Damage fields could straightforwardly be converted to seismic velocity maps for comparison with crustal tomography data. The distribution of damage along the model's free surface could also be used as a proxy for the erodibility of rocks exposed to weathering, and help understand variability in erosion rates across tectonically active landscapes \cite{Molnar2007,Gallen2015}
Finally, damage accumulation is intrinsically linked to an increase in porosity within the rock, and fracture connectivity is a primary control on fluid pathways. The evolving damage field could therefore be used as a proxy for the permeability of the brittle crust (\eg, \citeA{PerolBhat2016}) enabling coupled models of progressive brittle failure and fluid flow. This would enable the consideration of both poro-elastic feedbacks between fluid pressure and deformation, as well as alterations of mechanical properties through fluid-rock reactions.

\setcounter{section}{0}
\renewcommand\thesection{\Alph{section}}
\renewcommand\thesubsection{\thesection.\arabic{subsection}}

\begin{center}
\LARGE {\bf Appendix}
\end{center}

\section{Bayesian inversion procedure}
\label{ap:inversion}

\subsubsection{Mathematical description}

As illustrated in Section \ref{sec:triaxial_application}, our ``forward'' problem consists of predicting a vector of data points $\mathbf{d}$, e.g., a time series of stress or brittle creep strain rates, given a set of rock parameters (Table \ref{tab:parameters}) stored in a vector $\mathbf{m}$. This problem synthetically writes $\mathbf{d} = \mathbf{g(m)}$.
By contrast, the inverse problem consists of finding the  distribution of model parameters that best fits known experimental data $\mathbf{d_{obs}}$ with associated uncertainty. We adopt a least-squares approach in which all probability distributions quantifying uncertainties are Gaussian. Uncorrelated data uncertainty can thus be represented as a Gaussian distribution centered on $\mathbf{d_{obs}}$ with a diagonal covariance matrix $\mathbf{C}_{d}$. Similarly, a-priori knowledge on the distribution of model parameters can be assumed to follow a Gaussian distribution centered on $\mathbf{m}_{prior}$, with diagonal covariance matrix $\mathbf{C}_m$.

We seek the best-fitting model ($\mathbf{\tilde{m}}$) which minimizes the log-likelihood function: a measure of distance between $\mathbf{d_{obs}}$ and $\mathbf{d}$, weighted by data uncertainty and prior knowledge:

\begin{equation}
    \label{eq:log-likelihood}
    S(\mathbf{m}) = (\mathbf{g(m)-d_{obs}})^{T}\mathbf{C}_{d}^{-1}(\mathbf{g(m)-d_{obs}}) + (\mathbf{m - m_{prior}})^{T}\mathbf{C}_{m}^{-1}(\mathbf{m - m_{prior}}).
\end{equation}

The log-likelihood is then minimized using a Gauss-Newton iterative algorithm \cite{tarantola2005}. The iterative scheme, initialized at $\mathbf{m_0} = \mathbf{m_{prior}}$, reads :

\begin{equation}
\label{eq:model_estimate}
\mathbf{m}_{n+1}=\mathbf{m}_{n}-\kappa_{n}\left(\mathbf{G}_{n}^{t} \mathbf{C}_{\mathrm{D}}^{-1} \mathbf{G}_{n}+\mathbf{C}_{m}^{-1}\right)^{-1}\left(\mathbf{G}_{n}^{t} \mathbf{C}_{\mathrm{D}}^{-1}\left(\mathbf{g(m}_{n})-\mathbf{d}_{\mathrm{obs}}\right)+\mathbf{C}_{\mathrm{M}}^{-1}\left(\mathbf{m}_{n}-\mathbf{m}_{\text {prior }}\right)\right),
\end{equation}

where $n$ and $n+1$ refer to the current and next iteration, $\kappa_n \leq 1$ is the step multiplier, a hyperparameter that can be tuned to help convergence in case of strong non-linearities in the log-likelihood function. $\mathbf{G}$ is the Jacobian matrix:

\begin{equation}
\label{eq:jacobian}
G_{n}^{ij}=\left(\frac{\partial g_{i}}{\partial m_{j}}\right)_{\mathbf{m}_{n}}.
\end{equation}

The posterior model covariance matrix, a measure of the uncertainty on the inverted parameters, is then computed as

\begin{equation}
\label{eq:model_covariance}
\widetilde{\mathbf{C}}_{\mathrm{M}} \simeq\left(\mathbf{G}^{t} \mathbf{C}_{\mathrm{D}}^{-1} \mathbf{G}+\mathbf{C}_{\mathrm{M}}^{-1}\right)^{-1}=\mathbf{C}_{\mathrm{M}}-\mathbf{C}_{\mathrm{M}} \mathbf{G}^{t}\left(\mathbf{G} \mathbf{C}_{\mathrm{M}} \mathbf{G}^{t}+\mathbf{C}_{\mathrm{D}}\right)^{-1} \mathbf{G} \mathbf{C}_{\mathrm{M}}.
\end{equation}

\subsubsection{Implementation}

For each inversion, we assemble a data vector $\mathbf{d}_{obs}$ by concatenating data from several constant strain rate experiments conducted under different confining pressures, and several brittle creep experiments conducted under different imposed axial stresses. The constant strain rate data consists of time series of axial stress complemented by the peak axial stress and the corresponding axial strain that was reached in each experiment. By assigning a lower uncertainty (in $\mathbf{C}_{d}$) on the peak axial stress and strain relative to the uncertainty on the stress time series, we can assign more weight to this constraint. Doing so helps favor models that accurately predict the position of the peak stress. The brittle creep data for Darley Dale sandstone consist of concatenated time series of axial strain rate from experiments conducted under different axial stresses. Because such data was not available for Westerly granite, we instead concatenated measurements of the representative secondary creep strain rate from experiments conducted under different axial stresses.

At every step of the inversion algorithm, a vector of ``simulated data'' $\mathbf{d}_n = \mathbf{g(m}_{n})$ is built by simulating each individual experiment, and ordering the outputs to match the order of $\mathbf{d}_{obs}$. In order to simulate brittle creep, the axial stress is first raised by imposing a constant strain rate matching that of the experiment (e.g., loading up to the star in Figure \ref{0D_damaged_vs_plast}A as damage accumulates in in Figure \ref{0D_damaged_vs_plast}B). The damage state achieved by the sample is then used as initial condition for the brittle creep simulation. The axial stress is kept constant as we simulate the evolution of axial strain rates (See Section \ref{sec:triaxial_application}). 

To ensure that the inversion does not assign negative values to parameters which are inherently positive, we construct a model vector $\mathbf{m}$ that contains the logarithm of each SCAM parameter, namely: $G_0$, $\gamma$, $\mu$, $a$, $D_0$, $D_i$, $n$, $K_{IC}$, and $\dot{l}_0$. The associated uncertainties (prior or posterior) are thus log-normal distributions of the parameters, or Gaussian distributions of the logarithm of the parameters. If a positive parameter has a standard deviation $\bar{s}$ and a median $\bar{m}$, the distribution of its logarithm can be adequately represented by a Gaussian distribution centered on $\log(\bar{m})$ with variance $\bar{\sigma}^2 = \log(1 + \bar{s}^2/\bar{m}^2)$. We make use of these formula to assign uncertainties on the experimental data and on our prior knowledge of the model parameters.

\section{Derivation of the plastic viscosity} \label{ap:eta_plas}

Two situations have to be considered to construct the plastic viscosity term $\eta_{p}$ used in SCAM numerical simulations.
The first situation is stresses that lie above the plastic yield envelope at the onset of plastic behavior. This can happen because damage build-up may cause stresses to increase past the plastic yield stress ($\sigma_y$).
In this case, we require the plastic viscosity to relax excess stresses above $\sigma_y$ in one iteration. Equation \eqref{eq:plastic-elastic_constitutive_law} recast in term of $\dot{s}_{ij}$ thus becomes

\begin{equation}
\label{eq:stress_increment_plastic_constitutive_law}  
\dot{s}_{ij} = \frac{s_{ij}^{y} - s_{ij}}{\Delta t} = 2G_{0}\text{f}(D) \left( \dot{e}_{ij} - \frac{s_{ij}}{2\eta_{p}}\right),
\end{equation}

where $s_{ij}^{y}$ corresponds to any deviatoric stress tensor whose second invariant satisfies $s_{II}^{y}=\sigma_y$ . 
We approximate equation \eqref{eq:stress_increment_plastic_constitutive_law} by using $\sigma_y$, $s_{II}$ and the second invariant of the strain rate tensors $\dot{e}_{II} = \sqrt{J_2(\dot{\varepsilon})}$ :

\begin{equation}
\label{eq:stress_increment_plastic_constitutive_law2}   
\frac{\sigma_{y} - s_{II}}{\Delta t} = 2G_{0}\text{f}(D) \left( \dot{e}_{II} - \frac{s_{II}}{2\eta_{p}}\right).
\end{equation}

This yields the following closed form equation for $\eta_{p}$ :

\begin{equation}
\label{eq:overstress_plastic_viscosity}
\eta_p = \frac{s_{II}}{2\left(  \dot{e}_{II} - \frac{\sigma_{y} - s_{II}}{2G_{0}\text{f}(D) \Delta t}\right)}\ .
\end{equation}

The other possible situation involves stresses capped by the yield envelope (i.e., $s_{II}<\sigma_y$). In this case, we additionally require that $\eta_p$ be infinite if the elastic stress rate $2G_{0}\text{f}(D)\dot{e}_{II}$ is lower than the value required to reach the yield stress in one iteration ($\frac{\sigma_{y} - s_{II}}{\Delta t}$). This amounts to rewriting $\eta_p$:

\begin{equation}
\label{eq:plastic_viscosity_full2}
\eta_p = \frac{s_{II}}{2\left(  \dot{e}_{II} - \min{(\frac{\sigma_{y} - s_{II}}{2G_{0}\text{f}(D) \Delta t},\dot{e}_{II})}\right)\ } , 
\end{equation}

to ensure that the plastic viscosity is finite only when $s_{II} \geq \sigma_y$, or if an elastic stress increment suddenly brings $s_{II}$ above $\sigma_y$.

\section*{Acknowledgements}
Julia and MATLAB source codes for the 0-D and 2-D simulations are available at \url{https://zenodo.org/records/10464422}. This work was partially supported by an Emergence(s) Ville de Paris grant to J.-A.O. H.S.B. acknowledges the European Research Council Grant PERSISMO (Grant 865411) for partial support of this work.
We are grateful to Muriel Gerbault, Mike Heap, C\'ecile Prigent and Luc Lavier for their helpful comments on earlier versions of this work. We also thank our editor Boris Kaus as well as John Naliboff and an anonymous reviewer for their feedback.


\begin{thebibliography}{}

\bibitem [\protect \citeauthoryear {%
Aben%
, Brantut%
, Mitchell%
\BCBL {}\ \BBA {} David%
}{%
Aben%
\ \protect \BOthers {.}}{%
{\protect \APACyear {2019}}%
}]{%
AbenEtAl2019}
\APACinsertmetastar {%
AbenEtAl2019}%
\begin{APACrefauthors}%
Aben, F\BPBI M.%
, Brantut, N.%
, Mitchell, T\BPBI M.%
\BCBL {}\ \BBA {} David, E\BPBI C.%
\end{APACrefauthors}%
\unskip\
\newblock
\APACrefYearMonthDay{2019}{{\APACmonth{07}}}{}.
\newblock
{\BBOQ}\APACrefatitle {Rupture {{Energetics}} in {{Crustal Rock From
  Laboratory}}-{{Scale Seismic Tomography}}} {Rupture {{Energetics}} in
  {{Crustal Rock From Laboratory}}-{{Scale Seismic Tomography}}}.{\BBCQ}
\newblock
\APACjournalVolNumPages{Geophysical Research Letters}{46}{13}{7337--7344}.
\PrintBackRefs{\CurrentBib}

\bibitem [\protect \citeauthoryear {%
Arthur%
\ \BBA {} Dunstan%
}{%
Arthur%
\ \BBA {} Dunstan%
}{%
{\protect \APACyear {1977}}%
}]{%
ArthurDunstan1977}
\APACinsertmetastar {%
ArthurDunstan1977}%
\begin{APACrefauthors}%
Arthur, J.%
\BCBT {}\ \BBA {} Dunstan, T.%
\end{APACrefauthors}%
\unskip\
\newblock
\APACrefYearMonthDay{1977}{}{}.
\newblock
{\BBOQ}\APACrefatitle {Plastic Deformation and Failure in Granular Media}
  {Plastic deformation and failure in granular media}.{\BBCQ}
\newblock
\APACjournalVolNumPages{G\'eotechnique}{27}{1}{53--74}.
\PrintBackRefs{\CurrentBib}

\bibitem [\protect \citeauthoryear {%
Ashby%
\ \BBA {} Hallam%
}{%
Ashby%
\ \BBA {} Hallam%
}{%
{\protect \APACyear {1986}}%
}]{%
AshbyHallam1986}
\APACinsertmetastar {%
AshbyHallam1986}%
\begin{APACrefauthors}%
Ashby, M\BPBI F.%
\BCBT {}\ \BBA {} Hallam, S\BPBI D.%
\end{APACrefauthors}%
\unskip\
\newblock
\APACrefYearMonthDay{1986}{{\APACmonth{03}}}{}.
\newblock
{\BBOQ}\APACrefatitle {The Failure of Brittle Solids Containing Small Cracks
  under Compressive Stress States} {The failure of brittle solids containing
  small cracks under compressive stress states}.{\BBCQ}
\newblock
\APACjournalVolNumPages{Acta Metallurgica}{34}{3}{497--510}.
\PrintBackRefs{\CurrentBib}

\bibitem [\protect \citeauthoryear {%
Ashby%
\ \BBA {} Sammis%
}{%
Ashby%
\ \BBA {} Sammis%
}{%
{\protect \APACyear {1990}}%
}]{%
AshbySammis1990}
\APACinsertmetastar {%
AshbySammis1990}%
\begin{APACrefauthors}%
Ashby, M\BPBI F.%
\BCBT {}\ \BBA {} Sammis, C\BPBI G.%
\end{APACrefauthors}%
\unskip\
\newblock
\APACrefYearMonthDay{1990}{}{}.
\newblock
{\BBOQ}\APACrefatitle {The Damage Mechanics of Brittle Solids in Compression}
  {The damage mechanics of brittle solids in compression}.{\BBCQ}
\newblock
\APACjournalVolNumPages{}{133}{}{33}.
\PrintBackRefs{\CurrentBib}

\bibitem [\protect \citeauthoryear {%
Atkinson%
}{%
Atkinson%
}{%
{\protect \APACyear {1979}}%
}]{%
atkinson1979}
\APACinsertmetastar {%
atkinson1979}%
\begin{APACrefauthors}%
Atkinson, B\BPBI K.%
\end{APACrefauthors}%
\unskip\
\newblock
\APACrefYearMonthDay{1979}{}{}.
\newblock
{\BBOQ}\APACrefatitle {A Fracture Mechanics Study of Subcritical Tensile
  Cracking of Quartz in Wet Environments} {A fracture mechanics study of
  subcritical tensile cracking of quartz in wet environments}.{\BBCQ}
\newblock
\APACjournalVolNumPages{Pure and Applied Geophysics
  PAGEOPH}{117}{5}{1011--1024}.
\PrintBackRefs{\CurrentBib}

\bibitem [\protect \citeauthoryear {%
Atkinson%
}{%
Atkinson%
}{%
{\protect \APACyear {1984}}%
}]{%
Atkinson1984}
\APACinsertmetastar {%
Atkinson1984}%
\begin{APACrefauthors}%
Atkinson, B\BPBI K.%
\end{APACrefauthors}%
\unskip\
\newblock
\APACrefYearMonthDay{1984}{{\APACmonth{06}}}{}.
\newblock
{\BBOQ}\APACrefatitle {Subcritical Crack Growth in Geological Materials}
  {Subcritical crack growth in geological materials}.{\BBCQ}
\newblock
\APACjournalVolNumPages{Journal of Geophysical Research: Solid
  Earth}{89}{B6}{4077--4114}.
\PrintBackRefs{\CurrentBib}

\bibitem [\protect \citeauthoryear {%
Baud%
\ \BBA {} Meredith%
}{%
Baud%
\ \BBA {} Meredith%
}{%
{\protect \APACyear {1997}}%
}]{%
BaudMeredith1997}
\APACinsertmetastar {%
BaudMeredith1997}%
\begin{APACrefauthors}%
Baud, P.%
\BCBT {}\ \BBA {} Meredith, P.%
\end{APACrefauthors}%
\unskip\
\newblock
\APACrefYearMonthDay{1997}{}{}.
\newblock
{\BBOQ}\APACrefatitle {Damage Accumulation during Triaxial Creep of {{Darley
  Dale}} Sandstone from Pore Volumometry and Acoustic Emission} {Damage
  accumulation during triaxial creep of {{Darley Dale}} sandstone from pore
  volumometry and acoustic emission}.{\BBCQ}
\newblock
\APACjournalVolNumPages{International Journal of Rock Mechanics and Mining
  Sciences}{34}{3-4}{24--el}.
\PrintBackRefs{\CurrentBib}

\bibitem [\protect \citeauthoryear {%
Baud%
, Schubnel%
\BCBL {}\ \BBA {} Wong%
}{%
Baud%
\ \protect \BOthers {.}}{%
{\protect \APACyear {2000}}%
}]{%
BaudEtAl2000}
\APACinsertmetastar {%
BaudEtAl2000}%
\begin{APACrefauthors}%
Baud, P.%
, Schubnel, A.%
\BCBL {}\ \BBA {} Wong, T\BHBI f.%
\end{APACrefauthors}%
\unskip\
\newblock
\APACrefYearMonthDay{2000}{{\APACmonth{08}}}{}.
\newblock
{\BBOQ}\APACrefatitle {Dilatancy, Compaction, and Failure Mode in {{Solnhofen}}
  Limestone} {Dilatancy, compaction, and failure mode in {{Solnhofen}}
  limestone}.{\BBCQ}
\newblock
\APACjournalVolNumPages{Journal of Geophysical Research: Solid
  Earth}{105}{B8}{19289--19303}.
\PrintBackRefs{\CurrentBib}

\bibitem [\protect \citeauthoryear {%
Bhat%
, Rosakis%
\BCBL {}\ \BBA {} Sammis%
}{%
Bhat%
\ \protect \BOthers {.}}{%
{\protect \APACyear {2012}}%
}]{%
BhatEtAl2012}
\APACinsertmetastar {%
BhatEtAl2012}%
\begin{APACrefauthors}%
Bhat, H\BPBI S.%
, Rosakis, A\BPBI J.%
\BCBL {}\ \BBA {} Sammis, C\BPBI G.%
\end{APACrefauthors}%
\unskip\
\newblock
\APACrefYearMonthDay{2012}{{\APACmonth{05}}}{}.
\newblock
{\BBOQ}\APACrefatitle {A {{Micromechanics Based Constitutive Model}} for
  {{Brittle Failure}} at {{High Strain Rates}}} {A {{Micromechanics Based
  Constitutive Model}} for {{Brittle Failure}} at {{High Strain
  Rates}}}.{\BBCQ}
\newblock
\APACjournalVolNumPages{Journal of Applied Mechanics}{79}{3}{031016}.
\PrintBackRefs{\CurrentBib}

\bibitem [\protect \citeauthoryear {%
Bhat%
, Sammis%
\BCBL {}\ \BBA {} Rosakis%
}{%
Bhat%
\ \protect \BOthers {.}}{%
{\protect \APACyear {2011}}%
}]{%
BhatEtAl2011}
\APACinsertmetastar {%
BhatEtAl2011}%
\begin{APACrefauthors}%
Bhat, H\BPBI S.%
, Sammis, C.%
\BCBL {}\ \BBA {} Rosakis, A.%
\end{APACrefauthors}%
\unskip\
\newblock
\APACrefYearMonthDay{2011}{{\APACmonth{12}}}{}.
\newblock
{\BBOQ}\APACrefatitle {The {{Micromechanics}} of {{Westerley Granite}} at
  {{Large Compressive Loads}}} {The {{Micromechanics}} of {{Westerley Granite}}
  at {{Large Compressive Loads}}}.{\BBCQ}
\newblock
\APACjournalVolNumPages{Pure and Applied Geophysics}{168}{12}{2181--2198}.
\PrintBackRefs{\CurrentBib}

\bibitem [\protect \citeauthoryear {%
Brace%
\ \BBA {} Kohlstedt%
}{%
Brace%
\ \BBA {} Kohlstedt%
}{%
{\protect \APACyear {1980}}%
}]{%
BraceKohlstedt1980}
\APACinsertmetastar {%
BraceKohlstedt1980}%
\begin{APACrefauthors}%
Brace, W\BPBI F.%
\BCBT {}\ \BBA {} Kohlstedt, D\BPBI L.%
\end{APACrefauthors}%
\unskip\
\newblock
\APACrefYearMonthDay{1980}{{\APACmonth{11}}}{}.
\newblock
{\BBOQ}\APACrefatitle {Limits on Lithospheric Stress Imposed by Laboratory
  Experiments} {Limits on lithospheric stress imposed by laboratory
  experiments}.{\BBCQ}
\newblock
\APACjournalVolNumPages{Journal of Geophysical Research: Solid
  Earth}{85}{B11}{6248--6252}.
\PrintBackRefs{\CurrentBib}

\bibitem [\protect \citeauthoryear {%
Brace%
, Paulding%
\BCBL {}\ \BBA {} Scholz%
}{%
Brace%
\ \protect \BOthers {.}}{%
{\protect \APACyear {1966}}%
}]{%
BraceEtAl1966a}
\APACinsertmetastar {%
BraceEtAl1966a}%
\begin{APACrefauthors}%
Brace, W\BPBI F.%
, Paulding, B\BPBI W.%
\BCBL {}\ \BBA {} Scholz, C.%
\end{APACrefauthors}%
\unskip\
\newblock
\APACrefYearMonthDay{1966}{{\APACmonth{08}}}{}.
\newblock
{\BBOQ}\APACrefatitle {Dilatancy in the Fracture of Crystalline Rocks}
  {Dilatancy in the fracture of crystalline rocks}.{\BBCQ}
\newblock
\APACjournalVolNumPages{Journal of Geophysical Research}{71}{16}{3939--3953}.
\PrintBackRefs{\CurrentBib}

\bibitem [\protect \citeauthoryear {%
Brantut%
, Baud%
, Heap%
\BCBL {}\ \BBA {} Meredith%
}{%
Brantut%
\ \protect \BOthers {.}}{%
{\protect \APACyear {2012}}%
}]{%
BrantutEtAl2012}
\APACinsertmetastar {%
BrantutEtAl2012}%
\begin{APACrefauthors}%
Brantut, N.%
, Baud, P.%
, Heap, M\BPBI J.%
\BCBL {}\ \BBA {} Meredith, P\BPBI G.%
\end{APACrefauthors}%
\unskip\
\newblock
\APACrefYearMonthDay{2012}{{\APACmonth{08}}}{}.
\newblock
{\BBOQ}\APACrefatitle {Micromechanics of Brittle Creep in Rocks:
  {{MICROMECHANICS OF BRITTLE CREEP}}} {Micromechanics of brittle creep in
  rocks: {{MICROMECHANICS OF BRITTLE CREEP}}}.{\BBCQ}
\newblock
\APACjournalVolNumPages{Journal of Geophysical Research: Solid
  Earth}{117}{B8}{n/a-n/a}.
\PrintBackRefs{\CurrentBib}

\bibitem [\protect \citeauthoryear {%
Brantut%
, Heap%
, Meredith%
\BCBL {}\ \BBA {} Baud%
}{%
Brantut%
\ \protect \BOthers {.}}{%
{\protect \APACyear {2013}}%
}]{%
BrantutEtAl2013}
\APACinsertmetastar {%
BrantutEtAl2013}%
\begin{APACrefauthors}%
Brantut, N.%
, Heap, M.%
, Meredith, P.%
\BCBL {}\ \BBA {} Baud, P.%
\end{APACrefauthors}%
\unskip\
\newblock
\APACrefYearMonthDay{2013}{{\APACmonth{07}}}{}.
\newblock
{\BBOQ}\APACrefatitle {Time-Dependent Cracking and Brittle Creep in Crustal
  Rocks: {{A}} Review} {Time-dependent cracking and brittle creep in crustal
  rocks: {{A}} review}.{\BBCQ}
\newblock
\APACjournalVolNumPages{Journal of Structural Geology}{52}{}{17--43}.
\PrintBackRefs{\CurrentBib}

\bibitem [\protect \citeauthoryear {%
Buck%
}{%
Buck%
}{%
{\protect \APACyear {1991}}%
}]{%
Buck91}
\APACinsertmetastar {%
Buck91}%
\begin{APACrefauthors}%
Buck, W\BPBI R.%
\end{APACrefauthors}%
\unskip\
\newblock
\APACrefYearMonthDay{1991}{}{}.
\newblock
{\BBOQ}\APACrefatitle {Modes of continental lithospheric extension} {Modes of
  continental lithospheric extension}.{\BBCQ}
\newblock
\APACjournalVolNumPages{Journal of Geophysical Research: Solid
  Earth}{96}{B12}{20161-20178}.
\PrintBackRefs{\CurrentBib}

\bibitem [\protect \citeauthoryear {%
Budiansky%
\ \BBA {} O'connell%
}{%
Budiansky%
\ \BBA {} O'connell%
}{%
{\protect \APACyear {1976}}%
}]{%
BudianskyOconnell1976}
\APACinsertmetastar {%
BudianskyOconnell1976}%
\begin{APACrefauthors}%
Budiansky, B.%
\BCBT {}\ \BBA {} O'connell, R\BPBI J.%
\end{APACrefauthors}%
\unskip\
\newblock
\APACrefYearMonthDay{1976}{}{}.
\newblock
{\BBOQ}\APACrefatitle {Elastic Moduli of a Cracked Solid} {Elastic moduli of a
  cracked solid}.{\BBCQ}
\newblock
\APACjournalVolNumPages{International journal of Solids and
  structures}{12}{2}{81--97}.
\PrintBackRefs{\CurrentBib}

\bibitem [\protect \citeauthoryear {%
Byerlee%
}{%
Byerlee%
}{%
{\protect \APACyear {1967}}%
}]{%
Byerlee1967}
\APACinsertmetastar {%
Byerlee1967}%
\begin{APACrefauthors}%
Byerlee, J\BPBI D.%
\end{APACrefauthors}%
\unskip\
\newblock
\APACrefYearMonthDay{1967}{{\APACmonth{07}}}{}.
\newblock
{\BBOQ}\APACrefatitle {Frictional Characteristics of Granite under High
  Confining Pressure} {Frictional characteristics of granite under high
  confining pressure}.{\BBCQ}
\newblock
\APACjournalVolNumPages{Journal of Geophysical Research}{72}{14}{3639--3648}.
\PrintBackRefs{\CurrentBib}

\bibitem [\protect \citeauthoryear {%
Byerlee%
}{%
Byerlee%
}{%
{\protect \APACyear {1978}}%
}]{%
Byerlee1978}
\APACinsertmetastar {%
Byerlee1978}%
\begin{APACrefauthors}%
Byerlee, J\BPBI D.%
\end{APACrefauthors}%
\unskip\
\newblock
\APACrefYearMonthDay{1978}{}{}.
\newblock
{\BBOQ}\APACrefatitle {Friction of Rocks} {Friction of rocks}.{\BBCQ}
\newblock
\APACjournalVolNumPages{}{116}{}{12}.
\PrintBackRefs{\CurrentBib}

\bibitem [\protect \citeauthoryear {%
Carter%
, Anderson%
, Hansen%
\BCBL {}\ \BBA {} Kranz%
}{%
Carter%
\ \protect \BOthers {.}}{%
{\protect \APACyear {1981}}%
}]{%
CarterEtAl1981}
\APACinsertmetastar {%
CarterEtAl1981}%
\begin{APACrefauthors}%
Carter, N.%
, Anderson, D\BPBI A.%
, Hansen, F\BPBI D.%
\BCBL {}\ \BBA {} Kranz, R\BPBI L.%
\end{APACrefauthors}%
\unskip\
\newblock
\APACrefYearMonthDay{1981}{}{}.
\newblock
{\BBOQ}\APACrefatitle {Creep and {{Creep Rupture}} of {{Granitic Rocks}}}
  {Creep and {{Creep Rupture}} of {{Granitic Rocks}}}.{\BBCQ}
\newblock
\APACjournalVolNumPages{Washington DC American Geophysical Union Geophysical
  Monograph Series}{24}{}{61--82}.
\PrintBackRefs{\CurrentBib}

\bibitem [\protect \citeauthoryear {%
Charles%
}{%
Charles%
}{%
{\protect \APACyear {1958}}%
}]{%
Charles1958}
\APACinsertmetastar {%
Charles1958}%
\begin{APACrefauthors}%
Charles, R\BPBI J.%
\end{APACrefauthors}%
\unskip\
\newblock
\APACrefYearMonthDay{1958}{{\APACmonth{11}}}{}.
\newblock
{\BBOQ}\APACrefatitle {Static {{Fatigue}} of {{Glass}}. {{I}}} {Static
  {{Fatigue}} of {{Glass}}. {{I}}}.{\BBCQ}
\newblock
\APACjournalVolNumPages{Journal of Applied Physics}{29}{11}{1549--1553}.
\PrintBackRefs{\CurrentBib}

\bibitem [\protect \citeauthoryear {%
Costin%
}{%
Costin%
}{%
{\protect \APACyear {1983}}%
}]{%
Costin1983}
\APACinsertmetastar {%
Costin1983}%
\begin{APACrefauthors}%
Costin, L.%
\end{APACrefauthors}%
\unskip\
\newblock
\APACrefYearMonthDay{1983}{}{}.
\newblock
{\BBOQ}\APACrefatitle {A microcrack model for the deformation and failure of
  brittle rock} {A microcrack model for the deformation and failure of brittle
  rock}.{\BBCQ}
\newblock
\APACjournalVolNumPages{Journal of Geophysical Research: Solid
  Earth}{88}{B11}{9485-9492}.
\PrintBackRefs{\CurrentBib}

\bibitem [\protect \citeauthoryear {%
Costin%
}{%
Costin%
}{%
{\protect \APACyear {1985}}%
}]{%
Costin1985}
\APACinsertmetastar {%
Costin1985}%
\begin{APACrefauthors}%
Costin, L.%
\end{APACrefauthors}%
\unskip\
\newblock
\APACrefYearMonthDay{1985}{{\APACmonth{07}}}{}.
\newblock
{\BBOQ}\APACrefatitle {Damage Mechanics in the Post-Failure Regime} {Damage
  mechanics in the post-failure regime}.{\BBCQ}
\newblock
\APACjournalVolNumPages{Mechanics of Materials}{4}{2}{149--160}.
\PrintBackRefs{\CurrentBib}

\bibitem [\protect \citeauthoryear {%
Darot%
\ \BBA {} Gueguen%
}{%
Darot%
\ \BBA {} Gueguen%
}{%
{\protect \APACyear {1986}}%
}]{%
DarotGueguen1986}
\APACinsertmetastar {%
DarotGueguen1986}%
\begin{APACrefauthors}%
Darot, M.%
\BCBT {}\ \BBA {} Gueguen, Y.%
\end{APACrefauthors}%
\unskip\
\newblock
\APACrefYearMonthDay{1986}{}{}.
\newblock
{\BBOQ}\APACrefatitle {Slow Crack Growth in Minerals and Rocks: {{Theory}} and
  Experiments} {Slow crack growth in minerals and rocks: {{Theory}} and
  experiments}.{\BBCQ}
\newblock
\APACjournalVolNumPages{Pure and Applied Geophysics
  PAGEOPH}{124}{4-5}{677--692}.
\PrintBackRefs{\CurrentBib}

\bibitem [\protect \citeauthoryear {%
Deshpande%
\ \BBA {} Evans%
}{%
Deshpande%
\ \BBA {} Evans%
}{%
{\protect \APACyear {2008}}%
}]{%
DeshpandeEvans2008}
\APACinsertmetastar {%
DeshpandeEvans2008}%
\begin{APACrefauthors}%
Deshpande, V.%
\BCBT {}\ \BBA {} Evans, A.%
\end{APACrefauthors}%
\unskip\
\newblock
\APACrefYearMonthDay{2008}{{\APACmonth{10}}}{}.
\newblock
{\BBOQ}\APACrefatitle {Inelastic Deformation and Energy Dissipation in
  Ceramics: {{A}} Mechanism-Based Constitutive Model} {Inelastic deformation
  and energy dissipation in ceramics: {{A}} mechanism-based constitutive
  model}.{\BBCQ}
\newblock
\APACjournalVolNumPages{Journal of the Mechanics and Physics of
  Solids}{56}{10}{3077--3100}.
\PrintBackRefs{\CurrentBib}

\bibitem [\protect \citeauthoryear {%
Dey%
\ \BBA {} Wang%
}{%
Dey%
\ \BBA {} Wang%
}{%
{\protect \APACyear {1981}}%
}]{%
DeyWang1981}
\APACinsertmetastar {%
DeyWang1981}%
\begin{APACrefauthors}%
Dey, T\BPBI N.%
\BCBT {}\ \BBA {} Wang, C\BHBI Y.%
\end{APACrefauthors}%
\unskip\
\newblock
\APACrefYearMonthDay{1981}{}{}.
\newblock
{\BBOQ}\APACrefatitle {Some {{Mechanisms}} of {{Microcrack Growth}} and
  {{Interaction}} in {{Compressive Rock Failure}}} {Some {{Mechanisms}} of
  {{Microcrack Growth}} and {{Interaction}} in {{Compressive Rock
  Failure}}}.{\BBCQ}
\newblock
\APACjournalVolNumPages{International Journal of Rock Mechanics and Mining
  Sciences \textbackslash\& Geomechanics Abstracts}{18}{3}{199--209}.
\PrintBackRefs{\CurrentBib}

\bibitem [\protect \citeauthoryear {%
Duretz%
, Borst%
\BCBL {}\ \BBA {} Yamato%
}{%
Duretz%
\ \protect \BOthers {.}}{%
{\protect \APACyear {2021}}%
}]{%
DuretzEtAl2021}
\APACinsertmetastar {%
DuretzEtAl2021}%
\begin{APACrefauthors}%
Duretz, T.%
, Borst, R.%
\BCBL {}\ \BBA {} Yamato, P.%
\end{APACrefauthors}%
\unskip\
\newblock
\APACrefYearMonthDay{2021}{{\APACmonth{08}}}{}.
\newblock
{\BBOQ}\APACrefatitle {Modeling {{Lithospheric Deformation Using}} a
  {{Compressible Visco}}-{{Elasto}}-{{Viscoplastic Rheology}} and the
  {{Effective Viscosity Approach}}} {Modeling {{Lithospheric Deformation
  Using}} a {{Compressible Visco}}-{{Elasto}}-{{Viscoplastic Rheology}} and the
  {{Effective Viscosity Approach}}}.{\BBCQ}
\newblock
\APACjournalVolNumPages{Geochemistry, Geophysics, Geosystems}{22}{8}{}.
\PrintBackRefs{\CurrentBib}

\bibitem [\protect \citeauthoryear {%
Eppes%
\ \BBA {} Keanini%
}{%
Eppes%
\ \BBA {} Keanini%
}{%
{\protect \APACyear {2017}}%
}]{%
Eppes2017}
\APACinsertmetastar {%
Eppes2017}%
\begin{APACrefauthors}%
Eppes, M\BHBI C.%
\BCBT {}\ \BBA {} Keanini, R.%
\end{APACrefauthors}%
\unskip\
\newblock
\APACrefYearMonthDay{2017}{}{}.
\newblock
{\BBOQ}\APACrefatitle {Mechanical weathering and rock erosion by
  climate-dependent subcritical cracking} {Mechanical weathering and rock
  erosion by climate-dependent subcritical cracking}.{\BBCQ}
\newblock
\APACjournalVolNumPages{Reviews of Geophysics}{55}{2}{470-508}.
\PrintBackRefs{\CurrentBib}

\bibitem [\protect \citeauthoryear {%
Gallen%
, Clark%
\BCBL {}\ \BBA {} Godt%
}{%
Gallen%
\ \protect \BOthers {.}}{%
{\protect \APACyear {2015}}%
}]{%
Gallen2015}
\APACinsertmetastar {%
Gallen2015}%
\begin{APACrefauthors}%
Gallen, S\BPBI F.%
, Clark, M\BPBI K.%
\BCBL {}\ \BBA {} Godt, J\BPBI W.%
\end{APACrefauthors}%
\unskip\
\newblock
\APACrefYearMonthDay{2015}{01}{}.
\newblock
{\BBOQ}\APACrefatitle {{Coseismic landslides reveal near-surface rock strength
  in a high-relief, tectonically active setting}} {{Coseismic landslides reveal
  near-surface rock strength in a high-relief, tectonically active
  setting}}.{\BBCQ}
\newblock
\APACjournalVolNumPages{Geology}{43}{1}{11-14}.
\PrintBackRefs{\CurrentBib}

\bibitem [\protect \citeauthoryear {%
Gerya%
}{%
Gerya%
}{%
{\protect \APACyear {2010}}%
}]{%
Gerya2010}
\APACinsertmetastar {%
Gerya2010}%
\begin{APACrefauthors}%
Gerya, T.%
\end{APACrefauthors}%
\unskip\
\newblock
\APACrefYear{2010}.
\newblock
\APACrefbtitle {Introduction to {{Numerical Geodynamic Modelling}}}
  {Introduction to {{Numerical Geodynamic Modelling}}}.
\newblock
\APACaddressPublisher{}{{Cambridge}}.
\PrintBackRefs{\CurrentBib}

\bibitem [\protect \citeauthoryear {%
Gerya%
\ \BBA {} Yuen%
}{%
Gerya%
\ \BBA {} Yuen%
}{%
{\protect \APACyear {2003}}%
}]{%
GeryaYuen2003}
\APACinsertmetastar {%
GeryaYuen2003}%
\begin{APACrefauthors}%
Gerya, T.%
\BCBT {}\ \BBA {} Yuen, D\BPBI A.%
\end{APACrefauthors}%
\unskip\
\newblock
\APACrefYearMonthDay{2003}{{\APACmonth{12}}}{}.
\newblock
{\BBOQ}\APACrefatitle {Characteristics-Based Marker-in-Cell Method with
  Conservative Finite-Differences Schemes for Modeling Geological Flows with
  Strongly Variable Transport Properties} {Characteristics-based marker-in-cell
  method with conservative finite-differences schemes for modeling geological
  flows with strongly variable transport properties}.{\BBCQ}
\newblock
\APACjournalVolNumPages{Physics of the Earth and Planetary
  Interiors}{140}{4}{293--318}.
\PrintBackRefs{\CurrentBib}

\bibitem [\protect \citeauthoryear {%
Gratier%
, Dysthe%
\BCBL {}\ \BBA {} Renard%
}{%
Gratier%
\ \protect \BOthers {.}}{%
{\protect \APACyear {2013}}%
}]{%
GratierEtAl2013}
\APACinsertmetastar {%
GratierEtAl2013}%
\begin{APACrefauthors}%
Gratier, J\BHBI P.%
, Dysthe, D\BPBI K.%
\BCBL {}\ \BBA {} Renard, F.%
\end{APACrefauthors}%
\unskip\
\newblock
\APACrefYearMonthDay{2013}{}{}.
\newblock
{\BBOQ}\APACrefatitle {The {{Role}} of {{Pressure Solution Creep}} in the
  {{Ductility}} of the {{Earth}}'s {{Upper Crust}}} {The {{Role}} of {{Pressure
  Solution Creep}} in the {{Ductility}} of the {{Earth}}'s {{Upper
  Crust}}}.{\BBCQ}
\newblock
\BIn{} \APACrefbtitle {Advances in {{Geophysics}}} {Advances in
  {{Geophysics}}}\ (\BVOL~54, \BPGS\ 47--179).
\newblock
\APACaddressPublisher{}{{Elsevier}}.
\PrintBackRefs{\CurrentBib}

\bibitem [\protect \citeauthoryear {%
Hamiel%
, Liu%
, Lyakhovsky%
, {Ben-Zion}%
\BCBL {}\ \BBA {} Lockner%
}{%
Hamiel%
\ \protect \BOthers {.}}{%
{\protect \APACyear {2004}}%
}]{%
HamielEtAl2004}
\APACinsertmetastar {%
HamielEtAl2004}%
\begin{APACrefauthors}%
Hamiel, Y.%
, Liu, Y.%
, Lyakhovsky, V.%
, {Ben-Zion}, Y.%
\BCBL {}\ \BBA {} Lockner, D.%
\end{APACrefauthors}%
\unskip\
\newblock
\APACrefYearMonthDay{2004}{{\APACmonth{12}}}{}.
\newblock
{\BBOQ}\APACrefatitle {A Viscoelastic Damage Model with Applications to Stable
  and Unstable Fracturing} {A viscoelastic damage model with applications to
  stable and unstable fracturing}.{\BBCQ}
\newblock
\APACjournalVolNumPages{Geophysical Journal International}{159}{3}{1155--1165}.
\PrintBackRefs{\CurrentBib}

\bibitem [\protect \citeauthoryear {%
Heap%
, Baud%
, Meredith%
, Bell%
\BCBL {}\ \BBA {} Main%
}{%
Heap%
\ \protect \BOthers {.}}{%
{\protect \APACyear {2009}}%
}]{%
HeapEtAl2009}
\APACinsertmetastar {%
HeapEtAl2009}%
\begin{APACrefauthors}%
Heap, M\BPBI J.%
, Baud, P.%
, Meredith, P\BPBI G.%
, Bell, A\BPBI F.%
\BCBL {}\ \BBA {} Main, I\BPBI G.%
\end{APACrefauthors}%
\unskip\
\newblock
\APACrefYearMonthDay{2009}{{\APACmonth{07}}}{}.
\newblock
{\BBOQ}\APACrefatitle {Time-Dependent Brittle Creep in {{Darley Dale}}
  Sandstone} {Time-dependent brittle creep in {{Darley Dale}}
  sandstone}.{\BBCQ}
\newblock
\APACjournalVolNumPages{Journal of Geophysical Research}{114}{B7}{B07203}.
\PrintBackRefs{\CurrentBib}

\bibitem [\protect \citeauthoryear {%
Kachanov%
}{%
Kachanov%
}{%
{\protect \APACyear {1982}}%
{\protect \APACexlab {{\protect \BCnt {1}}}}}]{%
Kachanov1982a}
\APACinsertmetastar {%
Kachanov1982a}%
\begin{APACrefauthors}%
Kachanov, M\BPBI L.%
\end{APACrefauthors}%
\unskip\
\newblock
\APACrefYearMonthDay{1982{\protect \BCnt {1}}}{}{}.
\newblock
{\BBOQ}\APACrefatitle {A Microcrack Model of Rock Inelasticity Part {{I}}:
  {{Frictional}} Sliding on Microcracks} {A microcrack model of rock
  inelasticity part {{I}}: {{Frictional}} sliding on microcracks}.{\BBCQ}
\newblock
\APACjournalVolNumPages{Mechanics of Materials}{1}{1}{19--27}.
\PrintBackRefs{\CurrentBib}

\bibitem [\protect \citeauthoryear {%
Kachanov%
}{%
Kachanov%
}{%
{\protect \APACyear {1982}}%
{\protect \APACexlab {{\protect \BCnt {2}}}}}]{%
Kachanov1982c}
\APACinsertmetastar {%
Kachanov1982c}%
\begin{APACrefauthors}%
Kachanov, M\BPBI L.%
\end{APACrefauthors}%
\unskip\
\newblock
\APACrefYearMonthDay{1982{\protect \BCnt {2}}}{{\APACmonth{05}}}{}.
\newblock
{\BBOQ}\APACrefatitle {Microcrack Model of Rock Inelasticity Part {{III}}:
  {{Time-dependent}} Growth of Microcracks} {Microcrack model of rock
  inelasticity part {{III}}: {{Time-dependent}} growth of microcracks}.{\BBCQ}
\newblock
\APACjournalVolNumPages{Mechanics of Materials}{1}{2}{123--129}.
\PrintBackRefs{\CurrentBib}

\bibitem [\protect \citeauthoryear {%
Kachanov%
}{%
Kachanov%
}{%
{\protect \APACyear {1982}}%
{\protect \APACexlab {{\protect \BCnt {3}}}}}]{%
Kachanov1982b}
\APACinsertmetastar {%
Kachanov1982b}%
\begin{APACrefauthors}%
Kachanov, M\BPBI L.%
\end{APACrefauthors}%
\unskip\
\newblock
\APACrefYearMonthDay{1982{\protect \BCnt {3}}}{}{}.
\newblock
{\BBOQ}\APACrefatitle {A Microcrack Model of Rock Inelasticity Part {{II}}:
  {{Propagation}} of Microcracks} {A microcrack model of rock inelasticity part
  {{II}}: {{Propagation}} of microcracks}.{\BBCQ}
\newblock
\APACjournalVolNumPages{Mechanics of Materials}{1}{1}{29--41}.
\PrintBackRefs{\CurrentBib}

\bibitem [\protect \citeauthoryear {%
Kachanov%
}{%
Kachanov%
}{%
{\protect \APACyear {1993}}%
}]{%
Kachanov1993}
\APACinsertmetastar {%
Kachanov1993}%
\begin{APACrefauthors}%
Kachanov, M\BPBI L.%
\end{APACrefauthors}%
\unskip\
\newblock
\APACrefYearMonthDay{1993}{}{}.
\newblock
{\BBOQ}\APACrefatitle {Elastic {{Solids}} with {{Many Cracks}} and {{Related
  Problems}}} {Elastic {{Solids}} with {{Many Cracks}} and {{Related
  Problems}}}.{\BBCQ}
\newblock
\BIn{} \APACrefbtitle {Advances in {{Applied Mechanics}}} {Advances in
  {{Applied Mechanics}}}\ (\BVOL~30, \BPGS\ 259--445).
\newblock
\APACaddressPublisher{}{{Elsevier}}.
\PrintBackRefs{\CurrentBib}

\bibitem [\protect \citeauthoryear {%
Karabulut%
\ \BBA {} Bouchon%
}{%
Karabulut%
\ \BBA {} Bouchon%
}{%
{\protect \APACyear {2007}}%
}]{%
KarabulutBouchon2007}
\APACinsertmetastar {%
KarabulutBouchon2007}%
\begin{APACrefauthors}%
Karabulut, H.%
\BCBT {}\ \BBA {} Bouchon, M.%
\end{APACrefauthors}%
\unskip\
\newblock
\APACrefYearMonthDay{2007}{{\APACmonth{07}}}{}.
\newblock
{\BBOQ}\APACrefatitle {Spatial Variability and Non-Linearity of Strong Ground
  Motion near a Fault} {Spatial variability and non-linearity of strong ground
  motion near a fault}.{\BBCQ}
\newblock
\APACjournalVolNumPages{Geophysical Journal International}{170}{1}{262--274}.
\PrintBackRefs{\CurrentBib}

\bibitem [\protect \citeauthoryear {%
Karrech%
, {Regenauer-Lieb}%
\BCBL {}\ \BBA {} Poulet%
}{%
Karrech%
\ \protect \BOthers {.}}{%
{\protect \APACyear {2011a}}%
}]{%
KarrechEtAl2011a}
\APACinsertmetastar {%
KarrechEtAl2011a}%
\begin{APACrefauthors}%
Karrech, A.%
, {Regenauer-Lieb}, K.%
\BCBL {}\ \BBA {} Poulet, T.%
\end{APACrefauthors}%
\unskip\
\newblock
\APACrefYearMonthDay{2011a}{}{}.
\newblock
{\BBOQ}\APACrefatitle {A Damaged Visco-Plasticity Model for Pressure and
  Temperature Sensitive Geomaterials} {A damaged visco-plasticity model for
  pressure and temperature sensitive geomaterials}.{\BBCQ}
\newblock
\APACjournalVolNumPages{International Journal of Engineering
  Science}{49}{10}{1141--1150}.
\PrintBackRefs{\CurrentBib}

\bibitem [\protect \citeauthoryear {%
Kaus%
}{%
Kaus%
}{%
{\protect \APACyear {2010}}%
}]{%
Kaus2010}
\APACinsertmetastar {%
Kaus2010}%
\begin{APACrefauthors}%
Kaus, B\BPBI J.%
\end{APACrefauthors}%
\unskip\
\newblock
\APACrefYearMonthDay{2010}{{\APACmonth{03}}}{}.
\newblock
{\BBOQ}\APACrefatitle {Factors That Control the Angle of Shear Bands in
  Geodynamic Numerical Models of Brittle Deformation} {Factors that control the
  angle of shear bands in geodynamic numerical models of brittle
  deformation}.{\BBCQ}
\newblock
\APACjournalVolNumPages{Tectonophysics}{484}{1-4}{36--47}.
\PrintBackRefs{\CurrentBib}

\bibitem [\protect \citeauthoryear {%
Kranz%
}{%
Kranz%
}{%
{\protect \APACyear {1979}}%
}]{%
Kranz1979}
\APACinsertmetastar {%
Kranz1979}%
\begin{APACrefauthors}%
Kranz, R\BPBI L.%
\end{APACrefauthors}%
\unskip\
\newblock
\APACrefYearMonthDay{1979}{}{}.
\newblock
{\BBOQ}\APACrefatitle {Crack {{Growth}} and {{Development During Creep}} of
  {{Barre Granite}}} {Crack {{Growth}} and {{Development During Creep}} of
  {{Barre Granite}}}.{\BBCQ}
\newblock
\APACjournalVolNumPages{International Journal of Rock Mechanics and Mining
  Sciences \textbackslash\& Geomechanics Abstracts}{16}{1}{23--35}.
\PrintBackRefs{\CurrentBib}

\bibitem [\protect \citeauthoryear {%
Lavier%
\ \BBA {} Buck%
}{%
Lavier%
\ \BBA {} Buck%
}{%
{\protect \APACyear {2002}}%
}]{%
LavierBuck2002}
\APACinsertmetastar {%
LavierBuck2002}%
\begin{APACrefauthors}%
Lavier, L\BPBI L.%
\BCBT {}\ \BBA {} Buck, W\BPBI R.%
\end{APACrefauthors}%
\unskip\
\newblock
\APACrefYearMonthDay{2002}{}{}.
\newblock
{\BBOQ}\APACrefatitle {Half graben versus large-offset low-angle normal fault:
  Importance of keeping cool during normal faulting} {Half graben versus
  large-offset low-angle normal fault: Importance of keeping cool during normal
  faulting}.{\BBCQ}
\newblock
\APACjournalVolNumPages{Journal of Geophysical Research: Solid
  Earth}{107}{B6}{ETG--8}.
\PrintBackRefs{\CurrentBib}

\bibitem [\protect \citeauthoryear {%
Lavier%
, Buck%
\BCBL {}\ \BBA {} Poliakov%
}{%
Lavier%
\ \protect \BOthers {.}}{%
{\protect \APACyear {2000}}%
}]{%
LavierEtAl2000}
\APACinsertmetastar {%
LavierEtAl2000}%
\begin{APACrefauthors}%
Lavier, L\BPBI L.%
, Buck, W\BPBI R.%
\BCBL {}\ \BBA {} Poliakov, A\BPBI N\BPBI B.%
\end{APACrefauthors}%
\unskip\
\newblock
\APACrefYearMonthDay{2000}{{\APACmonth{10}}}{}.
\newblock
{\BBOQ}\APACrefatitle {Factors Controlling Normal Fault Offset in an Ideal
  Brittle Layer} {Factors controlling normal fault offset in an ideal brittle
  layer}.{\BBCQ}
\newblock
\APACjournalVolNumPages{Journal of Geophysical Research: Solid
  Earth}{105}{B10}{23431--23442}.
\PrintBackRefs{\CurrentBib}

\bibitem [\protect \citeauthoryear {%
Lemiale%
, M{\"u}hlhaus%
, Moresi%
\BCBL {}\ \BBA {} Stafford%
}{%
Lemiale%
\ \protect \BOthers {.}}{%
{\protect \APACyear {2008}}%
}]{%
LemialeEtAl2008}
\APACinsertmetastar {%
LemialeEtAl2008}%
\begin{APACrefauthors}%
Lemiale, V.%
, M{\"u}hlhaus, H\BHBI B.%
, Moresi, L.%
\BCBL {}\ \BBA {} Stafford, J.%
\end{APACrefauthors}%
\unskip\
\newblock
\APACrefYearMonthDay{2008}{{\APACmonth{12}}}{}.
\newblock
{\BBOQ}\APACrefatitle {Shear Banding Analysis of Plastic Models Formulated for
  Incompressible Viscous Flows} {Shear banding analysis of plastic models
  formulated for incompressible viscous flows}.{\BBCQ}
\newblock
\APACjournalVolNumPages{Physics of the Earth and Planetary
  Interiors}{171}{1-4}{177--186}.
\PrintBackRefs{\CurrentBib}

\bibitem [\protect \citeauthoryear {%
Le~Pourhiet%
}{%
Le~Pourhiet%
}{%
{\protect \APACyear {2013}}%
}]{%
lepourhiet2013}
\APACinsertmetastar {%
lepourhiet2013}%
\begin{APACrefauthors}%
Le~Pourhiet, L.%
\end{APACrefauthors}%
\unskip\
\newblock
\APACrefYearMonthDay{2013}{{\APACmonth{07}}}{}.
\newblock
{\BBOQ}\APACrefatitle {Strain Localization Due to Structural Softening during
  Pressure Sensitive Rate Independent Yielding} {Strain localization due to
  structural softening during pressure sensitive rate independent
  yielding}.{\BBCQ}
\newblock
\APACjournalVolNumPages{Bulletin de la Soci\'et\'e G\'eologique de
  France}{184}{4-5}{357--371}.
\PrintBackRefs{\CurrentBib}

\bibitem [\protect \citeauthoryear {%
Lockner%
}{%
Lockner%
}{%
{\protect \APACyear {1998}}%
}]{%
Lockner1998}
\APACinsertmetastar {%
Lockner1998}%
\begin{APACrefauthors}%
Lockner, D\BPBI A.%
\end{APACrefauthors}%
\unskip\
\newblock
\APACrefYearMonthDay{1998}{{\APACmonth{03}}}{}.
\newblock
{\BBOQ}\APACrefatitle {A Generalized Law for Brittle Deformation of
  {{Westerly}} Granite} {A generalized law for brittle deformation of
  {{Westerly}} granite}.{\BBCQ}
\newblock
\APACjournalVolNumPages{Journal of Geophysical Research: Solid
  Earth}{103}{B3}{5107--5123}.
\PrintBackRefs{\CurrentBib}

\bibitem [\protect \citeauthoryear {%
Lockner%
, Byerlee%
, Kuksenko%
, Ponomarev%
\BCBL {}\ \BBA {} Sidorin%
}{%
Lockner%
\ \protect \BOthers {.}}{%
{\protect \APACyear {1991}}%
}]{%
LocknerEtAl1991}
\APACinsertmetastar {%
LocknerEtAl1991}%
\begin{APACrefauthors}%
Lockner, D\BPBI A.%
, Byerlee, J\BPBI D.%
, Kuksenko, V.%
, Ponomarev, A.%
\BCBL {}\ \BBA {} Sidorin, A.%
\end{APACrefauthors}%
\unskip\
\newblock
\APACrefYearMonthDay{1991}{{\APACmonth{03}}}{}.
\newblock
{\BBOQ}\APACrefatitle {Quasi-Static Fault Growth and Shear Fracture Energy in
  Granite} {Quasi-static fault growth and shear fracture energy in
  granite}.{\BBCQ}
\newblock
\APACjournalVolNumPages{Nature}{350}{6313}{39--42}.
\PrintBackRefs{\CurrentBib}

\bibitem [\protect \citeauthoryear {%
Lyakhovsky%
, {Ben-Zion}%
\BCBL {}\ \BBA {} Agnon%
}{%
Lyakhovsky%
\ \protect \BOthers {.}}{%
{\protect \APACyear {1997}}%
}]{%
LyakhovskyEtAl1997}
\APACinsertmetastar {%
LyakhovskyEtAl1997}%
\begin{APACrefauthors}%
Lyakhovsky, V.%
, {Ben-Zion}, Y.%
\BCBL {}\ \BBA {} Agnon, A.%
\end{APACrefauthors}%
\unskip\
\newblock
\APACrefYearMonthDay{1997}{{\APACmonth{12}}}{}.
\newblock
{\BBOQ}\APACrefatitle {Distributed Damage, Faulting, and Friction} {Distributed
  damage, faulting, and friction}.{\BBCQ}
\newblock
\APACjournalVolNumPages{Journal of Geophysical Research: Solid
  Earth}{102}{B12}{27635--27649}.
\PrintBackRefs{\CurrentBib}

\bibitem [\protect \citeauthoryear {%
Manaker%
, Turcotte%
\BCBL {}\ \BBA {} Kellogg%
}{%
Manaker%
\ \protect \BOthers {.}}{%
{\protect \APACyear {2006}}%
}]{%
ManakerEtAl2006}
\APACinsertmetastar {%
ManakerEtAl2006}%
\begin{APACrefauthors}%
Manaker, D\BPBI M.%
, Turcotte, D\BPBI L.%
\BCBL {}\ \BBA {} Kellogg, L\BPBI H.%
\end{APACrefauthors}%
\unskip\
\newblock
\APACrefYearMonthDay{2006}{{\APACmonth{09}}}{}.
\newblock
{\BBOQ}\APACrefatitle {Flexure with Damage} {Flexure with damage}.{\BBCQ}
\newblock
\APACjournalVolNumPages{Geophysical Journal International}{166}{3}{1368--1383}.
\PrintBackRefs{\CurrentBib}

\bibitem [\protect \citeauthoryear {%
Marrett%
, Ortega%
\BCBL {}\ \BBA {} Kelsey%
}{%
Marrett%
\ \protect \BOthers {.}}{%
{\protect \APACyear {1999}}%
}]{%
MarrettEtAl1999}
\APACinsertmetastar {%
MarrettEtAl1999}%
\begin{APACrefauthors}%
Marrett, R.%
, Ortega, O\BPBI J.%
\BCBL {}\ \BBA {} Kelsey, C\BPBI M.%
\end{APACrefauthors}%
\unskip\
\newblock
\APACrefYearMonthDay{1999}{}{}.
\newblock
{\BBOQ}\APACrefatitle {Extent of Power-Law Scaling for Natural Fractures in
  Rock} {Extent of power-law scaling for natural fractures in rock}.{\BBCQ}
\newblock
\APACjournalVolNumPages{Geology}{27}{9}{799}.
\PrintBackRefs{\CurrentBib}

\bibitem [\protect \citeauthoryear {%
McBeck%
, Kandula%
, Aiken%
, Cordonnier%
\BCBL {}\ \BBA {} Renard%
}{%
McBeck%
\ \protect \BOthers {.}}{%
{\protect \APACyear {2019}}%
}]{%
McBeckEtAl2019}
\APACinsertmetastar {%
McBeckEtAl2019}%
\begin{APACrefauthors}%
McBeck, J.%
, Kandula, N.%
, Aiken, J\BPBI M.%
, Cordonnier, B.%
\BCBL {}\ \BBA {} Renard, F.%
\end{APACrefauthors}%
\unskip\
\newblock
\APACrefYearMonthDay{2019}{{\APACmonth{10}}}{}.
\newblock
{\BBOQ}\APACrefatitle {Isolating the {{Factors That Govern Fracture
  Development}} in {{Rocks Throughout Dynamic In Situ X}}-{{Ray Tomography
  Experiments}}} {Isolating the {{Factors That Govern Fracture Development}} in
  {{Rocks Throughout Dynamic In Situ X}}-{{Ray Tomography
  Experiments}}}.{\BBCQ}
\newblock
\APACjournalVolNumPages{Geophysical Research Letters}{46}{20}{11127--11135}.
\PrintBackRefs{\CurrentBib}

\bibitem [\protect \citeauthoryear {%
Men{\'e}ndez%
, Zhu%
\BCBL {}\ \BBA {} Wong%
}{%
Men{\'e}ndez%
\ \protect \BOthers {.}}{%
{\protect \APACyear {1996}}%
}]{%
MenendezEtAl1996}
\APACinsertmetastar {%
MenendezEtAl1996}%
\begin{APACrefauthors}%
Men{\'e}ndez, B.%
, Zhu, W.%
\BCBL {}\ \BBA {} Wong, T\BHBI F.%
\end{APACrefauthors}%
\unskip\
\newblock
\APACrefYearMonthDay{1996}{{\APACmonth{01}}}{}.
\newblock
{\BBOQ}\APACrefatitle {Micromechanics of Brittle Faulting and Cataclastic Flow
  in {{Berea}} Sandstone} {Micromechanics of brittle faulting and cataclastic
  flow in {{Berea}} sandstone}.{\BBCQ}
\newblock
\APACjournalVolNumPages{Journal of Structural Geology}{18}{1}{1--16}.
\PrintBackRefs{\CurrentBib}

\bibitem [\protect \citeauthoryear {%
Meyer%
, Kaus%
\BCBL {}\ \BBA {} Passchier%
}{%
Meyer%
\ \protect \BOthers {.}}{%
{\protect \APACyear {2017}}%
}]{%
Meyer2017}
\APACinsertmetastar {%
Meyer2017}%
\begin{APACrefauthors}%
Meyer, S\BPBI E.%
, Kaus, B\BPBI J\BPBI P.%
\BCBL {}\ \BBA {} Passchier, C.%
\end{APACrefauthors}%
\unskip\
\newblock
\APACrefYearMonthDay{2017}{}{}.
\newblock
{\BBOQ}\APACrefatitle {Development of branching brittle and ductile shear
  zones: A numerical study} {Development of branching brittle and ductile shear
  zones: A numerical study}.{\BBCQ}
\newblock
\APACjournalVolNumPages{Geochemistry, Geophysics,
  Geosystems}{18}{6}{2054-2075}.
\PrintBackRefs{\CurrentBib}

\bibitem [\protect \citeauthoryear {%
Molnar%
, Anderson%
\BCBL {}\ \BBA {} Anderson%
}{%
Molnar%
\ \protect \BOthers {.}}{%
{\protect \APACyear {2007}}%
}]{%
Molnar2007}
\APACinsertmetastar {%
Molnar2007}%
\begin{APACrefauthors}%
Molnar, P.%
, Anderson, R\BPBI S.%
\BCBL {}\ \BBA {} Anderson, S\BPBI P.%
\end{APACrefauthors}%
\unskip\
\newblock
\APACrefYearMonthDay{2007}{}{}.
\newblock
{\BBOQ}\APACrefatitle {Tectonics, fracturing of rock, and erosion} {Tectonics,
  fracturing of rock, and erosion}.{\BBCQ}
\newblock
\APACjournalVolNumPages{Journal of Geophysical Research: Earth
  Surface}{112}{F3}{}.
\PrintBackRefs{\CurrentBib}

\bibitem [\protect \citeauthoryear {%
Moresi%
, Dufour%
\BCBL {}\ \BBA {} M{\"u}hlhaus%
}{%
Moresi%
\ \protect \BOthers {.}}{%
{\protect \APACyear {2003}}%
}]{%
MoresiEtAl2003}
\APACinsertmetastar {%
MoresiEtAl2003}%
\begin{APACrefauthors}%
Moresi, L.%
, Dufour, F.%
\BCBL {}\ \BBA {} M{\"u}hlhaus, H\BHBI B.%
\end{APACrefauthors}%
\unskip\
\newblock
\APACrefYearMonthDay{2003}{{\APACmonth{01}}}{}.
\newblock
{\BBOQ}\APACrefatitle {A {{Lagrangian}} Integration Point Finite Element Method
  for Large Deformation Modeling of Viscoelastic Geomaterials} {A
  {{Lagrangian}} integration point finite element method for large deformation
  modeling of viscoelastic geomaterials}.{\BBCQ}
\newblock
\APACjournalVolNumPages{Journal of Computational Physics}{184}{2}{476--497}.
\PrintBackRefs{\CurrentBib}

\bibitem [\protect \citeauthoryear {%
Morley%
}{%
Morley%
}{%
{\protect \APACyear {1996}}%
}]{%
Morley1996}
\APACinsertmetastar {%
Morley1996}%
\begin{APACrefauthors}%
Morley, C\BPBI K.%
\end{APACrefauthors}%
\unskip\
\newblock
\APACrefYearMonthDay{1996}{{\APACmonth{01}}}{}.
\newblock
{\BBOQ}\APACrefatitle {Discussion of Potential Errors in Fault Heave Methods
  for Extension Estimates in Rifts, with Particular Reference to Fractal Fault
  Populations and Inherited Fabrics} {Discussion of potential errors in fault
  heave methods for extension estimates in rifts, with particular reference to
  fractal fault populations and inherited fabrics}.{\BBCQ}
\newblock
\APACjournalVolNumPages{Geological Society, London, Special
  Publications}{99}{1}{117--134}.
\PrintBackRefs{\CurrentBib}

\bibitem [\protect \citeauthoryear {%
Naliboff%
, Glerum%
, Brune%
, P{\'e}ron-Pinvidic%
\BCBL {}\ \BBA {} Wrona%
}{%
Naliboff%
\ \protect \BOthers {.}}{%
{\protect \APACyear {2020}}%
}]{%
Naliboff2020}
\APACinsertmetastar {%
Naliboff2020}%
\begin{APACrefauthors}%
Naliboff, J\BPBI B.%
, Glerum, A.%
, Brune, S.%
, P{\'e}ron-Pinvidic, G.%
\BCBL {}\ \BBA {} Wrona, T.%
\end{APACrefauthors}%
\unskip\
\newblock
\APACrefYearMonthDay{2020}{}{}.
\newblock
{\BBOQ}\APACrefatitle {Development of 3-D Rift Heterogeneity Through Fault
  Network Evolution} {Development of 3-d rift heterogeneity through fault
  network evolution}.{\BBCQ}
\newblock
\APACjournalVolNumPages{Geophysical Research Letters}{47}{13}{e2019GL086611}.
\newblock
\APACrefnote{e2019GL086611 2019GL086611}
\PrintBackRefs{\CurrentBib}

\bibitem [\protect \citeauthoryear {%
{Nemat-Nasser}%
\ \BBA {} Horii%
}{%
{Nemat-Nasser}%
\ \BBA {} Horii%
}{%
{\protect \APACyear {1982}}%
}]{%
Nemat-NasserHorii1982}
\APACinsertmetastar {%
Nemat-NasserHorii1982}%
\begin{APACrefauthors}%
{Nemat-Nasser}, S.%
\BCBT {}\ \BBA {} Horii, H.%
\end{APACrefauthors}%
\unskip\
\newblock
\APACrefYearMonthDay{1982}{{\APACmonth{08}}}{}.
\newblock
{\BBOQ}\APACrefatitle {Compression-Induced Nonplanar Crack Extension with
  Application to Splitting, Exfoliation, and Rockburst} {Compression-induced
  nonplanar crack extension with application to splitting, exfoliation, and
  rockburst}.{\BBCQ}
\newblock
\APACjournalVolNumPages{Journal of Geophysical Research: Solid
  Earth}{87}{B8}{6805--6821}.
\PrintBackRefs{\CurrentBib}

\bibitem [\protect \citeauthoryear {%
Olive%
, Behn%
, Mittelstaedt%
, Ito%
\BCBL {}\ \BBA {} Klein%
}{%
Olive%
\ \protect \BOthers {.}}{%
{\protect \APACyear {2016}}%
}]{%
OliveEtAl2016}
\APACinsertmetastar {%
OliveEtAl2016}%
\begin{APACrefauthors}%
Olive, J\BHBI A.%
, Behn, M\BPBI D.%
, Mittelstaedt, E.%
, Ito, G.%
\BCBL {}\ \BBA {} Klein, B\BPBI Z.%
\end{APACrefauthors}%
\unskip\
\newblock
\APACrefYearMonthDay{2016}{{\APACmonth{05}}}{}.
\newblock
{\BBOQ}\APACrefatitle {The Role of Elasticity in Simulating Long-Term Tectonic
  Extension} {The role of elasticity in simulating long-term tectonic
  extension}.{\BBCQ}
\newblock
\APACjournalVolNumPages{Geophysical Journal International}{205}{2}{728--743}.
\PrintBackRefs{\CurrentBib}

\bibitem [\protect \citeauthoryear {%
Pan%
, Naliboff%
, Bell%
\BCBL {}\ \BBA {} Jackson%
}{%
Pan%
\ \protect \BOthers {.}}{%
{\protect \APACyear {2022}}%
}]{%
Pan2022}
\APACinsertmetastar {%
Pan2022}%
\begin{APACrefauthors}%
Pan, S.%
, Naliboff, J.%
, Bell, R.%
\BCBL {}\ \BBA {} Jackson, C.%
\end{APACrefauthors}%
\unskip\
\newblock
\APACrefYearMonthDay{2022}{}{}.
\newblock
{\BBOQ}\APACrefatitle {Bridging Spatiotemporal Scales of Normal Fault Growth
  During Continental Extension Using High-Resolution 3D Numerical Models}
  {Bridging spatiotemporal scales of normal fault growth during continental
  extension using high-resolution 3d numerical models}.{\BBCQ}
\newblock
\APACjournalVolNumPages{Geochemistry, Geophysics,
  Geosystems}{23}{7}{e2021GC010316}.
\newblock
\APACrefnote{e2021GC010316 2021GC010316}
\PrintBackRefs{\CurrentBib}

\bibitem [\protect \citeauthoryear {%
Pan%
, Naliboff%
, Bell%
\BCBL {}\ \BBA {} Jackson%
}{%
Pan%
\ \protect \BOthers {.}}{%
{\protect \APACyear {2023}}%
}]{%
Pan2023}
\APACinsertmetastar {%
Pan2023}%
\begin{APACrefauthors}%
Pan, S.%
, Naliboff, J.%
, Bell, R.%
\BCBL {}\ \BBA {} Jackson, C.%
\end{APACrefauthors}%
\unskip\
\newblock
\APACrefYearMonthDay{2023}{}{}.
\newblock
{\BBOQ}\APACrefatitle {How Do Rift-Related Fault Network Distributions Evolve?
  Quantitative Comparisons Between Natural Fault Observations and 3D Numerical
  Models of Continental Extension} {How do rift-related fault network
  distributions evolve? quantitative comparisons between natural fault
  observations and 3d numerical models of continental extension}.{\BBCQ}
\newblock
\APACjournalVolNumPages{Tectonics}{42}{10}{e2022TC007659}.
\newblock
\APACrefnote{e2022TC007659 2022TC007659}
\PrintBackRefs{\CurrentBib}

\bibitem [\protect \citeauthoryear {%
Paterson%
\ \BBA {} Wong%
}{%
Paterson%
\ \BBA {} Wong%
}{%
{\protect \APACyear {2005}}%
}]{%
PatersonWong2005}
\APACinsertmetastar {%
PatersonWong2005}%
\begin{APACrefauthors}%
Paterson, M\BPBI S.%
\BCBT {}\ \BBA {} Wong, T\BHBI F.%
\end{APACrefauthors}%
\unskip\
\newblock
\APACrefYear{2005}.
\newblock
\APACrefbtitle {Experimental Rock Deformation: The Brittle Field} {Experimental
  rock deformation: The brittle field}\ (\BVOL~348).
\newblock
\APACaddressPublisher{}{{Springer}}.
\PrintBackRefs{\CurrentBib}

\bibitem [\protect \citeauthoryear {%
Perol%
\ \BBA {} Bhat%
}{%
Perol%
\ \BBA {} Bhat%
}{%
{\protect \APACyear {2016}}%
}]{%
PerolBhat2016}
\APACinsertmetastar {%
PerolBhat2016}%
\begin{APACrefauthors}%
Perol, T.%
\BCBT {}\ \BBA {} Bhat, H\BPBI S.%
\end{APACrefauthors}%
\unskip\
\newblock
\APACrefYearMonthDay{2016}{{\APACmonth{08}}}{}.
\newblock
{\BBOQ}\APACrefatitle {Micromechanics-{{Based Permeability Evolution}} in
  {{Brittle Materials}} at {{High Strain Rates}}} {Micromechanics-{{Based
  Permeability Evolution}} in {{Brittle Materials}} at {{High Strain
  Rates}}}.{\BBCQ}
\newblock
\APACjournalVolNumPages{Pure and Applied Geophysics}{173}{8}{2857--2868}.
\PrintBackRefs{\CurrentBib}

\bibitem [\protect \citeauthoryear {%
Poliakov%
\ \BBA {} Buck%
}{%
Poliakov%
\ \BBA {} Buck%
}{%
{\protect \APACyear {1998}}%
}]{%
PoliakovBuck1998}
\APACinsertmetastar {%
PoliakovBuck1998}%
\begin{APACrefauthors}%
Poliakov, A\BPBI N\BPBI B.%
\BCBT {}\ \BBA {} Buck, W\BPBI R.%
\end{APACrefauthors}%
\unskip\
\newblock
\APACrefYearMonthDay{1998}{}{}.
\newblock
{\BBOQ}\APACrefatitle {Mechanics of {{Stretching Elastic-Plastic-Viscous
  Layers}}: {{Applications}} to {{Slow-Spreading Mid-Ocean Ridges}}} {Mechanics
  of {{Stretching Elastic-Plastic-Viscous Layers}}: {{Applications}} to
  {{Slow-Spreading Mid-Ocean Ridges}}}.{\BBCQ}
\newblock
\BIn{} W.~Roger~Buck, P\BPBI T.~Delaney, J\BPBI A.~Karson\BCBL {}\ \BBA {}
  Y.~Lagabrielle\ (\BEDS), \APACrefbtitle {Geophysical {{Monograph Series}}}
  {Geophysical {{Monograph Series}}}\ (\BPGS\ 305--324).
\newblock
\APACaddressPublisher{{Washington, D. C.}}{{American Geophysical Union}}.
\PrintBackRefs{\CurrentBib}

\bibitem [\protect \citeauthoryear {%
Rackauckas%
\ \BBA {} Nie%
}{%
Rackauckas%
\ \BBA {} Nie%
}{%
{\protect \APACyear {2017}}%
}]{%
RackauckasNie2017}
\APACinsertmetastar {%
RackauckasNie2017}%
\begin{APACrefauthors}%
Rackauckas, C.%
\BCBT {}\ \BBA {} Nie, Q.%
\end{APACrefauthors}%
\unskip\
\newblock
\APACrefYearMonthDay{2017}{{\APACmonth{05}}}{}.
\newblock
{\BBOQ}\APACrefatitle {{{DifferentialEquations}}.Jl \textendash{} {{A
  Performant}} and {{Feature-Rich Ecosystem}} for {{Solving Differential
  Equations}} in {{Julia}}} {{{DifferentialEquations}}.jl \textendash{} {{A
  Performant}} and {{Feature-Rich Ecosystem}} for {{Solving Differential
  Equations}} in {{Julia}}}.{\BBCQ}
\newblock
\APACjournalVolNumPages{Journal of Open Research Software}{5}{1}{15}.
\PrintBackRefs{\CurrentBib}

\bibitem [\protect \citeauthoryear {%
Rudnicki%
\ \BBA {} Rice%
}{%
Rudnicki%
\ \BBA {} Rice%
}{%
{\protect \APACyear {1975}}%
}]{%
RudnickiRice1975}
\APACinsertmetastar {%
RudnickiRice1975}%
\begin{APACrefauthors}%
Rudnicki, J.%
\BCBT {}\ \BBA {} Rice, J.%
\end{APACrefauthors}%
\unskip\
\newblock
\APACrefYearMonthDay{1975}{{\APACmonth{12}}}{}.
\newblock
{\BBOQ}\APACrefatitle {Conditions for the Localization of Deformation in
  Pressure-Sensitive Dilatant Materials} {Conditions for the localization of
  deformation in pressure-sensitive dilatant materials}.{\BBCQ}
\newblock
\APACjournalVolNumPages{Journal of the Mechanics and Physics of
  Solids}{23}{6}{371--394}.
\PrintBackRefs{\CurrentBib}

\bibitem [\protect \citeauthoryear {%
Sammis%
\ \BBA {} Ashby%
}{%
Sammis%
\ \BBA {} Ashby%
}{%
{\protect \APACyear {1986}}%
}]{%
SammisAshby1986}
\APACinsertmetastar {%
SammisAshby1986}%
\begin{APACrefauthors}%
Sammis, C\BPBI G.%
\BCBT {}\ \BBA {} Ashby, M\BPBI F.%
\end{APACrefauthors}%
\unskip\
\newblock
\APACrefYearMonthDay{1986}{{\APACmonth{03}}}{}.
\newblock
{\BBOQ}\APACrefatitle {The Failure of Brittle Porous Solids under Compressive
  Stress States} {The failure of brittle porous solids under compressive stress
  states}.{\BBCQ}
\newblock
\APACjournalVolNumPages{Acta Metallurgica}{34}{3}{511--526}.
\PrintBackRefs{\CurrentBib}

\bibitem [\protect \citeauthoryear {%
Scholz%
, Dawers%
, Yu%
, Anders%
\BCBL {}\ \BBA {} Cowie%
}{%
Scholz%
\ \protect \BOthers {.}}{%
{\protect \APACyear {1993}}%
}]{%
ScholzEtAl1993}
\APACinsertmetastar {%
ScholzEtAl1993}%
\begin{APACrefauthors}%
Scholz, C.%
, Dawers, N.%
, Yu, J\BHBI Z.%
, Anders, M.%
\BCBL {}\ \BBA {} Cowie, P.%
\end{APACrefauthors}%
\unskip\
\newblock
\APACrefYearMonthDay{1993}{}{}.
\newblock
{\BBOQ}\APACrefatitle {Fault growth and fault scaling laws: Preliminary
  results} {Fault growth and fault scaling laws: Preliminary results}.{\BBCQ}
\newblock
\APACjournalVolNumPages{Journal of Geophysical Research: Solid
  Earth}{98}{B12}{21951--21961}.
\PrintBackRefs{\CurrentBib}

\bibitem [\protect \citeauthoryear {%
Spiegelman%
, May%
\BCBL {}\ \BBA {} Wilson%
}{%
Spiegelman%
\ \protect \BOthers {.}}{%
{\protect \APACyear {2016}}%
}]{%
SpiegelmanEtAl2016}
\APACinsertmetastar {%
SpiegelmanEtAl2016}%
\begin{APACrefauthors}%
Spiegelman, M.%
, May, D\BPBI A.%
\BCBL {}\ \BBA {} Wilson, C\BPBI R.%
\end{APACrefauthors}%
\unskip\
\newblock
\APACrefYearMonthDay{2016}{{\APACmonth{06}}}{}.
\newblock
{\BBOQ}\APACrefatitle {On the Solvability of Incompressible {{Stokes}} with
  Viscoplastic Rheologies in Geodynamics: {{ISSUES IN VISCOPLASTICITY}}} {On
  the solvability of incompressible {{Stokes}} with viscoplastic rheologies in
  geodynamics: {{ISSUES IN VISCOPLASTICITY}}}.{\BBCQ}
\newblock
\APACjournalVolNumPages{Geochemistry, Geophysics,
  Geosystems}{17}{6}{2213--2238}.
\PrintBackRefs{\CurrentBib}

\bibitem [\protect \citeauthoryear {%
Tada%
, Paris%
\BCBL {}\ \BBA {} Irwin%
}{%
Tada%
\ \protect \BOthers {.}}{%
{\protect \APACyear {1973}}%
}]{%
TadaEtAl1973}
\APACinsertmetastar {%
TadaEtAl1973}%
\begin{APACrefauthors}%
Tada, H.%
, Paris, P\BPBI C.%
\BCBL {}\ \BBA {} Irwin, G\BPBI R.%
\end{APACrefauthors}%
\unskip\
\newblock
\APACrefYearMonthDay{1973}{}{}.
\newblock
{\BBOQ}\APACrefatitle {The Stress Analysis of Cracks} {The stress analysis of
  cracks}.{\BBCQ}
\newblock
\APACjournalVolNumPages{Handbook, Del Research Corporation}{34}{}{635}.
\PrintBackRefs{\CurrentBib}

\bibitem [\protect \citeauthoryear {%
Tapponnier%
\ \BBA {} Brace%
}{%
Tapponnier%
\ \BBA {} Brace%
}{%
{\protect \APACyear {1976}}%
}]{%
TapponnierBrace1976}
\APACinsertmetastar {%
TapponnierBrace1976}%
\begin{APACrefauthors}%
Tapponnier, P.%
\BCBT {}\ \BBA {} Brace, W\BPBI F.%
\end{APACrefauthors}%
\unskip\
\newblock
\APACrefYearMonthDay{1976}{}{}.
\newblock
{\BBOQ}\APACrefatitle {Development.of {{Stress-Induced}}} {Development.of
  {{Stress-Induced}}}.{\BBCQ}
\newblock
\APACjournalVolNumPages{}{}{}{10}.
\PrintBackRefs{\CurrentBib}

\bibitem [\protect \citeauthoryear {%
Tarantola%
}{%
Tarantola%
}{%
{\protect \APACyear {2005}}%
}]{%
tarantola2005}
\APACinsertmetastar {%
tarantola2005}%
\begin{APACrefauthors}%
Tarantola, A.%
\end{APACrefauthors}%
\unskip\
\newblock
\APACrefYear{2005}.
\newblock
\APACrefbtitle {Inverse {{Problem Theory}} and {{Methods}} for {{Model
  Parameter Estimation}}} {Inverse {{Problem Theory}} and {{Methods}} for
  {{Model Parameter Estimation}}}.
\newblock
\APACaddressPublisher{}{{Society for Industrial and Applied Mathematics}}.
\PrintBackRefs{\CurrentBib}

\bibitem [\protect \citeauthoryear {%
Thomas%
, Bhat%
\BCBL {}\ \BBA {} Klinger%
}{%
Thomas%
\ \protect \BOthers {.}}{%
{\protect \APACyear {2017}}%
}]{%
ThomasEtAl2017}
\APACinsertmetastar {%
ThomasEtAl2017}%
\begin{APACrefauthors}%
Thomas, M\BPBI Y.%
, Bhat, H\BPBI S.%
\BCBL {}\ \BBA {} Klinger, Y.%
\end{APACrefauthors}%
\unskip\
\newblock
\APACrefYearMonthDay{2017}{{\APACmonth{06}}}{}.
\newblock
{\BBOQ}\APACrefatitle {Effect of {{Brittle Off-Fault Damage}} on {{Earthquake
  Rupture Dynamics}}} {Effect of {{Brittle Off-Fault Damage}} on {{Earthquake
  Rupture Dynamics}}}.{\BBCQ}
\newblock
\BIn{} M\BPBI Y.~Thomas, T\BPBI M.~Mitchell\BCBL {}\ \BBA {} H\BPBI S.~Bhat\
  (\BEDS), \APACrefbtitle {Geophysical {{Monograph Series}}} {Geophysical
  {{Monograph Series}}}\ (\BPGS\ 255--280).
\newblock
\APACaddressPublisher{{Hoboken, NJ, USA}}{{John Wiley \& Sons, Inc.}}
\PrintBackRefs{\CurrentBib}

\bibitem [\protect \citeauthoryear {%
Tsitouras%
, Simos%
, Psihoyios%
\BCBL {}\ \BBA {} Tsitouras%
}{%
Tsitouras%
\ \protect \BOthers {.}}{%
{\protect \APACyear {2009}}%
}]{%
TsitourasEtAl2009}
\APACinsertmetastar {%
TsitourasEtAl2009}%
\begin{APACrefauthors}%
Tsitouras, {\relax Ch}.%
, Simos, T\BPBI E.%
, Psihoyios, G.%
\BCBL {}\ \BBA {} Tsitouras, {\relax Ch}.%
\end{APACrefauthors}%
\unskip\
\newblock
\APACrefYearMonthDay{2009}{}{}.
\newblock
{\BBOQ}\APACrefatitle {Runge-{{Kutta Pairs}} of {{Orders}} 5(4) Using the
  {{Minimal Set}} of {{Simplifying Assumptions}}} {Runge-{{Kutta Pairs}} of
  {{Orders}} 5(4) using the {{Minimal Set}} of {{Simplifying
  Assumptions}}}.{\BBCQ}
\newblock
\BIn{} \APACrefbtitle {{{AIP Conference Proceedings}}} {{{AIP Conference
  Proceedings}}}\ (\BPGS\ 69--72).
\newblock
\APACaddressPublisher{{Rethymno, Crete (Greece)}}{{AIP}}.
\PrintBackRefs{\CurrentBib}

\bibitem [\protect \citeauthoryear {%
Vermeer%
\ \BBA {} De~Borst%
}{%
Vermeer%
\ \BBA {} De~Borst%
}{%
{\protect \APACyear {1984}}%
}]{%
VermeerDeBorst1984}
\APACinsertmetastar {%
VermeerDeBorst1984}%
\begin{APACrefauthors}%
Vermeer, P\BPBI A.%
\BCBT {}\ \BBA {} De~Borst, R.%
\end{APACrefauthors}%
\unskip\
\newblock
\APACrefYearMonthDay{1984}{}{}.
\newblock
{\BBOQ}\APACrefatitle {Non-Associated Plasticity for Soils, Concrete and Rock}
  {Non-associated plasticity for soils, concrete and rock}.{\BBCQ}
\newblock
\APACjournalVolNumPages{HERON}{29}{3}{}.
\PrintBackRefs{\CurrentBib}

\bibitem [\protect \citeauthoryear {%
Walsh%
}{%
Walsh%
}{%
{\protect \APACyear {1965a}}%
}]{%
Walsh1965a}
\APACinsertmetastar {%
Walsh1965a}%
\begin{APACrefauthors}%
Walsh, J\BPBI B.%
\end{APACrefauthors}%
\unskip\
\newblock
\APACrefYearMonthDay{1965a}{}{}.
\newblock
{\BBOQ}\APACrefatitle {The Effect of Cracks on the Compressibility of Rock}
  {The effect of cracks on the compressibility of rock}.{\BBCQ}
\newblock
\APACjournalVolNumPages{Journal of Geophysical Research}{70}{2}{381--389}.
\PrintBackRefs{\CurrentBib}

\bibitem [\protect \citeauthoryear {%
Walsh%
}{%
Walsh%
}{%
{\protect \APACyear {1965b}}%
}]{%
Walsh1965ba}
\APACinsertmetastar {%
Walsh1965ba}%
\begin{APACrefauthors}%
Walsh, J\BPBI B.%
\end{APACrefauthors}%
\unskip\
\newblock
\APACrefYearMonthDay{1965b}{}{}.
\newblock
{\BBOQ}\APACrefatitle {The Effect of Cracks on the Uniaxial Elastic Compression
  of Rocks} {The effect of cracks on the uniaxial elastic compression of
  rocks}.{\BBCQ}
\newblock
\APACjournalVolNumPages{Journal of Geophysical Research}{70}{2}{399--411}.
\PrintBackRefs{\CurrentBib}

\bibitem [\protect \citeauthoryear {%
Wang%
, Schubnel%
, Fortin%
, Gu{\'e}guen%
\BCBL {}\ \BBA {} Ge%
}{%
Wang%
\ \protect \BOthers {.}}{%
{\protect \APACyear {2013}}%
}]{%
WangEtAl2013}
\APACinsertmetastar {%
WangEtAl2013}%
\begin{APACrefauthors}%
Wang, X\BHBI Q.%
, Schubnel, A.%
, Fortin, J.%
, Gu{\'e}guen, Y.%
\BCBL {}\ \BBA {} Ge, H\BHBI K.%
\end{APACrefauthors}%
\unskip\
\newblock
\APACrefYearMonthDay{2013}{{\APACmonth{12}}}{}.
\newblock
{\BBOQ}\APACrefatitle {Physical Properties and Brittle Strength of Thermally
  Cracked Granite under Confinement: {{STRENGTH OF THERMALLY CRACKED GRANITE}}}
  {Physical properties and brittle strength of thermally cracked granite under
  confinement: {{STRENGTH OF THERMALLY CRACKED GRANITE}}}.{\BBCQ}
\newblock
\APACjournalVolNumPages{Journal of Geophysical Research: Solid
  Earth}{118}{12}{6099--6112}.
\PrintBackRefs{\CurrentBib}

\bibitem [\protect \citeauthoryear {%
Wawersik%
\ \BBA {} Brace%
}{%
Wawersik%
\ \BBA {} Brace%
}{%
{\protect \APACyear {1971}}%
}]{%
WawersikBrace1971}
\APACinsertmetastar {%
WawersikBrace1971}%
\begin{APACrefauthors}%
Wawersik, W\BPBI R.%
\BCBT {}\ \BBA {} Brace, W\BPBI F.%
\end{APACrefauthors}%
\unskip\
\newblock
\APACrefYearMonthDay{1971}{{\APACmonth{06}}}{}.
\newblock
{\BBOQ}\APACrefatitle {Post-Failure Behavior of a Granite and Diabase}
  {Post-failure behavior of a granite and diabase}.{\BBCQ}
\newblock
\APACjournalVolNumPages{Rock Mechanics}{3}{2}{61--85}.
\PrintBackRefs{\CurrentBib}

\bibitem [\protect \citeauthoryear {%
C.~Wu%
, Peng%
\BCBL {}\ \BBA {} {Ben-Zion}%
}{%
C.~Wu%
\ \protect \BOthers {.}}{%
{\protect \APACyear {2009}}%
}]{%
WuEtAl2009}
\APACinsertmetastar {%
WuEtAl2009}%
\begin{APACrefauthors}%
Wu, C.%
, Peng, Z.%
\BCBL {}\ \BBA {} {Ben-Zion}, Y.%
\end{APACrefauthors}%
\unskip\
\newblock
\APACrefYearMonthDay{2009}{{\APACmonth{01}}}{}.
\newblock
{\BBOQ}\APACrefatitle {Non-Linearity and Temporal Changes of Fault Zone Site
  Response Associated with Strong Ground Motion} {Non-linearity and temporal
  changes of fault zone site response associated with strong ground
  motion}.{\BBCQ}
\newblock
\APACjournalVolNumPages{Geophysical Journal International}{176}{1}{265--278}.
\PrintBackRefs{\CurrentBib}

\bibitem [\protect \citeauthoryear {%
X.~Wu%
, Baud%
\BCBL {}\ \BBA {} Wong%
}{%
X.~Wu%
\ \protect \BOthers {.}}{%
{\protect \APACyear {2000}}%
}]{%
WuEtAl2000}
\APACinsertmetastar {%
WuEtAl2000}%
\begin{APACrefauthors}%
Wu, X.%
, Baud, P.%
\BCBL {}\ \BBA {} Wong, T.%
\end{APACrefauthors}%
\unskip\
\newblock
\APACrefYearMonthDay{2000}{{\APACmonth{01}}}{}.
\newblock
{\BBOQ}\APACrefatitle {Micromechanics of Compressive Failure and Spatial
  Evolution of Anisotropic Damage in {{Darley Dale}} Sandstone} {Micromechanics
  of compressive failure and spatial evolution of anisotropic damage in
  {{Darley Dale}} sandstone}.{\BBCQ}
\newblock
\APACjournalVolNumPages{International Journal of Rock Mechanics and Mining
  Sciences}{37}{1-2}{143--160}.
\PrintBackRefs{\CurrentBib}

\bibitem [\protect \citeauthoryear {%
Zhang%
, Wong%
\BCBL {}\ \BBA {} Davis%
}{%
Zhang%
\ \protect \BOthers {.}}{%
{\protect \APACyear {1990}}%
}]{%
ZhangEtAl1990}
\APACinsertmetastar {%
ZhangEtAl1990}%
\begin{APACrefauthors}%
Zhang, J.%
, Wong, T\BHBI F.%
\BCBL {}\ \BBA {} Davis, D\BPBI M.%
\end{APACrefauthors}%
\unskip\
\newblock
\APACrefYearMonthDay{1990}{}{}.
\newblock
{\BBOQ}\APACrefatitle {Micromechanics of Pressure-Induced Grain Crushing in
  Porous Rocks} {Micromechanics of pressure-induced grain crushing in porous
  rocks}.{\BBCQ}
\newblock
\APACjournalVolNumPages{Journal of Geophysical Research}{95}{B1}{341}.
\PrintBackRefs{\CurrentBib}

\bibitem [\protect \citeauthoryear {%
Zoback%
\ \BBA {} Byerlee%
}{%
Zoback%
\ \BBA {} Byerlee%
}{%
{\protect \APACyear {1975}}%
}]{%
ZobackByerlee1975}
\APACinsertmetastar {%
ZobackByerlee1975}%
\begin{APACrefauthors}%
Zoback, M\BPBI D.%
\BCBT {}\ \BBA {} Byerlee, J\BPBI D.%
\end{APACrefauthors}%
\unskip\
\newblock
\APACrefYearMonthDay{1975}{{\APACmonth{02}}}{}.
\newblock
{\BBOQ}\APACrefatitle {The Effect of Microcrack Dilatancy on the Permeability
  of Westerly Granite} {The effect of microcrack dilatancy on the permeability
  of westerly granite}.{\BBCQ}
\newblock
\APACjournalVolNumPages{Journal of Geophysical Research}{80}{5}{752--755}.
\PrintBackRefs{\CurrentBib}

\end{thebibliography}

\end{document}